\catcode`\@=11
\newif\if@fewtab\@fewtabtrue

{\count255=\time\divide\count255 by 60
\xdef\hourmin{\number\count255}
\multiply\count255 by-60\advance\count255 by\time
\xdef\hourmin{\hourmin:\ifnum\count255<10 0\fi\the\count255}}
\def\ps@draft{\let\@mkboth\@gobbletwo
    \def\@oddhead{}
    \def\@oddfoot
       {\hbox to 7 cm{$\scriptstyle Draft\ version:\ \draftdate$
       \hfil}\hskip -7cm\hfil\rm\thepage \hfil}
    \def\@evenhead{}\let\@evenfoot\@oddfoot}

\def\ceqno{\global\@fewtabfalse
    \ifcase\@eqcnt \def\@tempa{& & &}\or \def\@tempa{& &}
      \or \def\@tempa{&}
      \or\def\@tempa{}\fi\@tempa
{\rm(\theequation)}}

\def\aeqno#1{\global\@fewtabfalse
    \ifcase\@eqcnt \def\@tempa{& & &}\or \def\@tempa{& &}
      \or \def\@tempa{&}
      \or\def\@tempa{}\fi\@tempa
{\rm(\theequation,#1)}}

\def\label#1{\ifnum\draftcontrol=1
 \global\def\draftnote{$\scriptstyle #1$}\fi
 \@bsphack\if@filesw {\let\thepage\relax
   \def\protect{\noexpand\noexpand\noexpand}%
\xdef\@gtempa{\write\@auxout{\string
      \newlabel{#1}{{\@currentlabel}{\thepage}}}}}\@gtempa
   \if@nobreak \ifvmode\nobreak\fi\fi\fi
  \@esphack}

\def\alabel#1#2{\label{#1}\global\@fewtabfalse
    \ifcase\@eqcnt \def\@tempa{& & &}\or \def\@tempa{& &}
      \or \def\@tempa{&}
      \or\def\@tempa{}\fi\@tempa
{\hbox to 3cm{\phantom{\rm(\theequation,#2)}
\draftnote \hfil}\hskip -3cm {\rm(\theequation,#2)}}}

\def\clabel#1{\label{#1}\global\@fewtabfalse
    \ifcase\@eqcnt \def\@tempa{& & &}\or \def\@tempa{& &}
      \or \def\@tempa{&}
      \or\def\@tempa{}\fi\@tempa
{\hbox to 3cm{\phantom{\rm(\theequation)}
\draftnote \hfil}\hskip -3cm{\rm(\theequation)}}}

\def\eqnarray{\def\draftnote{{}}\global\@fewtabtrue
\stepcounter{equation}\let\@currentlabel=\theequation
\global\@eqnswtrue
\global\@eqcnt\z@\tabskip\@centering\let\\=\@eqncr
$$\halign to \displaywidth\bgroup\@eqnsel\hskip\@centering\@eqcnt\z@
  $\displaystyle\tabskip\z@{##}$&\global\@eqcnt\@ne
  \hskip 1\arraycolsep \hfil${##}$\hfil
  &\global\@eqcnt\tw@ \hskip 1\arraycolsep
$\displaystyle\tabskip\z@{##}$
\hfil  \tabskip\@centering&\global\@eqcnt\thr@@\llap{##}\tabskip\z@
\cr}

\def\endeqnarray{\@@eqncr\egroup
      \global\advance\c@equation\m@ne$$\global\@ignoretrue}

\def\@eqnnum{\hbox to 3cm{\phantom{\rm(\theequation)} \draftnote
                         \hfil}\hskip -3cm {\rm(\theequation)}}

\def\@@eqncr{\let\@tempa\relax
    \ifcase\@eqcnt \def\@tempa{& & &}\or \def\@tempa{& &}
      \or \def\@tempa{&}
      \or\def\@tempa{}
\fi\@tempa
\if@eqnsw
\if@fewtab\@eqnnum\fi
\stepcounter{equation}\fi\global
\@eqnswtrue\global\@eqcnt\z@\global\@fewtabtrue\cr}

\def\draftcite#1{\ifnum\draftcontrol=1#1\else{}\fi}

\def\@lbibitem[#1]#2{\item{}\hskip -3cm \hbox to 2cm
{\hfil$\scriptstyle\draftcite{#2}$}\hskip
1cm[\@biblabel{#1}]\if@filesw
     {\def\protect##1{\string ##1\space}\immediate
      \write\@auxout{\string\bibcite{#2}{#1}}}\fi\ignorespaces}

\def\@bibitem#1{\item\hskip -3cm \hbox to 2cm
{\hfil $\scriptstyle\draftcite{#1}$}\hskip 1cm
\if@filesw \immediate\write\@auxout
       {\string\bibcite{#1}{\the\value{\@listctr}}}\fi\ignorespaces}

 \def\nsection#1{\section{#1}\setcounter{equation}{0}}

     \def\nappendix#1{\vskip 1cm\no{\Large\bf Appendix
         #1}\def\thesection{#1} \setcounter{equation}{0}}

\font\tendl=msbm10  scaled \magstep1
\font\sevendl=msbm7 scaled \magstep1
\font\fivedl=msbm5 scaled \magstep1
\font\tengl=eufm10  scaled \magstep1
\font\sevengl=eufm7 scaled \magstep1
\font\fivegl=eufm5 scaled \magstep1

\newfam\dlfam \def\dl{\fam\dlfam\tendl} 
\textfont\dlfam=\tendl \scriptfont\dlfam=\sevendl
\scriptscriptfont\dlfam=\fivedl
\newfam\glfam  
\textfont\glfam=\tengl \scriptfont\glfam=\sevengl
\scriptscriptfont\glfam=\fivegl

\def\draftdate{\number\month/\number\day/\number\year\ \ \ \hourmin }

\global\def\draftcontrol{0}
\catcode`\@=12
\def\tilde{\widetilde}
\def\hat{\widehat}

\documentstyle[11pt,epsf]{article}
\def\theequation{{\thesection.\arabic{equation}}}
\newcommand{\LM}{{L\hspace{-0.02cm}M}}
\newcommand{\IM}{{I\hspace{-0.03cm}M}}
\newcommand{\LG}{{L\hspace{-0.02cm}G}}
\newcommand{\IG}{{I\hspace{-0.02cm}G}}
\newcommand{\be}{\begin{eqnarray}}
\newcommand{\en}{\end{eqnarray}\vs 0.5 cm}

\newcommand{\no}{\noindent}
\newcommand{\vs}{\vskip}

\newcommand{\p}{\partial}

\newcommand{\NR}{{{\dl R}}}
\newcommand{\NA}{{{\dl A}}}
\newcommand{\NP}{{{\dl P}}}
\newcommand{\NC}{{{\dl C}}}
\newcommand{\NZ}{{{\dl Z}}}
\newcommand{\NH}{{{\dl H}}}
\newcommand{\NU}{{{\dl U}}}
\newcommand{\Ng}{{{\bf g}}}
\newcommand{\Nt}{{{\bf t}}}
\newcommand{\qq}{\begin{eqnarray}}

\newcommand{\WZ}{{W\hspace{-0.08cm}Z}}
\newcommand{\SL}{{S\hspace{-0.04cm}L}}
\newcommand{\ee}{{\rm e}}

\newcommand{\qqq}{\end{eqnarray}}

\newcommand{\tr}{{\mathrm{tr}}}

\newcommand{\CA}{{\cal A}}

\newcommand{\CC}{{\cal C}}
\newcommand{\CD}{{\cal D}}
\newcommand{\CE}{{\cal E}}

\newcommand{\CG}{{\cal G}}
\newcommand{\CH}{{\cal H}}

\newcommand{\CK}{{\cal K}}
\newcommand{\CL}{{\cal L}}

\newcommand{\CN}{{\cal N}}
\newcommand{\CO}{{\cal O}}
\newcommand{\CP}{{\cal P}}

\newcommand{\CU}{{\cal U}}
\newcommand{\CV}{{\cal V}}
\newcommand{\CW}{{\cal W}}

\newcommand{\CY}{{\cal Y}}
\newcommand{\CZ}{{\cal Z}}

\newcommand{\m}{\hspace{0.025cm}}

\newcommand{\hf}{{_1\over^2}}

\renewcommand{\thepage}{\hfill \arabic{page} \hfill}

\newcommand{\sk}{{\mathsf k}}
\newcommand{\ssk}{{_\sk\over^4}}
\newcommand{\si}{{\mathsf i}}
\begin{document}

\title{\bf Abelian \m and \m non-Abelian \m branes\\ 
in WZW models and gerbes}

\author{\ \\Krzysztof Gaw\c{e}dzki\footnote{membre du C.N.R.S.}\\ 
\ \\ Laboratoire de Physique,
ENS-Lyon,\\46, All\'ee d'Italie, F-69364 Lyon, France
}
\date{ }
\maketitle
\thispagestyle{empty}
\begin{center}
\end{center}

\begin{abstract}
\vskip 0.3cm \noindent We discuss how gerbes may be used to
set up a consistent Lagrangian approach to the WZW models with 
boundary. The approach permits to study in detail possible 
boundary conditions that restrict the values of the fields
on the worldsheet boundary to brane submanifolds in the target 
group. Such submanifolds are equipped with an additional geometric
structure that is summarized in the notion of a gerbe module
and includes a twisted Chan-Paton gauge field. Using the geometric 
approach, we present a complete classification of the branes 
that conserve the diagonal current-algebra symmetry in the WZW 
models with simple, compact but not necessarily simply connected 
target groups. Such symmetric branes are supported by a discrete 
series of conjugacy classes in the target group and may carry 
Abelian or non-Abelian twisted gauge fields. The latter situation 
occurs for the conjugacy classes with fundamental group $\,\NZ_2
\hspace{-0.06cm}\times\hspace{-0.06cm}\NZ_2\,$ in $\,SO(4n)/\NZ_2$. 
\,The branes supported by such conjugacy classes have to be 
equipped with a projectively flat twisted $\,U(2)\,$ gauge field 
in one of the two possible WZW models differing by discrete torsion. 
We show how the geometric description of branes leads to explicit 
formulae for the boundary partition functions and boundary operator 
product coefficients in the WZW models with non-simply connected 
target groups.
\end{abstract}
\vskip 1.5cm

\nsection{Introduction}
\label{sec:intro}

Boundary conformal field theories play an important role in the 
study of two- or $(1+1)$-dimensional critical phenomena in finite 
geometries and in the understanding of branes in string theory. 
In the latter context much of the original work has been 
done in the Lagrangian approach where one considers tense strings 
moving in flat or nearly flat space-times, with ends restricted 
to special submanifolds called Dirichlet branes or D-branes 
\cite{Polch}. From the worldsheet point of view,
this amounts to the study of conformally invariant boundary sigma 
models with flat or nearly flat target spaces and with fields 
on the worldsheet boundary restricted to take values in the 
brane submanifolds of the target. The latter are chosen 
in a way that assures that the boundary theory still has (half of) 
the conformal symmetry. Such boundary sigma models constitute, at 
least perturbatively, examples of boundary conformal 
field theories. On the non-perturbative level, the field theories 
of this type may be analyzed in algebraic terms with the powerful 
2d conformal symmetry toolkit \cite{DiFMS}, as initiated in \cite{Cardy}. 
In this way, boundary conformal field theory provides a formulation 
of the stringy or $D$-geometry \cite{Dougl1}. Understanding how 
this geometry modifies the standard geometry requires also 
a translation of as many of its aspects as possible to a more 
standard Lagrangian geometric language. This program has been 
implemented to a certain degree for the Calabi-Yau sigma models 
and the $N=2$ supersymmetric boundary conformal field theories, see 
\cite{Dougl1,Dougl2,Aspin}, where, due to supersymmetry, the relation 
between the stringy and the classical geometry may be traced a long 
way and is reflected in the mirror symmetry \cite{Konts}.
\vskip 0.3cm

Similar question may be studied for the Wess-Zumino-Witten (WZW) model,
a conformally invariant sigma model with a group manifold as the target
\cite{WZW}. Here, it is the rich current-algebra symmetry that allows 
to trace the relation between the stringy and the classical geometry. In 
the WZW model, one may access the stringy geometry via the exact 
solution of the corresponding boundary conformal field theory. On 
the other hand, one may interpret this solution within the Lagrangian 
approach. In particular, the WZW theories provide an important laboratory 
for studying stringy geometry of curved D-branes with non-trivial 
fluxes of the Kalb-Ramond $\,B$-field \cite{Schom}. In the simplest 
case, the so called (maximally) symmetric D-branes in the WZW models
with compact targets correspond to a discrete series of 
the integrable conjugacy classes in the group manifold \cite{KlimS,ASchom}. 
These are the conjugacy classes that contain elements 
$\,\ee^{2\pi\si\lambda/\sk}\,$ where $\,\sk\,$ is the level of the model 
(related to its coupling constant) and $\,\lambda\,$ is the highest weight 
of a chiral current-algebra primary field of the theory. For the so called 
diagonal WZW models with simply connected target groups, there is a unique 
integrable conjugacy class for each $\,\lambda\,$ and a unique maximally
symmetric conformal boundary condition associated to it \cite{ASchom}, 
see also \cite{Ist}. This results in a one-to-one correspondence between 
the primary fields and the boundary conditions, as predicted in \cite{Cardy}.  
The situation is more complex for the non-diagonal WZW models 
corresponding to non-simply connected target groups. The set of 
maximally symmetric conformal boundary conditions and the exact solution 
for the boundary model have been described in \cite{FHSSW,FRSIII,Walch} 
in terms of the so called simple current structure \cite{KSchell}, see 
also \cite{BPPZ,PZ,GG1}. This description exposes a more complicated 
relation between the integrable conjugacy classes and the boundary 
conditions, with occurrence of non-trivial multiplicities of the latter 
for a fixed conjugacy class and with a possibility of spontaneous 
generation of non-Abelian gauge-type structure \cite{FKLLSW}. In 
particular, ref.\,\,\cite{FRSIII}stressed the role of the finite 
group cohomology in the boundary model. The aim of the present paper 
is to provide a geometric picture of the classification of the symmetric 
branes for the WZW models with non-simply connected targets by carefully 
setting up the Lagrangian approach to such models in the presence 
of boundaries.    
\vskip 0.3cm

The characteristic feature of the WZW model in the Lagrangian formulation
is the presence of the coupling to a topologically non-trivial Kalb-Ramond 
background 2-form field $\,B\,$ on the target group. \,This field is defined 
locally, only with its exterior derivative $\,H=dB\m$, \,equal to the standard 
invariant 3-form on the group, making sense globally. As noted in 
\cite{WZW}, 
the fact that the 3-form $\,H\,$ is not exact so that there is no global 
$\,B\,$ field allows a consistent definition of the classical amplitudes 
in the bulk theory (i.e.\,\,on worldsheets without boundary) only if 
the periods of the 3-form $\,H\,$ are in $\,2\pi\NZ$. \,This leads to 
the quantization of the the coupling constant of the WZW model that
was analyzed in \cite{WZW} for simply connected target groups and
in \cite{FGK} for non-simply connected ones. The careful Lagrangian 
formulation of the theory permitted a direct calculation of the quantum 
spectra of the bulk WZW models \cite{GW,FGK}. 
\vskip 0.3cm

An idea of how to define the bulk amplitudes for a general sigma model
coupled to a closed 3-form $\,H\,$ on the target with periods in 
$\,2\pi\NZ\,$ was proposed in \cite{Alva}. This proposal was reformulated 
in terms of the $3^{\rm rd}$ degree Deligne cohomology in \cite{Gaw0}. 
The recent years have brought a realization that such cohomology 
classifies geometric objects over the target manifolds called (Abelian) 
bundle gerbes with connection \cite{Murray,Chatt,Hitchin}\footnote{see 
\cite{Gir} and \cite{Bryl} for the categorial avatars of bundle gerbes}. 
The closed 3-form $\,H\,$ provides the curvature form of the gerbe. 
In the presence of boundaries, the $\,B$-field ambiguities lead to new 
phenomena and result in the quantization of possible D-brane boundary 
conditions first observed in \cite{KlimS} and \cite{ASchom} for the case 
of the WZW models with simply connected targets. The geometric language 
of gerbes appears quite useful in dealing with the intricacies 
of the open string amplitudes in the presence of topologically 
non-trivial $\,B$-fields. In particular, the Lagrangian description 
of the Chan-Paton coupling of the ends of the open string to a non-Abelian 
Yang-Mills field on the D-brane leads in a natural way to the notion 
of gerbe modules, as first noted in ref.\,\,\cite{Kapust} inspired
by \cite{FreWit}, see also \cite{CJM} (in \cite{Kapust} such modules
were viewed as modules over the Azumaya algebras). The gerbe modules 
also appear in a definition \cite{BCMMS} of the twisted $\,K$-theory 
groups containing the Ramond-Ramond charges of the supersymmetric D-branes 
\cite{WittKT,Moore}.   
\vskip 0.3cm

The construction of the bundle gerbes with curvature proportional to the 
invariant 3-form over the $\,SU(2)\,$ group may be traced back to 
\cite{Gaw0} and predates the geometric definition of the notion. The bundle 
gerbes over $\,SU(N)\,$ were constructed in \cite{Chatt} and, independently
but later, in \cite{GR1}. The extension of this construction to other simple, 
compact, simply connected groups required overcoming additional difficulties 
and was achieved in \cite{Meinr}, see \cite{Bryl1,Behr} for a related work.
The bundle gerbes over non-simply connected groups were described in 
\cite{Gaw0} for the $\,SO(3)\,$ group, in 
\cite{GR1} for the groups covered by $\,SU(N)\,$ and in \cite{GR2} for all 
non-simply connected, simple, compact Lie groups. As is well known,
see e.g.\,\,\cite{Kost}, pushing down a line bundle from manifold 
$\,M\,$ to its quotient by free action of a finite group $\,Z\,$ requires 
solving a cohomological equation $\,\delta W=V\,$ that expresses a 2-cocycle 
$\,V\,$ on group $\,Z\,$ as a boundary, with the cohomology class 
of $\,V\,$ representing the obstruction. Similar considerations for gerbes 
\cite{Sharpe,RT} lead to a cohomological equation $\,\delta V=U\,$ in one 
degree higher, with the obstruction to pushing down a gerbe to the 
quotient space represented by the cohomology class of the 3-cycle 
$\,U\,$ on $\,Z$. \,Calculating the 3-cocycles $\,U\,$ and solving 
the equation $\,\delta V=U\,$ whenever possible for all quotients 
$\,G'\,$ of simply connected, simple, compact groups $\,G\,$ by subgroups 
$\,Z\,$ of their center was the essence of the work done in \cite{GR2}. 
We shall crucially depend in the present paper on those results. 
In \cite{GR1}, the gerbes over $\,SU(N)\,$ and its quotients were used 
to classify the symmetric D-branes of the WZW models with these groups as 
the target. Such branes consisted of an integrable conjugacy class in 
the group equipped with additional geometric structure which included 
an Abelian gauge field on the conjugacy class. Pushing down the line 
bundle carrying the gauge field from a conjugacy class in $\,SU(N)\,$ 
to a conjugacy class in the quotient group required solving a 
cohomological equation $\,\delta W=V\m$, \,as discussed above.  
\vskip 0.3cm

Below, we shall extend the geometric classification of the symmetric 
D-branes to WZW models with target groups that are simple, compact but 
not necessarily simply connected. As in the $\,SU(N)\,$ case, 
such branes are supported by integrable conjugacy classes. 
The general case will require to include in the additional structure 
a projectively flat twisted non-Abelian gauge field on the conjugacy 
class. The adequate formulation is based on the notion of gerbe modules, 
also called twisted vector bundles \cite{Kapust,Mackaay,BCMMS}.
More exactly, we shall classify, up to isomorphism, the possible gerbe 
modules over the integrable conjugacy classes $\,\CC\,$ with the gerbe 
over $\,\CC\,$ obtained from the gerbe over the group by restriction. 
For simply connected groups there is a unique 1-dimensional gerbe module 
corresponding to a symmetric brane over each integrable conjugacy class. 
All higher dimensional ones are its direct sums. The construction
of the gerbe modules over integrable conjugacy classes $\,\CC'\,$
in the non-simply connected quotient groups $\,G'=G/Z\,$ leads to a 
cohomological equation $\,\delta\CW=\CV\,$ expressing a 2-cocycle 
$\,\CV\,$ on the fundamental group of the conjugacy class $\,\CC'\,$ 
as a boundary, similarly as for pushing to quotients usual line 
bundles. The cohomology class of $\,\CV\,$ is the obstruction to the 
existence of a 1-dimensional gerbe module over $\,\CC'\,$ giving rise 
to a symmetric brane. If the obstruction class is trivial then such 
1-dimensional modules exist and different ones differ by tensor
multiplication by flat line bundles over $\,\CC'$. \,Higher dimensional 
modules are then direct sums of 1-dimensional ones. If the obstruction 
class is non-trivial, there are still matrix-valued solutions of the 
equation $\,\delta\CW=\CV\,$ and they give rise to higher dimensional 
gerbe modules over the conjugacy classes $\,\CC'\,$ that do not 
decompose into a direct sum of 1-dimensional ones. This mimics the 
situation for line bundles which may always be pushed down to vector 
bundles over regular discrete quotients. The matrix-valued solutions 
$\,\CW\,$ give rise to symmetric branes carrying a projectively 
flat twisted non-Abelian gauge field. Such a situation occurs for the 
conjugacy classes with fundamental group $\,\NZ_2\hspace{-0.06cm}\times
\hspace{-0.06cm}\NZ_2\,$ in the $\,SO(4n)/\NZ_2\,$ groups. These are 
groups that admit two non-equivalent gerbes distinguished by discrete 
torsion \cite{Vafa,Sharpe}, as first realized in \cite{FGK} where 
the ambiguity was identified as periodic vacua. For one of those 
gerbes, the possible gerbe modules leading to symmetric branes 
are direct sums of 2-dimensional modules that cannot be decomposed 
further. Similar spontaneous enhancement of the gauge symmetry has been 
previously observed at orbifold fixed points of the Calabi-Yau sigma 
models in \cite{Dougl,DouglF}. Within the algebraic approach based on 
simple current structure to non-diagonal boundary WZW and coset models, 
it was related in \cite{FKLLSW} to the presence of projective 
representations of the simple current symmetries. We recover such 
representations in the description of the 2-dimensional gerbe modules 
corresponding to the D-branes in $\,SO(4n)/\NZ_2$. 
\vskip 0.3cm

The geometric constructions, although based on classical considerations, 
lead through geometric quantization to simple expressions for the boundary 
partition functions and the boundary operator product coefficients in the 
quantum WZW models with non-simply connected target groups, as noted 
in \cite{GR1}. Such expressions result from a realization of the open 
string states of the model with a non-simply connected target group 
$\,G'=G/Z\,$ as the states of the model with the covering group $\,G\,$ 
target that are invariant under an action of the fundamental group $\,Z$. 
\,The existence of such an action and the above picture become clear 
in the geometric realization of the states of the models, see Sect.\,9.2 of 
\cite{GR1}. Although they may be, and will be, formulated within a more 
standard algebraic approach, they employ the cohomological structures 
that appear naturally in the geometric classification of the branes 
discussed here showing the pertinence of the latter for the quantum
theory. 
\vskip 0.3cm
  
The paper is organized as follows. In Sect.\,\ref{sec:abgbm}, we
present the notions of bundle gerbes and gerbe modules in a
relatively pedestrian fashion, with stress on the local description. 
How those notions are employed to define the contributions 
to the action functional of the topologically non-trivial 
$\,B\,$ field and the Chan-Paton coupling to the non-Abelian 
gauge fields on the branes in the presence of such a $\,B\,$ field 
is the subject of Sect.\,\ref{sec:wzaf}. The next two sections are
the most technical ones. In Sect.\,\ref{sec:wzwmgcg} we recall 
briefly the construction of the gerbes over Lie groups relevant 
for the treatment of the $\,B\,$ field contributions in the WZW 
models. We discuss the case of simple compact groups, both
simply connected \cite{Meinr} and non-simply connected \cite{GR2}.
The symmetric branes and their geometric classification are studied
in Sect.\,\ref{sec:sbwzwm}, first for the simply connected groups
(in Sect.\,\ref{sec:sbscg}) and then for the non-simply connected ones 
(in Sect.\,\ref{sec:sbnscg}). The latter discussion, culminating in
considerations leading to the cohomological equation 
$\,\delta\CW=\CV\,$ constitutes the main part of the present 
work. We have not found yet a more concise way to describe the 
solutions $\,\CW\,$ for all cases with trivial cohomology class
of $\,\CV\,$ and have opted for tabulating the results in Appendix A
where the complete list of symmetric branes carrying Abelian gauge 
fields is given. In Sect.\,\ref{sec:nasb}, we discuss 
the special case of symmetric branes in $\,SO(4n)/\NZ_2\,$ that carry 
a twisted non-Abelian gauge field. Sect.\,\ref{sec:bpfop}
contains a discussion of the boundary space of states in the WZW models,
of the boundary partition functions and the boundary operator product.
It completes the discussion of Sects.\,9 and 10 of \cite{GR1}.
Conclusions summarize the results of the paper and draw perspectives 
for the future research. Appendix B provides a direct check that 
the action of elements of the group center on the multiplicity 
spaces of the boundary theory with simply connected target is well 
defined and Appendix C shows how the general formulae of 
Sect.\,\ref{sec:bpfop} permit to obtain the boundary partition 
functions and the boundary operator product coefficients for the WZW
model with the $\,SO(3)\,$ target.
\vskip 0.3cm

{\bf Acknowledgements.} \ This work has stemmed from research conducted
initially in collaboration with Nuno Reis. It was partially done in 
framework of the European contract EUCLID/HPRN-CT-2002-00325.

\nsection{Abelian bundle gerbes and gerbe modules}
\label{sec:abgbm}

\subsection{Local data}
\label{sec:ld}

We start by a quick introduction to Abelian bundle gerbes and gerbe modules.
It is simplest to specify such objects by local data. For the bundle gerbes, 
the local approach was developed long before the geometric approach 
of \cite{Murray,MurrS}, see \cite{Alva,Gaw0}. For bundle-gerbe modules 
\cite{BCMMS}, also called Azumaya algebra modules or twisted vector bundles, 
the local approach was discussed in \cite{Kapust}, see also \cite{Mackaay}. 
\vskip 0.3cm

Let $\,(O_i)\,$ be a good open covering 
of a manifold $\,M\,$ (i.e. such that $\,O_i\,$ and all non-empty 
intersections $\,O_{i_1\dots i_n}=O_{i_1}\cap\cdots\cap O_{i_{n}}\,$ are
contractible). \,We shall call a family $\,(B_i,A_{ij},g_{ijk})\,$
local data for an Abelian hermitian bundle gerbe with connection 
(in short, {\bf gerbe local data}) if $\,B_i\,$ are 2-forms on 
$\,O_i\m$, $\,A_{ij}=-A_{ji}\,$
are 1-forms on $\,O_{ij}\,$ and $\,g_{ijk}=g_{\sigma(i)\sigma(j)\sigma(k)}
^{sgn(\sigma)}\,$ are $\,U(1)\,$ valued functions on $\,O_{ijk}\,$
such that
\qq
&&{\hbox to 6cm{$B_j-\,B_i\ =\ dA_{ij}$\hfill}{\rm on}\ \ O_{ij}\,,}
\alabel{ldg}{a}\cr
&&{\hbox to 6cm{$A_{jk}-\,A_{ik}+\,A_{ij}\ =\ i\,g_{ijk}^{-1}dg_{ijk}$\hfill}
{\rm on}\ \ O_{ijk}\,,}\aeqno{b}\cr
&&{\hbox to 6cm{$g_{jkl}\,g_{ikl}^{-1}\,g_{ijl}\,g_{ijk}^{-1}\ =\ 1$\hfill}
{\rm on}\ \ O_{ijkl}\,.}\aeqno{c}\cr
\qqq
\vskip -0.5cm
\noindent The global closed 3-form $\,H\,$ equal to $\,dB_i\,$ on 
$\,O_i\,$ is called the {\bf curvature} of the corresponding data.
Local data with curvature $\,H\,$ exist if and only if
the 3-periods of $\,H\,$ are in $\,2\pi\NZ$. \,Writing 
$\,g_{ijk}=\ee^{2{\pi\si} f_{ijk}}\,$ for real-valued functions
$\,f_{ijk}\,$ one obtains an integer-valued 3-cocycle
\qq
n_{ijkl}\ =\ f_{jkl}-\,f_{ikl}+\,f_{ijl}-\,f_{ijk}\quad\ {\rm on}\ \,O_{ijkl}
\nonumber
\qqq
that defines a cohomology class in $\,H^3(M,\NZ)\m$, 
\,called the {\bf Dixmier-Douady} (DD) {\bf class} of the gerbe local data.
The image of the DD-class in $\,H^3(M,\NR)\,$ coincides with the de Rham 
cohomology class of $\,H$. 
\vskip 0.3cm

One says that two families $\,(B_i,A_{ij},g_{ijk})\,$ and
$\,(B'_i,A'_{ij},g'_{ijk})\,$ of gerbe local data are {\bf equivalent} 
if
\qq
B'_i\ &=&B_i+\,d\pi_i\,,\alabel{eq}{a}\cr
A'_{ij}\,&=&A_{ij}+\,\pi_j-\,\pi_i-\,i\,\chi_{ij}^{-1}d\chi_{ij}\,,
\aeqno{b}\cr
g'_{ijk}&=&g_{ijk}\,\chi_{ik}\,\chi_{jk}^{-1}\,\chi_{ij}^{-1}\,,
\aeqno{c}\cr
\qqq
\vskip -0.5cm
\noindent 
where $\,\pi_i\,$ are 1-forms on $\,O_i\,$ and
$\,\chi_{ij}=\chi_{ji}^{-1}\,$ are $\,U(1)$-valued functions
on $\,O_{ij}$. \,The DD-classes 
of equivalent local data coincide but one may have
non-equivalent local data that correspond to the same DD-class.
\vskip 0.3cm

Given $\,(B_i,A_{ij},g_{ijk})\,$ as above, we shall call a family
$\,(\Pi_i,G_{ij})\,$ local data for an $N$-dimensional 
hermitian gerbe module with connection (in short: {\bf gerbe-module local 
data}) if
$\,\Pi_i\,$ are $\,u(N)$-valued\footnote{$u(N)\,$ denotes the Lie algebra of
hermitian $N\times N$ matrices} 1-forms on $\,O_i\,$ and $\,G_{ij}\,$
are $\,U(N)$-valued maps on $\,O_{ij}\,$ such that
\qq
&&\hbox to 7.7cm{$\Pi_j-\,G_{ij}^{-1}\Pi_i\,G_{ij}\,-\,i\,
G_{ij}^{-1}dG_{ij}+A_{ij}\ =\ 0$\hfill}{\rm on}\ \ O_{ij}\,,\alabel{gm0}{a}\cr
&&\hbox to 7.7cm{$G_{ik}\,G_{jk}^{-1}\,G_{ij}^{-1}\,g_{ijk}\ =\ 1$\hfill}
{\rm on}\ \ O_{ijk}\,.\aeqno{b}\cr
\qqq
\vskip -0.5cm
\noindent Recall that local data of a hermitian vector bundle with 
connection satisfy similar relations
with $\,A_{ij}=0\,$ and $\,g_{ijk}=1$. \,Two families of gerbe-module 
local data $\,(\Pi_i,G_{ij})\,$ and $\,(\Pi'_i,G'_{ij})\,$
are said to be {\bf equivalent} if
\qq
\Pi'_{i}\,&=&H_i^{-1}\,\Pi_{i}\,H_i\,+\,i\,H_i^{-1}dH_i\,,\alabel{ebm}{a}
\cr
G'_{ij}&=&H_i^{-1}\,G_{ij}\,H_j\aeqno{b}\cr
\qqq
\vskip -0.5cm
\noindent for $\,U(N)$-valued functions $\,H_i\,$ on $\,O_i$.
Given gerbe-module local data $\,(\Pi_i,G_{ij})\,$ relative 
to gerbe local data $\,(B_i,A_{ij},g_{ijk})$ and equivalent
local gerbe data $\,(B'_i,A'_{ij},g'_{ijk})\m$, \,the family
$\,(\Pi_i-\pi_i,\,G_{ij}\chi_{ij}^{-1})\,$ provides a gerbe-module
local data relative to $\,(B'_i,A'_{ij},g'_{ijk})$. \,We shall call
such data {\bf induced} from $\,(\Pi_i,G_{ij})\,$ by the equivalence
of local gerbe data. 
\vskip 0.3cm

Gerbe-module local data exist only if and only if the DD-class
of the associated gerbe local data is pure torsion, i.e.\,\,if
the curvature $\,H\,$ is an exact form, a very stringent requirement. 
The only if part may be be easily deduced from (\ref{gm0}) by taking 
its trace and exterior derivative to infer that $\,B_i+{1\over N}\,
\tr\,\Pi_i\,$ define a global form on $\,M\,$ whose exterior derivative 
is equal to $\,H$. \,The if part may be inferred from the discussion in
\cite{BCMMS}. 
\vskip 0.3cm

It is sometimes convenient to reduce the structure group
of a gerbe module (as for the standard vector bundles) 
from $\,U(N)\,$ to a compact group $\,G\,$ assumed here 
to contain a distinguished central $\,U(1)\,$ subgroup. 
This is done by requiring in the definitions above that $\,\Pi_i\,$ 
are $\,\Ng$-valued forms, where $\,\Ng\,$ is the Lie algebra of $\,G\m$,
\,and $\,G_{ij}\,$ and $\,H_{i}\,$ are $\,G$-valued. One may
then return to the previous situation by composition with an $N$-dimensional
unitary representation $\,U\,$ of $\,G\,$ acting as identity on the
$\,U(1)\,$ subgroup of $\,G$. 

\subsection{Geometric definitions}
\label{sec:gd}

One may construct geometric objects from local data $\,(B_i,A_{ij},g_{ijk})$. 
\,Let $\,Y=\mathop{\sqcup}\limits_i O_i\,$ be the disjoint union of $\,O_i\,$ 
and $\,\pi\,$ be the natural map $\,(m,i)\mapsto m\,$ from $\,Y\,$ to $\,M$. 
\,Local 2-forms $\,B_i\,$ define a global 2-form $\,B\,$ on $\,Y\,$ such that 
\qq
dB\ =\ \pi^*H\,.
\label{g1}
\qqq
Denote by $\,Y^{[n]}\,$ the n-fold fiber product $\,Y\times_M\cdots\times_MY$.
Clearly, $\,Y^{[n]}=\mathop{\sqcup}\limits_{(i_1,\dots,i_n)}
O_{i_1\dots i_n}\,$ and we shall use the notation $\,(m,i_1,\dots,i_n)\,$
for its elements. Consider the trivial line bundle $\,L=Y^{[2]}\times\NC\,$ 
over $\,Y^{[2]}\,$ with the hermitian structure inherited from $\,\NC$. 
\,It may be equipped with a (unitary) connection with local connection 
forms equal to $\,A_{ij}\,$ on $\,O_{ij}$. \,The curvature form $\,F\,$ 
of this connection satisfies
\qq
F\ =\ p_2^*B\,-\,p_1^*B
\label{g2}
\qqq
where $\,p_i\,$ denote the projections in $\,Y^{[n]}\,$ on the 
$\,i^{\,\rm th}\,$ factor. Finally, $\,L\,$ may be equipped with a groupoid 
multiplication
\qq
L_{(y,y')}\times L_{(y',y'')}\ \mathop{\longrightarrow}^\mu\ L_{(y,y'')}
\label{g3}
\qqq
defined by $\,\mu((m,i,j;z),(m,j,k;z'))=(m,i,k;\,g_{ijk}(m)\,z\,z')$.
\vskip 0.3cm

In general, $\,\CG=(Y,B,L,\mu)\,$ is called an Abelian hermitian bundle gerbe
with connection over $\,M\,$ (in short: a {\bf gerbe}), with curvature
$\,H\m$, \,if $\,\pi:Y\rightarrow X\,$ is a surjective submersion, $\,B\,$
is a 2-form on $\,Y\,$ satisfying (\ref{g1}) called the gerbe's {\bf curving} 
and \,$L\,$ is a hermitian line bundle
with connection over $\,Y^{[2]}\,$ satisfying (\ref{g2}) and equipped with
a bilinear {\bf groupoid multiplication} (\ref{g3}) preserving hermitian 
structure and parallel transport. In particular, the unity of the groupoid 
multiplication defines the canonical isomorphisms $\,L_{(y,y)}\cong\NC\,$ 
and $\,L_{(y',y)}\cong L_{(y,y')}^*$. See \cite{Murray} for a more formal 
definition. The most trivial example of a gerbe is the
{\bf unit gerbe} $\,(M,0,M\times\NC,\cdot\,)\,$ with vanishing curvature 
and with the groupoid structure in the trivial line bundle $\,L\,$ over 
$\,M^{[2]}\cong M\,$ determined by the product in $\,\NC$. 
\vskip 0.3cm

As we have seen, there is a gerbe corresponding to each family of local
data. Conversely, given a gerbe $\,\CG\m$, \,one may associate to it
(non-canonically) a family of local data. In order to do this,
one chooses local maps $\,\sigma_i:O_i\rightarrow Y\,$ such that 
$\,\pi\circ\sigma_i=id\,$ and local sections $\,s_{ij}:\sigma_{ij}(O_{ij})
\rightarrow L\m$, \,where $\,\sigma_{ij}(m)=(\sigma_i(m),\sigma_j(m))\m$,
\,such that $\,s_{ji}\circ\sigma_{ji}=(s_{ij}\circ\sigma_{ij})^{-1}$.
Then the relations
\qq
&&{B}_i\,=\,\sigma_i^*{B}\,,\alabel{gtold0}{a}\cr
&&\sigma_{ij}^*(\nabla s_{ij})\,=\,{_1\over^\si}
{A}_{ij}\,s_{ij}\circ\sigma_{ij}\,,\aeqno{b}\cr
&&\mu\circ(s_{ij}\circ\sigma_{ij} \otimes
s_{jk}\circ\sigma_{jk})\,=\, g_{ijk}\,\,s_{ik}\circ\sigma_{ik}
\,.\aeqno{c}\cr \qqq 
\vskip -0.5cm 
\noindent determine gerbe local data $\,(B_i,A_{ij},g_{ijk})\,$
with curvature $\,H$. \,Different choices of $\,\sigma_i\,$ and 
$\,s_{ij}\,$ lead to equivalent local data. We shall call the DD-class 
associated to such local data the DD-class of the gerbe.
\vskip 0.3cm

Similarly, one may associate a geometric object to gerbe-module
local data $\,(\Pi_i,G_{ij})\,$ corresponding to gerbe local data
$\,(B_i,A_{ij},g_{ijk})$. \,Again with $\,Y=\mathop{\sqcup}\limits_i
O_i\m$, \,one considers the $N$-dimensional hermitian vector bundle  
$\,E=\mathop{\sqcup}\limits_iO_i\times\NC^N\,$ over $\,Y\,$ equipped
with a (unitary) connection given over $\,O_i\,$ by the
local connection 1-forms $\,\Pi_i$. \,Let 
\qq
\rho:\,L\otimes p_2^*E\ \longrightarrow\ p_1^*E
\nonumber
\qqq
be the isomorphism of the vector bundles over 
$\,Y^{[2]}\,$ such that $\,\rho((m,i,j;z)\otimes(m,j;v))=(m,i;G_{ij}(m)v)$.
\,Isomorphism $\,\rho\,$ is associative w.r.t. the groupoid multiplication 
$\,\mu\,$:
\qq
\rho\circ(\mu\otimes id)\ =\ \rho\circ({id}\otimes\rho)\,.
\label{vbi}
\qqq
\vskip 0.3cm

More generally, given a gerbe $\,\CG=(Y,B,L,\mu)\m$, \,a 
pair $\,(E,\rho)=\CE\,$
is called an $N$-dimensional hermitian $\,\CG$-module with connection
(in short: a $\,\CG${\bf-module}) if $\,E\,$ is an $N$-dimensional
vector bundle with (unitary) connection over $\,Y\,$ and $\,\rho\,$
is a vector bundle isomorphism as above, preserving hermitian structure and 
connection and satisfying (\ref{vbi}). 
Two $\CG$-modules $\,(E,\rho)\,$ and $\,(E',\rho')$ are called 
{\bf isomorphic} if there exists an isomorphism 
$\,\iota:E\rightarrow E'\,$ of hermitian 
vector bundles with connection such that
\qq
p_1^*\iota\circ\rho\ =\ \rho'\circ({id}\otimes p_2^*\iota)\,.
\nonumber
\qqq
The unit-gerbe modules coincide with hermitian vector bundles with connection 
over $\,M$.
\vskip 0.3cm

As we have seen, there is a $\,\CG$-module associated to a family 
$\,(\Pi_i,G_{ij})\,$ of gerbe-module local data corresponding to gerbe local 
data $\,(B_i,A_{ij},g_{ijk})\m$, \,where gerbe $\,\CG\,$ is constructed from 
the latter. \,Conversely, given a gerbe $\,\CG\,$ and a $\,\CG$-module
$\,\CE=(E,\rho)\m$, \,upon a choice of sections $\,\sigma_i\,$ and $\,s_{ij}\,$
for the former and of orthonormal bases $\,(e^a_i)\,$ of sections 
of $\,E\,$ over $\,\sigma_i(O_i)\m$, \,the relations 
\qq
&&\sigma_i^*\nabla e^a_i\ =\ {_1\over^\si}\,\Pi^{ba}_i\otimes e_i^b\,,
\alabel{gm}{a}\cr
&&\rho\circ(s_{ij}\circ\sigma_{ij}\otimes e^a_j\circ\sigma_j)\ 
=\ G^{ba}_{ij}\,e_i^b\circ\sigma_i\aeqno{b}\cr
\qqq
\vskip -0.6cm
\noindent define local gerbe-module data $\,(\Pi_i,G_{ij})$. 
\,Isomorphic $\,\CG$-modules give rise to equivalent local data
and conversely. 
\vskip 0.3cm

To reduce the structure group from $\,U(N)\,$ to a compact group
$\,G\,$ containing a central $\,U(1)\,$ subgroup, \,we may consider 
pairs $\,(P,\rho_{_P})\,$ where
$\,P\,$ is a principal $\,G$-bundle with connection over $\,Y\,$ 
and  
\qq
\rho_{_P}:\,\SL\times_{Y^{[2]}}p_2^*P\ \longrightarrow\ p_1^*P
\nonumber
\qqq
is an isomorphism of principal $G$-bundles with connection over 
$\,Y^{[2]}$, \,with $\,\SL\,$ denoting the circle subbundle of $\,L$. 
\,We assume that over $\,Y^{[3]}$,
\qq
\rho_{_P}\circ(\mu\times id)\ =\ \rho_{_P}\circ({id}\times\rho_{_P})\,.
\nonumber
\qqq
This implies that for each $\,y\in Y\m$, \,there is an action 
of $\,U(1)\cong \SL_{(y,y)}\,$ on $\,P_y\,$ and we shall require
that it coincide with the restriction to $\,U(1)\,$ of the action of $\,G\,$
on $\,P_y$. \,We shall call such pairs $\,(P,\rho_{_P})\,$ twisted
$\,G$-bundles. \,Given an $N$-dimensional unitary representation $\,U\,$ 
of $\,G\,$ acting as identity on the $\,U(1)\,$ subgroup,
the pair $\,(E,\rho)\m$, \,where $\,E=P\times_G\NC^N\,$ is the associated 
vector bundle and $\,\rho\,$ is naturally induced by $\,\rho_{_P}\m$, 
\,forms a $\,\CG${\bf-module associated to} $\,(P,\rho_{_P})$.
\,We shall demand that the isomorphisms $\,\iota\,$ of the associated 
$\,\CG$-modules be induced from the isomorphisms of the principal bundles 
that intertwine the maps $\rho_{_P}$. \,As for usual bundles, each 
$N$-dimensional gerbe module may be viewed as associated to its orthogonal 
frame bundle which is a twisted $\,U(N)$-bundle. 
\vskip 0.3cm

There are natural operations on gerbes and gerbe modules.
One defines the {\bf dual gerbe} $\,\CG^*=(Y,-B,L^*, (\mu^*)^{-1})\,$ 
with curvature $\,-H\,$ and the {\bf tensor product} of two gerbes 
$$\,\CG\otimes\CG'=(Y\times_MY',\,p^*B+{p'}^*B',\,(p^{[2]})^*L
\otimes({p'}^{[2]})^*L',\,\mu\otimes\mu')\,$$
with curvature $\,H+H'$ where $\,p,\,p'\,$ stand for
the projections of $\,Y\times_MY'\,$ onto $\,Y\,$ and $\,Y'$, 
\,respectively. Note that the tensor product of $\,\CG\,$ with 
the unit gerbe may be naturally identified with $\,\CG$.
\,For $\,f:N\rightarrow M\m$, $\,f^*\CG\,$ will denote
the {\bf pullback gerbe} (with curvature $\,f^*H$) and for $\,N\subset M\m$,
$\,\CG|_N\,$ the {\bf restriction} of $\,\CG\,$ to $\,N\,$ (the pullback 
by the embedding map). Similarly, for a $\,\CG$-module $\,\CE=(E,\rho)\m$, 
\,one may define a dual $\,\CG^*$-module 
$\,\CE^*=(E^*,(\rho^*)^{-1})\,$ as well as the pullbacks and the 
restrictions of $\,\CE$. \,If $\,\CE'=(E',\rho')\,$ is
another $\,\CG$-module then one may form the {\bf direct sum} $\,\CG$-module
$\,\CE\oplus\CE'=(E\oplus E',\rho\oplus\rho')$. \,Moreover, if 
$\,\CE'=(E',\rho')\,$ is a $\,\CG'$-module, then 
$\,\CE\otimes\CE'=(p^*E\otimes{p'}^*E',\rho\otimes\rho')\,$ 
is a $\,(\CG\otimes\CG')$-module,
the {\bf tensor product} of $\,\CE\,$ and $\,\CE'$. \,If there
exists a 1-dimensional $\,\CG$-module $\,\CE_1\,$ then any $\,\CG$-module
is isomorphic to a tensor product of $\,\CE_1\,$ by a unit-gerbe
module (i.e.\,\,by a vector bundle over $\,M$). We shall often use this 
property below.
\vskip 0.3cm

\,Given two gerbes $\,\CG,\ \CG'\,$ over $\,M\m$, \,a 1-dimensional 
$\,(\CG\otimes{\CG'}^*)$-module $\,\CN=(N,\nu)\,$ such that the curvature 
of the connection of $\,N\,$ is equal to the 2-form $\,{p'}^*B'-p^*B\,$ 
on $\,Y\times_MY'$ is called a {\bf stable isomorphism} between $\,\CG\,$ and
$\,\CG'\,$ and the gerbes for which such a module exists are said to be 
{\bf stably isomorphic} \cite{MurrS}. 
Given a stable isomorphism between $\,\CG\,$ and $\,\CG'$,
any $\,\CG'\,$ module $\,(E',\rho')\,$ induces
in a natural way a $\,\CG$-module $\,(E,\rho)\,$ where
$\,E_y=N_{(y,y')}\otimes E_{y'}\,$ for $\,(y,y')\in Y\times_MY'\,$
(the latter spaces are identified for different $\,y'\,$ with the use 
of the $\,\CG\otimes\CG'^*\,$ module structure on $\,\CN$).
One may this way compare gerbe modules over stably isomorphic gerbes.
In particular, two gerbe modules induced by the same stable isomorphism 
are themselves isomorphic if and only if the original gerbe modules are. 
\vskip 0.3cm

Gerbes over $\,M\,$ with curvature $\,H\,$ exist if and only if
the periods of $\,H\,$ are in $\,2\pi\NZ$. \,Two gerbes are stably 
isomorphic if and only if their local data are equivalent. 
For $\,H=0\m$, \,the set of stable isomorphism classes
forms a group under tensor product that may be identified
with the cohomology group $\,H^2(M,U(1))\,$ by considering
gerbes corresponding to local data $\,(0,0,u_{ijk})\m$, \,where
$\,u_{ijk}\,$ is a $\,U(1)$-valued $\check{\rm C}$ech 2-cocycle.
Two stable isomorphism classes of gerbes with the same curvature
differ by tensor multiplication by a class of zero curvature,
i.e. by an element of $\,H^2(M,U(1))$.
\vskip 0.3cm

$\CG$-modules exist if and only if the DD-class of $\,\CG\,$
is torsion i.e. if the curvature $\,H\,$ of $\,\CG\,$ is exact.
Two $\,\CG$-modules are isomorphic if and only if their local 
data associated to the same local data of $\,\CG\,$ are equivalent.

\nsection{Wess-Zumino action functional}
\label{sec:wzaf}

Let $\,\Sigma\,$ denote a 2-dimensional compact oriented surface and
let $\,M\,$ be a manifold. In 2d sigma models, $\,\Sigma\,$ plays
the role of a 2d spacetime or the string worldsheet and  $\,M\,$ that 
of the target space, with classical fields $\,\phi\,$ mapping 
$\,\Sigma\,$ to $\,M$. \,In such models that appear in conformal 
field theory or in string theory, for example in the celebrated WZW model 
\cite{WZW}, one would like to consider the Wess-Zumino type 
contributions to the action functional of fields $\,\phi\,$ that may 
symbolically be written as
\qq
S_{\WZ}(\phi)\ =\ \int\limits_\Sigma\phi^*d^{-1}\hspace{-0.08cm}H
\label{WZ}
\qqq
where $\,H\,$ is a closed 3-form on $\,M$. \,The problem with such
topological\footnote{i.e.\,\,independent of the metric on $\,\Sigma$} 
Wess-Zumino terms of the action is that the (Kalb-Ramond) 2-form 
$\,B=d^{-1}\hspace{-0.03cm}H\,$ such that $\,dB=H\,$ does not exist 
globally if $\,H\,$ is not exact and even if it is, $\,B\,$
is not unique. This is at this place that the gerbes and bundle gerbes 
show their utility \cite{Gaw0,Kapust}. 

\subsection{Closed string amplitudes}
\label{sec:csa}

Let $\,\CG\,$ be a gerbe with curvature $\,H$. \,Then $\,\phi^*\CG\,$
is a gerbe over $\,\Sigma\,$ (with curvature vanishing for dimensional
reasons) and its stable equivalence class defines a cohomology class in
$\,H^2(\Sigma,U(1))$. \,For $\,\Sigma\,$ without boundary (such surfaces 
describe bulk 2d space-time geometry or closed string worldsheets), 
$\,H^2(\Sigma,U(1))=U(1)\,$ and one may define the closed string 
amplitude $\,\CA(\phi)=\ee^{\,\si\m S_{WZ}(\phi)}\,$ as equal to 
$\,\phi^*\CG\,$ viewed as an element of $\,U(1)\,$ in a direct 
generalization of the notion of holonomy in line bundles \cite{Gaw0}, 
see also \cite{Picken}. The amplitude defined this way depends 
only on the stable isomorphism class of $\,\CG$. \,Given local 
data $\,(B_i,A_{ij},g_{ijk})\,$ and a sufficiently fine triangulation
of $\,\Sigma\,$ with each triangle $\,c\,$ contained in some
subset $\,O_{i_c}\m$, \,one has \cite{Alva,Gaw0}
\qq
\CA(\phi)\ =\ \exp\Big[i\sum\limits_c\int\limits_c\phi^*{B}_{i_c}
+i\sum\limits_{b\subset c}\int\limits_b\phi^*{A}_{i_ci_b}\Big]
\prod\limits_{v\in b\subset c}g_{i_ci_bi_v}(\phi(v))\,,
\label{Ampl} 
\qqq
where edges $\,b\,$ and vertices $\,v\,$ of the triangulation are contained 
in $\,O_{i_b}\,$ and $\,O_{i_v}\m$, \,respectively. In the product over 
vertices on the right hand side, the convention is assumed that inverts the 
entry if the orientation of $\,v\,$ inherited from $\,c\,$ via $\,b\,$ is 
negative (i.e. if $\,b\,$ is oriented from $\,v\,$ to another vertex). 
This formula makes explicit the corrections needed to add to the sum of local
contribution $\,\int\limits_c\phi^*B_{i_c}\,$ to the action for closed
surfaces $\,\Sigma$.

\subsection{Open string amplitudes: coupling to the Yang-Mills field}
\label{sec:asacymf}

Surfaces with boundary describe 2d space-time geometry with boundaries or 
open-string worldsheets (with closed strings propagating in the handles). 
Suppose that $\,\partial\Sigma 
=\mathop{\sqcup}\limits_s\ell^s\,$ where $\,\ell^s\,$ are the boundary loops 
that we shall, for convenience, parametrize by the standard circle $\,S^1$. 
Applying the same formula (\ref{Ampl}) to define the amplitude $\,\CA(\phi)\,$
as in the closed string case, one observes that now the right hand side 
changes upon a change of the triangulation and the assignments $\,i_b\,$ 
and $\,i_v\,$ on the boundary. These changes may be summarized by saying that
\qq
\ee^{\,\si\m S_{\WZ}(\phi)}\ \in\ \mathop{\otimes}\limits_s\CL_{\phi|_{\ell^s}}
\nonumber
\qqq
where $\,\CL\,$ is a hermitian line bundle with connection over the loop 
space $\,\LM\,$ determined canonically by the local data 
$\,(B_i,A_{ij},g_{ijk})\,$ of gerbe $\,\CG\,$ \cite{Gaw0}. In order to 
obtain number-valued
open-string amplitudes, one may couple the ends of the open string 
to the (twisted) Yang-Mills field represented by a $\,\CG$-module 
$\,\CE\,$ with local data $\,(\Pi_i,G_{ij})$. \,The coupling
is done by multiplying the amplitude $\,\ee^{\,\si\m S_{\WZ}(\phi)}\,$ 
by the traces of the appropriately defined holonomy of the boundary loops, 
see \cite{Kapust,Mackaay,CJM}. More precisely, define for a (pointed) loop 
$\,\varphi:S^1\rightarrow M\,$ its holonomy by
\qq
\CH(\varphi)\ =\ \CP\,\prod\limits_{v\in b}G_{i_vi_b}(\varphi(v))\,
\exp\Big[\,i\int\limits_{b}\varphi^*\Pi_{i_b}\Big]
\label{hol}
\qqq
for a sufficiently fine triangulation (split) of $\,S^1\,$ by intervals 
$\,b\,$ with vertices $\,v\,$ such that $\,\varphi(b)\subset O_{i_b}\,$ and 
$\,\varphi(v)\in O_{i_v}$. \,The operator $\,\CP\,$ orders the terms
and the exponetial path-wise (i.e. in agreement with the standard orientation 
of $\,S^1$) from right to left, starting from the vertex $\,{1}$. \,Again, 
the matrix $\,G_{i_vi_b}\,$ should be inverted if $\,v\,$ has negative 
orientation. Below, we shall use the same expression
to define the parallel transport along open lines. For loops, let 
$\,\CW(\phi)=\tr\,\CH(\phi)\,$ be the corresponding Wilson loop 
``observable''. \,Note that $\,\CW(\varphi)\,$ does not change if we use 
in its definition local data 
$\,(\Pi'_i,G'_{ij})\,$ equivalent to $\,(\Pi_i,G_{ij})$. \,A straightforward 
check shows that under the change of the triangulation and its indexing, 
$\,\CW(\varphi)\,$ transforms as an elements of $\,\CL_\varphi^{-1}$. \,It 
follows that 
\qq
\CA(\phi)\ =\ \ee^{\,\si\m S_\WZ(\phi)}\,
\prod\limits_s \CW(\phi|_{\ell^s})\,,
\label{actWL}
\qqq
is independent of the choice of the indexed triangulation of $\,\Sigma\m$,
\,provided that in the definition of $\,\CW(\phi|_{\ell^s})\,$ one uses the 
indexed triangulations of the boundary loops induced from
the indexed triangulation of $\,\Sigma$. 
\,The amplitude  $\,\CA(\phi)\,$ depends only on the 
stable isomorphism class of $\,\CG\,$ and does not change if we use 
equivalent or induced local data for $\,\CE$. \,If $\,\CE\,$
is an associated gerbe module, then the holonomy $\,\CH\,$ may be viewed 
as taking values in group $\,G$. \,We shall use then the notation 
$\,\CW_{_U}\,$ for the Wilson loop obtained by taking the trace of
$\,\CH\,$ in representation $\,U\,$ of $\,G$.
\vskip 0.4cm

\subsection{Branes and general open string amplitudes}
\label{sec:bgosa}

In the above definition of $\,\CA(\phi)\,$ one may use 
holonomies with respect to different $\,\CG$-modules for different boundary 
components. As we have seen above, existence of $\,\CG$-modules puts
a strong restriction on gerbe $\,\CG\,$ requiring that its curvature
be an exact form. Even if not satisfied globally, such a property
might hold for the restriction of $\,\CG\,$ to a submanifold of $\,M$.
\,We shall call a pair $\,\CD=(D,\CE)\m$, \,where $\,D\,$ is a 
submanifold of $\,M\,$ and $\,\CE\,$ is an $\,N$-dimensional 
$\,\CG|_D$-modules, a $\,\CG${\bf-brane} 
(supported by $\,D\m$, \,of rank $\,N$). \,Two $\,\CG$-branes will
be called isomorphic if they have the same support $\,D\,$ and if
the corresponding $\,\CG|_D$-modules are isomorphic. \,One may apply 
the above construction of amplitude $\,\CA(\phi)\,$ if for each boundary 
component of $\,\Sigma\,$ we are given a $\,\CG$-brane $\,\CD^s\,$ 
with support $\,D^s$, provided that we impose on maps $\,\phi\,$ 
the boundary conditions
\qq
\phi(\ell^s)\ \subset\ D^s\,.
\label{bcb}
\qqq
\vskip 0.3cm
 
We shall need to define amplitudes $\,\CA(\phi)\,$ in a still more general
situations \cite{Ist,GR1}. Let $\,\varphi:[0,\pi]\rightarrow M\,$ be an 
open line extending between submanifolds $\,D^0\,$ and $\,D^1$, 
\,i.e.\,\,such that
\qq
\varphi(0)\,\in\,D^0\,,\qquad\varphi(\pi)\,\in\,D^1\,.
\nonumber
\qqq
We shall denote by $\,\IM_{\hspace{-0.05cm}D^0}^{D^1}\,$ the space 
of such lines.
Given $\,N_s$-dimensional $\,\CG|_{D^s}$-modules $\,\CE^s$, $\,s=0,1\m$, \,one 
may naturally define a hermitian vector bundle with connection 
$\CE_{\hspace{-0.02cm}\CD^0}^{\CD^1}\,$ over 
$\,\IM_{\hspace{-0.05cm}D^0}^{D^1}\,$
with typical fiber $\,\NC^{N_0}\otimes\overline{\NC^{N_1}}$. 
\,This is done in a way that is a straightforward generalization of the 
construction from \cite{GR1} where the case of 1-dimensional modules 
has been treated, see Sects.\,\,7 and 11.2 therein. 
Suppose now that the map $\,\phi:\Sigma\to M\,$ satisfies the boundary 
conditions (\ref{bcb}) for $\,\ell^s\,$ being closed disjoint sub-intervals 
of the boundary loops of $\,\Sigma\m$, \,see Fig.\,\,1. 

\leavevmode\epsffile[-105 -20 215 185]{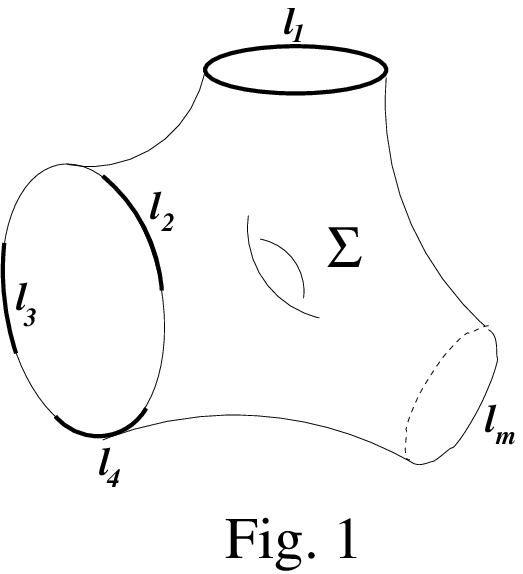}

\noindent Then
\qq
\CA(\phi)\,&=&\,\ee^{\,\si\m S_\WZ(\phi)}\,\,\Big({\mathop{\prod}\limits_s}'
\,\CW(\phi|_{\ell^s}\Big)\otimes\Big(\CP\,\,{\mathop{\otimes}\limits_s}''\, 
\CH(\phi|_{\ell^s})\Big)\cr\cr
&\in&\,\ \Big(
\mathop{\otimes}\limits_m\CL_{\phi|_{\ell^m}}\Big)\otimes
\Big(\CP\mathop{\otimes}\limits_{(s_-,s_+)}\,
(\CE_{\hspace{-0.02cm}\CD^{s_-}}^{\CD^{s_+}})_{\phi|_{\ell^{s_-s_+}}}\Big)\,,
\label{gma}
\qqq
where the product $\,\prod'\,$ is over 
 $\,\ell^s\,$ that cover entirely a boundary component and the tensor
product $\,\otimes''\,$ over the proper sub-intervals of boundary 
components, with the parallel transport $\,\CH\,$ given by the same 
formula (\ref{hol}) as for closed loops. In the second line,
$\,\ell^m\,$ run through the boundary loops without labeled 
sub-intervals and pairs $\,(s_-,s_+)\,$ label subsequent pairs 
of proper sub-intervals of the boundary loops separated by unlabeled
intervals $\,\ell^{s_-s_+}$. \,This is the higher 
rank version of the relation (7.11) of \cite{GR1}.

\subsection{Purely classical approach}
\label{sec:pca}

In the last incarnation, the topological amplitudes $\,\CA(\phi)\m$, 
\,multiplied by the more standard non-topological\footnote{i.e.\,\,depending
on the metric of $\,\Sigma$} contributions, enter the functional integral 
expressions for general quantum amplitudes of the boundary two-dimensional 
field theory with the Wess-Zumino term (\ref{WZ}) in the action. By 
themselves, they are, however, objects belonging to a mixed classical-quantum 
realm, the parallel transport or holonomy $\,\CH\,$ being operator-valued, 
i.e. a quantum-mechanical concept. One may go back one step and obtain
$\,\CA(\phi)\,$ from entirely classical amplitudes. This discussion
is somewhat parenthetical with respect to the main discourse of the 
paper and may be safely omitted by going directly to 
Sect.\,\ref{sec:wzwmgcg}. 
\vskip 0.3cm

Let us first recall that the standard holonomy in
the principal $\,G$-bundle $\,P\,$ over $\,M\,$ may be obtained by 
quantizing a mechanical system with the action 
$\,S_{\hspace{-0.03cm}_P}\,$ given by the connection form. More exactly, let
$\,\varphi\,$ be a loop in $\,M\,$ and 
$\,t\mapsto\Phi(t)\,$ its lift to a loop in $\,P$. 
\,In a local trivialization where $\,P=O\times G\m$,
$\,\Phi(t)=(\varphi(t),\gamma(t))\,$ and 
\qq
S_{\hspace{-0.03cm}_P}(\Phi)\ 
=\ \int\tr\,\lambda\left(i\,\gamma^{-1}d\gamma\,+
\,\gamma^{-1}(\varphi^*A)\,\gamma\right)\,,
\nonumber
\qqq
where $\,A\,$ is the local connection form with values in the Lie algebra 
$\,{\bf g}\,$ of $\,G$. \,Above, $\,\tr\,$ denotes a bilinear $\,ad$-invariant 
form on $\,{\bf g}\,$ and $\,\lambda\,$ is a highest weight in the Cartan 
subalgebra $\,\Nt\subset{\bf g}\,$ defining a unitary representation 
$\,U_\lambda\,$ of $\,G$. \,Note that the expression for the action 
is invariant under the local gauge transformations
\qq
\gamma\ \mapsto\ (\varphi^*h)^{-1}\gamma\,,\qquad 
A\ \mapsto\ h^{-1}A\,h\,+\,i\,h^{-1}dh
\nonumber
\qqq
induced by a change of trivialization of $\,P\,$ so that the action 
is globally defined. The Wilson loop $\,\CW_{_{\hspace{-0.03cm}
U_\lambda}}(\varphi)\,$ 
in representation $\,U_\lambda\,$ may be expressed as a path integral 
(that may be given a precise sense with the help of a Feynman-Kac type
formula) over the lifts $\,\Phi\m$: 
\qq
\CW_{_{\hspace{-0.03cm}U_\lambda}}(\varphi)\ 
=\ \int\ee^{\,\si\m S_{\hspace{-0.01cm}P}(\Phi)}\,\,D_\varphi\Phi
\nonumber
\qqq
where, locally, the measure $\,D_\varphi\Phi=D\gamma$. 
\vskip 0.3cm

The same approach may be generalized to the case of Wilson loops
in gerbe modules. Let $\,\CE\,$ be a $\,\CG$-module associated
to a twisted $\,G$-bundle $\,(P,\rho_{_P})$. \,Observe that 
$\,P/U(1)\,$ descends to a $\,G'$-bundle $\,P'\,$ over 
$\,M\,$ for $\,G'=G/U(1)$.
\,Let $\,\varphi\,$ be a loop in $\,M\,$ and $\,\Phi'\,$ its lift
to a loop in $\,P'$. \,Choose local data $\,(\Pi_i,G_{ij})\,$
for $\,\CE$. \,They provide local trivializations 
$\,P'|_{O_i}\cong O_i\times G'$. \,For a triangulation of $\,S^1\,$ 
by intervals $\,b\,$ with common vertices $\,v\,$ such that
$\,\varphi(b)\subset O_{i_b}\,$ and $\,\varphi(v)\in O_{i_v}\m$,
\,choose maps $\,\gamma_b:b\mapsto G\,$ such that 
$\,(\varphi|_b,\gamma_bU(1))\,$ represent $\,\Phi'|_b\,$ in the local
trivialization of $\,P'|_{O_{i_b}}$. \,Then for two subsequent 
intervals $\,b_-,b_+\,$ with common vertex $\,v\m$,
\qq
\gamma_{b_+}(v)\,\,u_v\ =\ \,G_{i_{b_+}i_v}(\varphi(v))
\,\,G_{i_vi_{b_-}}(\varphi(v))\,\,\gamma_{b_-}(v)
\nonumber
\qqq
for some $\,u_v\in U(1)$. \,Define
\qq
\ee^{\,\si\m S_P(\Phi')}\ =\ \exp\Big[i\m\sum\limits_b\int\limits_b
\tr\,\lambda\left(i\,\gamma_b^{-1}d\gamma_b+\,\gamma_b^{-1}
(\varphi^*\Pi_{i_b})\,\gamma_b\right)\Big]\,\prod\limits_{v} u_v\,,
\nonumber
\qqq
for the highest weight $\,\lambda\,$ satisfying $\,\tr\,\lambda\,
{1\over\si}{d\over{dt}}\Big|_{0}\ee^{\si t}=1$.
\,It is easy to check that the above expression is independent of the
choice of $\,\gamma_b$. \,On the other hand, under the change of indexed 
triangulations, it behaves as an element 
in $\,\CL_\varphi^{-1}$. \,As for the untwisted case, the path integral 
of $\,\ee^{\,\si\m S_P(\Phi')}\,$ over the lifts $\,\Phi'\,$ reproduces 
the Wilson loop $\,\CW_{_{\hspace{-0.03cm}U_\lambda}}(\varphi)$.
\vskip 0.3cm

Let now $\,\phi:\Sigma\to M\,$ and
$\,(P^s,\rho_{_{\hspace{-0.03cm}P^s}})\,$ be twisted $\,G^s$-bundles. 
Then the purely classical amplitude 
\qq
\CA_{cl}\left(\phi,({\Phi'}^s)\right)\ =\ \ee^{\,\si\m S_\WZ(\phi)}\,
\prod\limits_s\ee^{\,\si\m S_{P^s}({\Phi'}^s)}\,,
\nonumber
\qqq
where $\,{\Phi'}^s\,$ are lifts to $\,{P'}^s\,$ of the boundary loops
$\,\phi|_{\ell^s}\m$, \,is unambiguously determined. If the twisted 
bundles $\,(P^s,\rho_{_{\hspace{-0.03cm}P^s}})\,$ are defined 
only over submanifold 
$\,D^s\subset M\m$, \,one should impose the boundary conditions (\ref{bcb})
on the map $\,\phi$. \,The path integral over the lifts $\,{\Phi'}^s\,$ of the 
classical amplitudes $\,\CA_{cl}\left(\phi,({\Phi'}^s)\right)\,$ reproduces 
the mixed amplitudes (\ref{actWL}).
It is not difficult to generate similar path-integral representations  
also for the general mixed amplitudes (\ref{gma}). Unlike
the mixed formalism, the purely classical one lends itself to
canonical quantization. This is the main reason why it is useful
to consider it.

\nsection{WZW models and gerbes over compact groups}
\label{sec:wzwmgcg}

In the WZW models of conformal field theory 
\cite{WZW}, the classical (Euclidean) fields $\,\phi\,$ map 2-dimensional 
Riemann surfaces into the manifold of a Lie group $\,G\,$ that we shall
assume here to be connected, compact and simple.  The action functional
is taken to be
\qq
S(\phi)\ =\ {_\sk\over^{4\pi}}\int\limits_\Sigma\tr\,(\phi^{-1}\p \phi)
(\phi^{-1}\bar\p \phi)
\ +\ \int\limits_\Sigma\phi^*d^{-1}\hspace{-0.04cm}H
\nonumber
\qqq
where the closed 3-form $\,H\,$ on $\,G\,$ entering the Wess-Zumino term
of the action is given by 
\qq
H\ =\ {_\sk\over^{12\pi}}\,\tr\,(g^{-1}dg)^3\,.
\label{Gcurv}
\qqq
The bilinear form $\,\tr\,$ on the Lie algebra $\,\Ng\,$ will
be normalized so that for simply connected groups $\,G\,$ the 3-form $\,H\,$ 
has periods in $\,2\pi\NZ\,$ if and only if $\,\sk\,$ (called the
level of the model) is an integer. As discussed in the previous section,
in order to fully define the amplitudes $\,\ee^{\,\si\m S(\phi)}\,$ for closed 
surfaces $\,\Sigma\,$ we shall need a gerbe $\,\CG_\sk\,$ over $\,G\,$ with 
curvature 3-form $\,H$.

\subsection{Gerbes over simply connected groups}
\label{sec:gscg}

If $\,G\,$ is a (connected, compact, simple) simply connected group, 
a gerbe over $\,G\,$ with curvature
$\,H\,$ given by (\ref{Gcurv}) exists if and only if $\,\sk\,$ is an integer 
and it is unique up to stable isomorphisms since $\,H^2(G,U(1))=\{1\}$. 
\,Clearly one may take $\,\CG_\sk=\CG_1^{^{\otimes\sk}}$. \,The basic gerbe 
$\,\CG_1\,$ over simply connected groups $\,G\,$ has been constructed 
in \cite{Meinr}. We shall describe gerbes $\,\CG_\sk\,$ (dropping below 
the subscript $\,\sk$) borrowing on the discussion in Sect.\,\,2 
of \cite{GR2} and employing similar notations. 
\vskip 0.3cm

Let $\,\CA_{_W}\,$ denote 
the positive Weyl alcove in the Cartan subalgebra $\,\Nt\,$ of the 
Lie algebra $\,\Ng\,$ of $\,G\,$ that will be identified with its dual 
$\,\Ng^*\,$ using the bilinear form defined by $\,\tr$. \,The normalization 
of $\,\tr\,$ renders the length squared of the long roots equal to 
$\,2$. $\,\CA_{_W}\,$ is a simplex with vertices $\,\tau_i,\ i\in R
\equiv\{0,1,\dots,r\}\m$, \,where $\,r\,$ is the rank 
of $\,\Ng$. \,In particular, 
$\,\tau_0=0\,$ and $\,\tau_i={1\over k_i^\vee}\lambda_i\,$ for $\,i\not=0\,$
with $\,\lambda_i\,$ the simple weights and $\,k_i^\vee\,$ the dual
Kac labels. \,For $\,i\in R\m$, \,let
\qq
\CA_i=\{\,\tau\in\CA_{_W}\ |\ \tau=\sum\limits_{j}s_j\tau_j\ {\rm with}
\ s_i>0\,\}\,,\quad\, 
O_i=\{\,h\,\ee^{\,2{\pi\si}\,\tau}\,h^{-1}\,|\,\,h\in G,\ 
\tau\in\CA_i\,\}
\nonumber
\qqq
and, for $\,I\subset R\m$, $\,\CA_I=\mathop{\cap}\limits_{i\in I}\CA_i\,$
and $\,\,O_I=\mathop{\cap}\limits_{i\in I}O_i$. Subsets $\,\CO_i\,$ of 
$\,G\,$ are open and $\,\CO_I\,$ are composed of elements
$\,h\,\ee^{\,2{\pi\si}\,\tau}h^{-1}\,$ with $\,h\in G\,$ and $\,\tau
\in\CA_I$. \,The expressions
\qq
B_i\ =\ {_\sk\over^{4\pi}}\,\tr\,(h^{-1}dh)\,\ee^{2{\pi\si}\tau}\,
(h^{-1}dh)\,\ee^{-2{\pi\si}\tau}\,+\ \si\m\sk\,\tr\,(\tau-\tau_i)(h^{-1}dh)^2
\nonumber
\qqq
define (smooth) 2-forms on $\,O_i\,$ such that $\,dB_i=H|_{O_i}$.
\,An important role is played by the subgroups
\qq 
G_I\ =\ \{\,h\in G\,\,|\,\,h\,\ee^{\,2{\pi\si}\,\tau}\,h^{-1}
=\ee^{\,2{\pi\si}\,\tau}\ \,{\rm for}\ \,
\tau\in\CA_I\setminus\mathop{\cup}\limits_{i\notin I}\CA_{i}\,\}
\nonumber
\qqq
and the (smooth) maps
\qq
\CO_I\,\ni\,g\,=\,h\,\ee^{2{\pi\si}\tau}\m h^{-1}\ \mathop{\longrightarrow}
\limits^{\rho_I}\ \ hG_I\,\in\,G/G_I
\nonumber
\qqq
well defined because the adjoint action stabilizers of $\,\ee^{\,2{\pi\si}\,\tau}\,$
for $\,\tau\in\CA_I\,$ are contained in $\,G_I$. \,One introduces the principal 
$\,G_I$-bundles $\,\pi_I:P_I\to O_I\,$
\qq
P_I\ =\ \{\,(g,h)\in O_I\times G\,\,|\,\,\rho_I(g)=hG_I\,\}\,.
\nonumber
\qqq
For the gerbes $\,\CG=(Y,B,L,\mu)\m$, \,one sets
\qq
Y\ =\ \mathop{\sqcup}\limits_{i\in R}P_i
\nonumber
\qqq
with $\,\pi:Y\to G\,$ restricting to $\,\pi_i\,$ on $\,P_i\,$ and the 2-form
$\,B\,$ restricting to $\,\pi_i^*B_i$. \,Let
\qq
\hat Y_{i_1..i_n}\ =\ P_I\times G_{i_1}\times\cdots\times G_{i_n}
\ \quad{\rm and}\ \quad 
Y_{i_1..i_n}\ =\ \hat Y_{i_1..i_n}/G_I
\nonumber
\qqq
for $\,I=\{i_1,\dots,i_n\}\,$ and $\,G_I\,$ acting on $\,\hat Y_{i_1..i_n}$ 
diagonally by the right multiplication. The fiber power $\,Y^{[n]}\,$ of $\,Y\,$ 
may be identified with the disjoint union of $\,Y_{i_1..i_n}\,$
by assigning to the $\,G_I$-orbit of $\,((g,h),\gamma_1,..,\gamma_n)\,$ the
$\,n$-tuple $\,(y_1,..,y_n)\in Y^{[n]}\,$ with $\,y_m=(g,h\gamma_m^{-1})\,$: 
\qq
Y^{[n]}\ \cong\ \mathop{\sqcup}\limits_{(i_1,..,i_n)}Y_{i_1..i_n}\,.
\nonumber
\qqq
\vskip 0.3cm

The construction of the line bundle $\,L\,$ over $\,Y^{[2]}\,$ uses more 
detailed properties of the stabilizer groups $\,G_I$. \,For $\,I\subset 
J\subset R\m$, \,$G_I\supset G_J\,$ and 
the smallest of those groups, $\,G_R\m$, 
\,coincides with the Cartan subgroup $\,T\,$ of $\,G$. \,Groups $\,G_I\,$ 
are connected but not necessarily simply connected. 
Let $\,\Ng_I\supset\Nt\,$ denote 
the Lie algebra of $\,G_I\,$ and let $\,\ee_I\,$ be the exponential map 
from $\,\si\Ng_I\,$ to the universal cover $\,\tilde G_I$. \,One has
\qq
G_I\ =\ \tilde G_I/\CZ_I\ \quad{\rm for}\quad\ \CZ_I\ =\ 
\ee_I^{2{\pi\si}Q^{\hspace{-0.04cm}^\vee}}  
\nonumber
\qqq
where $\,Q^{\hspace{-0.04cm}^\vee}\subset\Nt\,$ is the coroot 
lattice of $\,\Ng$. \,Let $\,\chi_i:\CZ_i\to U(1)\,$ be the character 
defined by
\qq
\chi_i(\ee_i^{2{\pi\si}q})\ =\ \ee^{2{\pi\si}\,{\mathrm tr}\,\tau_i q}
\label{chii}
\qqq
for $\,q\in Q^{\hspace{-0.04cm}^\vee}\,$ and $\,\chi_{ij}:\tilde G_{ij}
\to U(1)\,$ be the 1-dimensional representation of $\,\tilde G_{ij}\,$ 
given by the formula
\qq
\chi_{ij}(\tilde\gamma)\ =\ \exp\Big[{_1\over^\si}\int\limits_{\tilde\gamma} 
a_{ij}\Big]
\label{chiij}
\qqq
where $\,a_{ij}=\si\,\tr\,(\tau_j-\tau_i)(\gamma^{-1}d\gamma)\,$ is a closed
1-form on $\,G_{ij}$. \,Let $\,\hat L_{ij}\,$ denote the line bundle over
$\,\hat Y_{ij}\,$ with the fiber over $\,((g,h),\gamma,\gamma')\,$ composed
of the equivalence classes $\,[\tilde\gamma,\tilde\gamma',u]_{ij}\,$
with respect to the relation
\qq
(\tilde\gamma,{\tilde\gamma}',\m u)
\ \ \mathop{\sim}\limits_{^{ij}}\ \ (\tilde\gamma\zeta,
{\tilde\gamma}'\zeta',\m\chi_i(\zeta)^{\sk}\chi_j(\zeta')^{{-\sk}}\,u)
\nonumber
\qqq
for $\,\tilde\gamma\in\tilde G_i\m$, $\,\tilde\gamma'\in\tilde G_j\,$
projecting to $\,\gamma\in G_i\,$ and $\,\gamma'\in G_j\m$, \,respectively,
and $\,u\in\NC\m$, $\,\zeta\in \CZ_i\m$, $\,\zeta'\in \CZ_j$. \,We shall 
twist the flat structure of $\,\hat L_{ij}\,$ by the connection form
$\,A_{ij}=\si\m\sk\,\tr\,(\tau_j-\tau_i)(h^{-1}dh)$. \,The right action of 
$\,G_{ij}\,$ on $\,\hat Y_{ij}\,$ lifts to the action on $\,\hat L_{ij}\,$
defined by
\qq
((g,h),\,[\tilde\gamma,{\tilde\gamma}',u]_{{ij}})\ \ \longmapsto\ 
\ ((g,h\gamma''),\,[\tilde\gamma{\tilde\gamma}'',{\tilde\gamma}'
{\tilde\gamma}'',\m\chi_{ij}({\tilde\gamma}'')^{{-\sk}}\,u]_{{ij}})
\nonumber
\qqq
for $\,\gamma''\in G_{ij}\,$ and $\,{\tilde\gamma}''\,$ its lift
to $\,\tilde G_{ij}\,$ (the right hand side is independent on the choice
of the latter). The hermitian structure and the connection 
of $\,\hat L_{ij}\,$ descend to the quotient bundle $\,\hat L_{ij}/G_{ij}
=L_{ij}\,$ over $\,Y_{ij}\,$ and the line bundle $\,L\,$ over $\,Y^{[2]}\,$ 
for the gerbe $\,\CG\,$ is taken as equal to $\,L_{ij}\,$ when restricted 
to $\,Y_{ij}$. \,Note that the curvature of $\,L_{ij}\,$ is equal to
the lift to $\,Y_{ij}\,$ of the 2-form $\,B_j-B_i\,$ on $\,O_{ij}\m$, 
\,as required. In general, unlike for the $\,SU(N)\,$ groups, there is
no line bundle over $\,O_{ij}\,$ with such curvature, hence the need
of a more complicated construction of the gerbe \cite{Meinr}.
\vskip 0.3cm

The groupoid multiplication $\,\mu\,$ of $\,\CG\,$ is defined as follows.
Let $\,((g,h),\gamma,\gamma',\gamma'')\in
\hat Y_{ijk}\,$ represent $\,(y,y',y'')\in Y^{[3]}\,$ with
$\,y=(g,h\gamma^{-1})\m$, $\,y'=(g,h{\gamma'}^{-1})\,$
and $\,y''=(g,h{\gamma''}^{-1})\,$ and let
\qq
\hat\ell_{ij}\ =\ [\tilde\gamma,\tilde\gamma',u]_{ij}\,,\quad\ 
\hat\ell_{jk}\ =\ [\tilde\gamma',\tilde\gamma'',u']_{jk}\,,\quad\ 
\hat\ell_{ik}\ =\ [\tilde\gamma,\tilde\gamma'',uu']_{ij} 
\nonumber
\qqq
induce the elements $\,\ell_{ij}\in L_{(y,y')}\m$, 
$\,\ell_{jk}\in L_{(y',y'')}\,$ and $\,\ell_{ik}\in 
L_{(y,y'')}$. \,Then
\qq
\mu(\ell_{ij},\ell_{jk})\ =\ \ell_{ik}\,.
\nonumber
\qqq
This ends the description of Meinrenken's gerbes $\,\CG=(Y,B,L,\mu)\,$
over the simply connected group $\,G$.  
 
\subsection{Gerbes over non-simply connected groups}
\label{sec:gnscg}

Let $\,G'=G/Z\m$, \,with $\,Z\,$ a subgroup of the center of $\,G\m$, \,be
a non-simply connected quotient of a simply connected group $\,G$.
As mentioned in Introduction, to push down the gerbes $\,\CG\,$ from
$\,G\,$ to $\,G'\,$ one has to solve a cohomological equation.
Let us start by identifying simple cohomological objects related
to the pair $\,(G,Z)$. 
\vskip 0.3cm

Choose for each $\,z\in Z\,$ an element $\,w_z\in G\,$ that normalizes 
the Cartan subgroup $\,T\subset G\,$ such that for $\,\tau\in\CA_{_W}\m$,
\qq
z\,e^{2{\pi\si}\tau}\ =\ w_z^{-1}\ee^{2{\pi\si}\m z\tau}\,w_z
\label{wz}
\qqq
for some $\,z\tau\in\CA_{_W}$. \,For $\,z=1\,$ we shall take $\,w_z=1$.
\,The map 
\qq
\tau\ \longmapsto\ z\tau\ = \ w_z\,\tau\,w_z^{-1}\,+\,\tau_{z0}
\nonumber
\qqq
is an affine transformation of $\,\CA_{_W}\,$ that permutes the vertices:
$\,z\tau_i\equiv\tau_{zi}$. \,Let $\,b_{z,z'}\in\Nt\,$ be such that
\qq
w_z\,w_{z'}\,w_{zz'}^{-1}\ =\ c_{z,z'}\ =\ \ee^{2{\pi\si}\,b_{z,z'}}\,. 
\label{bzz'}
\qqq
We shall take $\,b_{z,1}=0=b_{1,z'}$. \,The Cartan subgroup valued 
chain $c=(c_{z,z'})\,$ is a 2-cocycle on group 
$\,Z\,$:
\qq
(\delta c)_{z,z'z''}\ =\ (w_z\m c_{z',z''}w_z^{-1})\, c_{zz',z''}^{\,-1}\,
c_{z,z'z''}\,c_{z,z'}^{,-1}\ =\ 1\,, 
\nonumber
\qqq
see Appendix A of \cite{GR1} for a brief summary on finite group 
cohomology. 
\vskip 0.3cm

The 3-form $\,H\,$ descends to a 3-form $\,H'\,$ on $\,G'$. \,For (integer)
$\,\sk\,$ for which $\,H'\,$ has periods in $\,2\pi\NZ\,$ there exist
a gerbe $\,\CG'=(Y',B',L',\mu')\,$ over $\,G'\,$ with curvature $\,H'$.
\,To describe its explicit construction, we shall follow \cite{GR2}. One 
takes $\,Y'=Y=\sqcup P_i\,$ with the natural projection $\,\pi'\,$ on 
$\,G'\,$ and $\,B'=B$. \,For $\,y=(g,h)\in P_i\,$ let $\,zy\,$ denote 
the element $\,(zg,hw_z^{-1})\in P_{zi}\,$. \,The fiber product space 
$\,{Y'}^{[n]}\,$ is composed of $n$-tuples $\,(y,y',\dots,y^{(n-1)})\,$ 
such that for some $\,z,z',\dots,z^{(n-2)}\in Z\,$  
\qq
(y,zy',\cdot\cdot,z(z'(\cdot\cdot(z^{(n-2)}y^{(n-1)})\cdot\cdot)))
\ \ \in\ \ Y^{[n]}\,. 
\nonumber
\qqq
One may then identify
\qq
{Y'}^{[n]}\,\ \cong\mathop{\sqcup}\limits_{(z,z',\cdot\cdot,z^{(n-2)})\,\in\,
Z^{n-1}}Y^{[n]}\ \ \cong\mathop{\sqcup}\limits_{(z,z',
\cdot\cdot,z^{(n-2)})\,\in
\,Z^{n-1}}\,\mathop{\sqcup}\limits_{(i_1,..,i_n)}Y_{i_1..i_n}\,.
\nonumber
\qqq
Let $\,L'\,$ be the line bundle over $\,{Y'}^{[2]}\,$
that restricts to $\,L\,$ on each component $\,Y^{[2]}\,$ in the 
above identification, i.e.\,\,to $\,L_{ij}\,$ 
on $\,Y_{ij}\subset Y^{[2]}$.
\,It remains to describe the groupoid multiplication $\,\mu'$.
\vskip 0.3cm

Let $\,(y,y',y'')\in{Y'}^{[3]}\,$ be such that 
$\,(y,zy',z(z'y''))\in Y^{[3]}$. 
\,Identifying the triple $\,(y,zy',z(z'y''))\,$ with the $\,G_{ijk}$-orbit of
$\,((g,h),\gamma,\gamma',\gamma'')\in\hat Y_{ijk}\,$ for some $(i,j,k)\,$
and employing the notations $\,i_z\equiv z^{-1}i\m$, $\,\gamma_z\equiv 
w_z^{-1}\gamma\,w_z\in G_{i_z}\,$ for $\,\gamma\in G_i\m$, \,we 
have 
\qq
&&\hbox to 11cm{$\displaystyle{y\ =\ (g,\m h\m\gamma^{-1})\,,
\quad\ z\m y'\ =\ (g,\m h\m{\gamma'}^{-1})\,,\quad\ 
z(z'y'')\ =\ (g,\m h\m{\gamma''}^{-1})\,,}$\hfill}\label{wmr0}\\ \cr
&&\hbox to 5.5cm{$\displaystyle{y'\ =\ (z^{-1}g,\m h\,w_z
\m{\gamma'_z}^{-1})\,,}
$\hfill}y''\ =\ ((zz')^{-1}g,h\,w_zw_{z'}(\gamma''_z)_{z'}^{-1})\,,
\label{wmr1}\\ \cr
&&\hbox to 5.5cm{$\displaystyle{z'y''\ = 
\ (z^{-1}g,\m h\,w_z\m{\gamma''_z}^{-1})\,,}$\hfill}(zz')y'' 
\ =\ (g,\m h\,(c_{z,z'}^{\,-1}\gamma'')^{-1})\,.
\label{wmr2}
\qqq
Recalling the explicit description of the line bundles $\,L_{ij}\,$
as quotients of bundles $\,\hat L_{ij}\m$,
\,consider the equivalence classes
\qq
\hat\ell_{ij}\ =\ [\tilde\gamma,
{\tilde\gamma}',\m u]_{{ij}}\,,\quad
\hat\ell_{j_zk_z}\ =\ 
[{\tilde\gamma}'_{z},{\tilde\gamma}''_{z},\m u']_{j_zk_z}\,,\quad
\hat\ell_{ik}\ =\ [\tilde\gamma,{\tilde c}_{z,z'}^{\,-1}{\tilde\gamma}'',
\m u\m u']_{ik}\,,
\label{hatlll}
\qqq
with $\,\tilde\gamma\in\tilde G_i\,$ projecting to $\,\gamma\in G_i\,$ and
$\,\tilde\gamma_z\equiv w_z^{-1}\tilde\gamma\,w_z\in\tilde G_{i_z}\,$ etc.
and with $\,\tilde c_{z,z'}=\ee_I^{2{\pi\si}\,b_{z,z'}}\hspace{-0.1cm}
\in\tilde G_I\m$. \,The above classes determine the elements
\qq
\ell_{ij}\in L_{(y,zy')}=L'_{(y,y')}\,,\ \ 
\ell_{j_zk_z}\in L_{(y',z'y'')}=L'_{(y',y'')}\,,\ \ 
\ell_{ik}\in L_{(y,(zz')y'')}=L'_{(y,y'')}\quad 
\label{lll}
\qqq
The groupoid multiplication $\,\mu'\,$ in $\,L'\,$ is then defined by
specifying the product of $\,\ell_{ij}\,$ and $\,\ell_{j_zk_z}\,$
via the relation \cite{GR2} 
\qq
\mu'\m(\m\ell_{ij},\,\ell_{j_zk_z})\ =\ 
\ee^{2\pi\si\m\sk\,\tr\,\tau_k\m b_{z,z'}}\,
V_{z,z'}\,\,\ell_{ik}\,,
\label{mup}
\qqq
where $\,V\,=\,(V_{z,z'})\,$ solves the cohomological equation
\qq
\delta V\ =\ U\m.
\label{ce}
\qqq
Explicitly, $\,(\delta V)_{z,z',z''}=V_{z',z''}\,V_{zz',z''}^{\,-1}
\,V_{z,z'z''}\,V_{z,z'}^{\,-1}\,$ and $\,U=(U_{z,z',z''})\,$ is a 3-cocycle 
on $\,Z\,$ with values in $\,U(1)\,$ defined by\footnote{To obtain more
concise expressions, we have multiplied the 3-cocycle 
$\,U^\sk\,$ with $\,U\,$ given by Eqs.\,(3.29) or (3.30) in \cite{GR2} 
by the coboundary $\,\delta V'\,$ with $\,V'_{z,z'}
=\ee^{-2\pi\si\m\sk\,\tr\,\tau_{zz'0}\m b_{z,z'}}$}
\qq
U_{z,z',z''}\ =\ 
\ee^{-2\pi\si\m\sk\,\tr\,\tau_{_{\hspace{-0.02cm}0_z}}\m b_{z',z''}}
\nonumber
\qqq
with $\,\tau_{0_z}=\tau_{z^{-1}0}=z^{-1}\tau_0$.
\vskip 0.3cm

The cohomology class $\,[U]\in H^3(Z,U(1))\,$ of $\,U\,$ does not
depend on the ambiguity in the choice of $\,w_z\,$ nor of $\,b_{z,z'}$.
\,The changes
\qq
w_z\ \mapsto\ \ee^{2\pi\si\,a_z}w_z\,,\qquad
b_{z,z'}\ \mapsto\ b_{z,z'}\,+\,w_z\m a_{z'}w_z^{-1}\,-\,
a_{zz'}\,+\,a_z\,+\,q_{z,z'}
\label{changes}
\qqq
with $\,a_z\in \Nt\,$ and $\,q_{z,z'}\in Q^{\hspace{-0.04cm}^\vee}\,$
induces the transformation
\qq
U\ \longmapsto\ (\delta V')\,U\qquad{\rm for}\qquad V'_{z,z'}\,=\,
\ee^{2\pi\si\m\sk\,\tr\,\tau_{_{\hspace{-0.02cm}0_z}}a_{z'}}\,.
\nonumber
\qqq
The cohomological equation (\ref{ce}) has a solution only for $\,\sk\,$ 
for which the cohomology class $\,[U]\,$ is trivial. 
The resulting set of levels $\,\sk\,$ coincides with the one for which
the periods of the 3-form 
$\,H'\,$ are in $\,2\pi\NZ\,$ \cite{FGK,GR2}. For all cases
except when $\,G'=Spin(4n)/({\NZ_2\hspace{-0.06cm}\times\hspace{-0.06cm}
\NZ_2})=SO(4n)/\NZ_2$ for $\,n=2,3,...\m$, \,we have $\,H^2(Z,U(1))=\{1\}\,$
and any two solutions $\,V\,$ differ by a coboundary $\,\delta W\,$ with 
$\,(\delta W)_{z,z'}=W_{z'}W_{zz'}^{-1}W_z\,$ and lead to 
stably isomorphic gerbes. \,In the exceptional case, $\,[U]\,$ is trivial 
if $\,\sk\in\NZ\,$ for $\,n\,$ even and $\,\sk\in2\NZ\,$ 
for $\,n\,$ odd. In this case $\,H^2({\NZ_2\hspace{-0.06cm}
\times\hspace{-0.06cm}\NZ_2},U(1))\cong\NZ_2\,$ and 
there are two classes of non-equivalent solutions of Eq.\,\,(\ref{ce}) 
with all the solutions within one class differing by coboundaries 
$\,\delta W$. \,Such solutions lead to 
gerbes that fall into two different stable-isomorphism classes. 
In all cases one may choose $\,V_{z,z'}\,$ so that $\,V_{z,1}\equiv1
\equiv V_{1,z'}\,$ which, in particular, assures that
on $\,L'|_{Y^{[2]}}=L\,$ the groupoid product $\,\mu'\,$ 
coincides with $\,\mu$.

\nsection{Symmetric branes in the WZW models}
\label{sec:sbwzwm}

In the open-string version of the WZW model, one may restrict the string ends 
to move on branes $\,D\,$ in $\,G\,$ and couple them to Yang-Mills fields on
$\,D$. \,This is done using the concept of $\,\CG$-branes as discussed
in Sect.\,\ref{sec:asacymf}. One would like, however, to assure 
that the boundary
version of the theory still possesses rich symmetry. The fundamental symmetry 
of the bulk WZW model is provided by two chiral copies of the current algebra
$\,\hat\Ng\,$ associated to the Lie algebra $\,\Ng\,$ of $\,G$. \,The simplest 
family of branes (called symmetric) assures that the diagonal
current algebra is not broken in the presence of boundaries. 
Such requirement restricts the brane supports $\,D\subset G\,$ to coincide 
with a conjugacy class in $\,G\,$  \cite{ASchom} and imposes further conditions 
on the $\,\CG$-branes supported by $\,D\,$ that we shall describe now. 
For simply connected 
$\,G\m$, \,the conjugacy classes may be labeled by the elements $\,\tau\,$ 
of the Weyl alcove $\,\CA_{_W}\,$ with
\qq
\CC_\tau\ =\ \{\,h\,\ee^{2{\pi\si}\tau}\,h^{-1}\ |\ h\in G\,\}\,.
\nonumber
\qqq
being the class corresponding to $\,\tau$. \,When restricted to 
$\,\CC_\tau\m$, \,the 3-form $\,H\,$ becomes exact. 
In particular, $\,H|_{\CC_\tau}=dQ_\tau\,$
where
\qq
Q_\tau\ =\ =\ {_\sk\over^{4\pi}} \,\tr\,(h^{-1}dh) \,\ee^{2\pi
\si\tau}(h^{-1}dh)\,\ee^{-2{\pi\si}\tau} 
\nonumber
\qqq
is a (smooth) 2-form on $\,\CC_\tau$. \,Let $\,\CG=(Y,B,L,\mu)\,$ be a gerbe 
over $\,G\,$ with curvature $\,H\,$ as described above. Recall that a 
$\,\CG$-brane $\,\CD\,$ supported by $\,D\,$ is a pair $\,(D,\CE)\,$ where 
$\,\CE\,$ is a $\,\CG|_D$-module. Such module determines, in turn, a vector 
bundle $\,E\,$ with connection over $\,Y_D=\pi^{-1}(D)\,$ for 
$\,\pi\,$ denoting the projection from $\,Y\,$ to $\,G$. 
\,With $\,D=\CC_\tau\m$, \,the additional restriction, imposed 
by the conservation of the diagonal current algebra, fixes the curvature
of bundle $\,E\,$ to be the scalar 2-form
\qq
F\ =\ \pi^*Q_\tau\ -\ B|_{Y_D}\,.
\label{F}
\qqq
Note that this restricts very strongly the (twisted) Yang-Mills fields 
to which the ends of the string may be coupled without breaking the diagonal 
current-algebra symmetry. 
\vskip 0.3cm

The conjugacy classes in a non-simply connected group $\,G'=G/Z\,$ 
may in turn be labelled by the $\,Z$-orbits $\,[\tau]\,$ in the 
Weyl alcove $\,\CA_{_W}\,$ with
\qq
\CC'_{[\tau]}\ =\ \CC_\tau Z
\nonumber
\qqq
for any $\,\tau\in[\tau]$. \,Let $\,Z_{[\tau]}\subset Z\,$ denotes 
the stabilizer subgroup of any $\,\tau\in[\tau]$. \,Note that 
$\,Z_{[\tau]}\,$ is composed of $\,z\in Z\,$
such that $\,z\CC_\tau=\CC_\tau$. \,It follows that
\qq
\CC'_{[\tau]}\ \cong\ C_\tau/Z_{[\tau]}
\nonumber
\qqq
for $\,\tau\in[\tau]$. \,In particular, $\,Z_{[\tau]}\,$ is the 
fundamental group of $\,\CC'_{[\tau]}\,$ since the conjugacy classes
$\,\CC_\tau\,$ are simply connected. The 2-forms $\,Q_\tau\,$ project 
to a unique 2-form $\,Q'_{[\tau]}\,$ on $\,\CC'_{[\tau]}$. \,If $\,\CG'
=(Y',B',L',\mu')\,$ is a gerbe over $\,G'\,$ as described above then
the symmetric $\,\CG'$-branes supported by $\,D'=\CC'_{[\tau]}\,$ are 
restricted by fixing the curvature of the vector bundle $\,E'\,$ over 
$\,Y'_{D'}={\pi'}^{-1}(\CC'_{[\tau]})\,$ to be
\qq
F'\ =\ {\pi'}^*Q'_{[\tau]}\ -\ B'|_{Y'_{D'}}\,.
\label{F'}
\qqq
\vskip 0.3cm

The main aim of this paper is to classify such symmetric $\,\CG$- and 
$\,\CG'$-branes up to isomorphism. For rank 1 $\,\CG$-(or $\,\CG'$-)branes, 
the curvature restriction may be reformulated by stating that the
branes provide a stable isomorphism between the gerbes $\,\CG|_{D}\,$ and 
$\,\CK=(D,Q_\tau,D\times\NC,\,\cdot\,)\m$, \,or between $\,\CG'|_{D'}\,$ and 
$\,\CK'=(D',Q'_{[\tau]},D'\times\NC,\,\cdot\,)\m$, \,and the notion of a 
rank 1 symmetric $\,\CG$- or $\CG'$-branes is equivalent to the 
one introduced in \cite{GR1} and studied there in detail for groups 
$\,SU(N)\,$ and their non-simply connected quotients. Admitting 
rank $N$ branes does not change much in the latter story, as we 
shall see below, but is necessary if we want to account properly for 
the conformal boundary conditions of general WZW models.

\subsection{Symmetric branes for simply connected groups}
\label{sec:sbscg}

Let $\,D=\CC_\tau\,$ be the conjugacy class in the simply connected group  
$\,G\,$ and $\,Y_D\,$ be defined as above. Clearly
\qq
Y_D\ =\ \mathop{\sqcup}\limits_i P_i|_{\CC_\tau}
\label{yd}
\qqq
where $\,P_i|_{\CC_\tau}\,$ is non-empty if and only if $\,\tau\in\CA_i$.
Let $\,G_\tau\,$ denote the adjoint-action isotropy subgroup of 
$\,\ee^{2{\pi\si}\tau}\,$ so that $\,\CC_\tau\cong G/G_\tau$. $\,G_\tau\,$
is connected and $\,T\subset G_\tau\subset G_i\,$ if $\,\tau\in\CA_i\,$
($\m G_\tau\,$ is is one of groups $\,G_I\,$ with $\,I\ni i$). 
\,We shall denote by $\,\Ng_\tau\,$ the Lie algebra of $\,G_\tau$. 
\,The space
\qq
P_\tau\ =\ \{\,(g,h)\in\CC_\tau\times G\,\,|\,\,g=h\,\ee^{2{\pi\si}\tau}
\,h^{-1}\,\}
\nonumber
\qqq
equipped with the projection on the first factor and the right action of
$\,G_\tau\,$ on the second factor becomes a principal $\,G_\tau$-bundle 
over $\,\CC_\tau$. \,Of course, $\,P_\tau\cong G$. \,Note that if 
$\,\tau\in\CA_i\,$ then $\,P_\tau\subset P_i|_{\CC_\tau}$. \,Besides, we may 
identify 
\qq
P_i|_{\CC_\tau}\ \cong\ Y_{i\tau}\ =\ 
\hat Y_{i\tau}/G_\tau\ \quad{\rm for}\ \quad
\hat Y_{i\tau}\ =\ P_\tau\times G_i 
\label{ident1}
\qqq
assigning to the orbit of $\,((g,h),\gamma)\in\hat Y_{i\tau}\,$ 
under the right diagonal action of $\,G_\tau\,$ the element 
$\,(g,h\gamma^{-1})\in P_i$.  
\vskip 0.3cm

Let us start by describing the rank 1 symmetric $\,\CG$-branes 
$\,(\CC_\tau,\CE_1)\,$  supported by the conjugacy classes $\,\CC_\tau\,$ 
such that $\,\sk\tau=\lambda\,$ is a weight. Weights $\,\lambda=\sk\tau\,$
for $\,\tau\in\CA_{_W}\,$ are often called integrable \cite{Kac} 
(at level $\,\sk$) \,and we shall also call the corresponding conjugacy
classes {\bf integrable}. \,The construction of the line 
bundle $\,E\,$ over $\,Y_D\,$ such that $\,\CE_1=(E,\rho)\,$ 
will resemble that of the line bundle $\,L\,$ in Sect.\,\ref{sec:gscg}. 
Let $\,\tilde G_\tau\,$ denote the universal covering group of 
$\,G_\tau=\tilde G_\tau/\CZ_\tau\,$ where $\,\CZ_\tau
=\ee_\tau^{2{\pi\si}Q^{\hspace{-0.04cm}^\vee}}\,$ for $\,\ee_\tau\,$ 
standing for the exponential map from $\,\si\Ng_\tau\,$ to $\,\tilde G_\tau$. 
\,Consider the 1-form $\,a_{i\tau}=\si\,\tr\,(\tau-\tau_i)
(\gamma^{-1}d\gamma)\,$ on $\,G_\tau$. \,It is easy to see that 
the adjoint action of $\,G_\tau\,$ preserves $\,\tau-\tau_i$. \,As
a result,  $\,a_{i\tau}\,$ is a closed form. By the formula
\qq
\chi_{i\tau}(\tilde\gamma)\ =\ \exp\Big[{_1\over^\si}\int\limits_{\tilde\gamma} 
a_{i\tau}\Big]
\nonumber
\qqq
it defines a 1-dimensional representation $\,\chi_{i\tau}:\tilde 
G_\tau\to U(1)\m$, \,compare (\ref{chiij}). The embedding $\,\Ng_\tau\subset 
\Ng_i\,$ induces the canonical homomorphism 
$\,\tilde G_\tau\to\tilde G_i\,$ that maps $\,\CZ_\tau\,$ to
$\,\CZ_i\,$ sending $\,\zeta_\tau=\ee_\tau^{2{\pi\si}q}\in\CZ_\tau\,$ to
$\,\zeta_i=\ee_i^{2{\pi\si}q}\in\CZ_i\,$ for $\,q\in Q^{\hspace{-0.04cm}^\vee}$. 
\,Note that 
\qq
\chi_{i\tau}(\tilde\gamma\zeta_\tau)^{\sk}\ =\ \chi_{i\tau}(\tilde\gamma)
^{\sk}\,\chi_i(\zeta_i)^{{-\sk}}\,\ee^{2\pi\si\m\sk\,\tr\,\tau q}
\nonumber
\qqq
and the last factor drops out if $\,\sk\tau\,$ is a weight.
\vskip 0.3cm

Let $\,\hat L_{i}\,$ denote the flat line bundle over $\,G_i\,$ 
with the fiber over $\,\gamma\,$ composed of the equivalence classes
$\,[\tilde\gamma,u]_{i}\,$ with respect to the equivalence 
relation
\qq
(\tilde\gamma,v)\ \ \mathop{\sim}_{^{i}}\ \ (\tilde\gamma\zeta,
\chi_i(\zeta)^{^{\sk}}v)
\label{eri}
\qqq
for $\,\tilde\gamma\in\tilde G_i\,$ projecting to $\,\gamma\m$, 
$\,\zeta\in\CZ_i\,$ and $\,v\in\NC\m$, \,see (\ref{chii}). 
We shall denote by $\,\hat L_{i\tau}\,$ the pullback 
of $\,\hat L_i\,$ to $\,\hat Y_{i\tau}=P_{\tau}\times G_i\,$ with the flat 
structure twisted by the connection form $\,A_{i\tau}=\,\si\m\sk\,\tr\,
(\tau-\tau_i)(h^{-1}dh)$. The right action of $\,G_\tau\,$ on $\,
\hat Y_{i\tau}\,$ lifts to the action on $\,\hat L_{i\tau}\,$ defined by
\qq
((g,h),\,[\tilde\gamma,v]_i)\ \ \longmapsto\ \ 
((g,h\gamma_\tau),\,[\tilde\gamma
\tilde\gamma_\tau,\chi_{i\tau}(\tilde\gamma_\tau)^{{-\sk}}v]_i)
\label{mapt}
\qqq
for $\,\tilde\gamma_\tau\in\tilde G_\tau\,$ projecting to $\,\gamma_\tau
\in G_\tau$. \,For $\,\sk\tau\,$ a weight, the right hand side does 
not depend on the choice of the lift $\,\tilde\gamma_\tau$. \,The hermitian 
structure and the connection on $\,\hat L_{i\tau}\,$ descend to the 
quotient bundle $\,\hat L_{i\tau}/G_\tau=L_{i\tau}\,$ over $\,Y_{i\tau}\m$.
\,The curvature form $\,F_{i\tau}\,$ of $\,L_{i\tau}\,$ is given 
by the relation: 
\qq
\hat\iota^*F_{i\tau}\ =\ i\m\sk\,\tr\,(\tau_i-\tau)(h^{-1}dh)^2
\nonumber
\qqq
with $\,\hat\iota\,$ standing for the projection from $\,\hat Y_{i\tau}\,$
to $\,Y_{i\tau}$. \,Let $\,E\,$ be a hermitian line bundle 
with connection over $\,Y_D\,$ 
coinciding with $\,L_{i\tau}\,$ over $\,Y_{i\tau}\m$, \,see (\ref{yd}) and 
(\ref{ident1}). Note that $\,E\,$ satisfies the curvature constraint 
(\ref{F}), as required. Besides, there exist a bundle isomorphism
\qq
\rho:\,L|_{Y_D^{[2]}}\otimes p_2^*E\ \longrightarrow\ p_1^*E 
\label{rhod}
\qqq 
satisfying relation (\ref{vbi}). It is defined in the following way. Let, 
for $\,\tau\in\CA_{ij}\m$, $\,(g,h)\in P_\tau\m$, $\,\tilde\gamma
\in\tilde G_i\,$ projecting to $\,\gamma\in G_i\,$ and
$\,\tilde\gamma'\in\tilde G_j\,$ projecting to $\,\gamma'\in G_j\m$,
\qq
\hat\ell_{ij}\ =\ [\tilde\gamma,\tilde\gamma',u]_{ij}\,,\ \quad 
\hat\ell_{i\tau}\ =\ [\tilde\gamma,uv']_i\,,\ \quad\hat\ell_{j\tau}\ 
= [\tilde\gamma',v']_{j\tau}
\nonumber
\qqq
induce, respectively, the elements $\,\ell_{ij}\in L_{(y,y')}\m$, 
$\,\ell_{i\tau}\in E_y\,$ and $\,\ell_{j\tau}\in E_{y'}\,$ for 
$\,y=(g,h\gamma^{-1})\,$ and $\,y'=(g,h{\gamma'}^{-1})$. \,Then
\qq
\rho(\ell_{ij}\otimes\ell_{j\tau})\ =\ \ell_{i\tau}\,.
\nonumber
\qqq
We infer that if $\,\sk\tau\,$ is a weight then $\,(E,\rho)\,$ defines 
a 1-dimensional $\,\CG|_D$-module $\,\CE_1$. \,Besides, the curvature of 
bundle $\,E\,$ satisfies the constraint (\ref{F}). Consequently, 
$\,\CD_1=(\CC_\tau,\CE_1)\,$ is a symmetric rank 1 $\,\CG$-brane 
supported by $\,D=\CC_\tau$. \,General symmetric $\,\CG$-branes supported 
by the integrable conjugacy classes are isomorphic to the ones obtained 
by tensoring $\,E\,$ with a trivial $\,N$-dimensional bundle (all flat 
vector bundles over $\,\CC_\tau\,$ are trivial up to isomorphism).
In other words, they correspond to rank $\,N\,$ $\CG|_D$-modules 
\qq
\CE_N\ =\ \mathop{\CE_1\oplus\dots\oplus\CE_1}\limits_{N\ {\rm terms}}\,.
\nonumber
\qqq
A rank  $\,N\,$ $\,\CG$-brane $\,\CD_N=(\CC_\tau,\CE_N)\,$ obtained 
this way is called 
a stack of $\,N\,$ $\,\CG$-branes $\,\CD_1\,$ in physicists' language.
\vskip 0.3cm

The non-integrable conjugacy classes $\,D=\CC_\tau\m$, \,i.e.\,\,such that 
$\,\sk\tau\,$ is not a weight, support no symmetric $\,\CG$-branes,
as we shall show now. Existence of such a brane implies that $\,Y|_D\,$ 
admits a vector bundle $\,E\,$ with curvature $\,F\,$ given by (\ref{F}), 
i.e.\,\,restricting on each $\,Y_{i\tau}\,$ to the scalar 2-form 
$\,F_{i\tau}$. \,The periods of $\,F_{i\tau}\,$ over 2-spheres 
in $\,Y_{i\tau}\,$ have then to belong to $\,2\pi\NZ$. \,Let 
$\,\alpha_j^\vee\,$ for $\,j=1,\dots,r\,$ be the $\,j^{\rm th}$ 
simple coroot of $\,\Ng\,$ and $\,e_{\pm\alpha_j}\,$ be the corresponding
step generators. Consider for each $\,j\not=i\,$ the $\,su(2)\,$ subalgebra 
of $\,\Ng\,$ which is the real form of the subalgebra of $\,\Ng^\NC\,$ 
generated by $\,\alpha_j^\vee\,$ and $\,e_{\pm\alpha_j}$.
\,Such an $\,su(2)\,$ subalgebra exponentiates to a subgroup 
$\,SU(2)\subset G_i\subset G\,$ whose diagonal embedding into 
$\,G\times G_i\cong\hat Y_{i\tau}\,$ induces the map from 
$\,S^2\cong SU(2)/U(1)\,$ to $\,Y_{i\tau}$. \,The period of $\,F_{i\tau}\,$ 
on such a 2-cycle is equal to 
\qq
2\pi\sk\,\tr(\tau-\tau_i)\alpha_j^\vee\,=\,2\pi\sk\,\tr\,\tau\alpha_j^\vee\,.
\nonumber
\qqq
If $\,\tau\,$ is not equal to a vertex of the Weyl alcove $\,\CA_{_W}\,$ 
then $\,Y_{i\tau}\,$ is not empty (i.e. $\tau\in\CA_i$) for at least two 
values of $\,i\,$ and we infer that $\,\sk\tau\,$ must be a weight. 
If $\,\tau\,$ is a vertex of $\,\CA_{_W}\,$
then it belongs to the unique $\,\CA_i\,$ for which $\,\tau=\tau_i\,$
and the 2-form $\,F_{i\tau}\,$ vanishes. In this case $\,Y_D\cong P_{\tau_i}
\cong G\,$ and $\,\CG$-brane's vector bundle $\,E\,$ over $\,Y_D\,$
must be flat and hence isomorphic to a trivial $\,N$-dimensional bundle
since $\,G\,$ is simply connected. On the other hand, 
$\,Y_D^{[2]}\cong\{\,(h\gamma^{-1},h)\ |\ h\in G\,,\ \gamma\in G_i\,\}\cong
G\times G_i\,$ with the line bundle $\,L|_{Y_D^{[2]}}\,$ isomorphic 
to the pullback of $\,\hat L_i\,$ from $\,G_i$. \,The isomorphism 
(\ref{rhod}) maps now flat bundles and is given by a constant map 
$\,{\mathsf R}:G\times\tilde G_i\longrightarrow GL(N)\,$
such that 
\qq
{\mathsf R}(h,\tilde\gamma\zeta)\ =\ \chi_i(\zeta)^{-\sk}\,
{\mathsf R}(h,\tilde\gamma)\,. 
\nonumber
\qqq
Clearly, such a constant map exists only if $\,\sk\tau_i\,$ is a weight 
so that the right hand side is $\,\zeta$-independent.  
\vskip 0.3cm

\noindent In summary, in the case of simply connected groups, 
the symmetric $\,\CG$-branes are determined up to isomorphism by giving 
their support, an integrable conjugacy class $\,\CC_\tau\m$, \,and their 
rank $\,N$.

\subsection{Symmetric branes for non-simply connected groups}
\label{sec:sbnscg}

Let $\,G'=G/Z\,$ be a non-simply connected group and $\,\CG'=
(Y',B',L',\mu')\,$ be a gerbe over it described in Sect.\,\ref{sec:gnscg}.
We shall look here for symmetric $\,\CG'$-branes supported by 
the conjugacy classes $\,\CC'_{[\tau]}\subset G'$. \,This again
leads to a problem in finite group $\,Z\,$ cohomology, as already
noted in \cite{GR1} for the quotient groups of $\,SU(N)$. \,For 
$\,D'=\CC'_{[\tau]}\m$,
\qq
Y'_{D'}\ =\ \mathop{\sqcup}\limits_{\tau\in[\tau]}\mathop{\sqcup}
\limits_{i}P_i|_{\CC_\tau}\,.
\label{ydp}
\qqq
Any symmetric $\,\CG'$-brane $\,\CD'=(D',\CE')\,$ supported by 
$\,\CC'_{[\tau]}\,$ induces by restriction symmetric $\,\CG$-branes 
supported by $\,\CC_\tau\,$ for $\,\tau\in[\tau]\,$ (note that 
$\,Y_D\subset Y'_{D'}\,$ for $\,D=\CC_\tau\,$ etc.). It follows that 
only integrable conjugacy classes $\,\CC'_{[\tau]}\,$ for which
$\,\sk\tau\,$ is a weight may  support symmetric $\,\CG'$-branes 
(if $\,\sk\tau\,$ is a weight for some 
$\,\tau\in[\tau]\,$ then it is for all $\,\tau\in[\tau]\,$). 
Since now on we shall then assume that $\,\sk\tau\,$ is a weight 
for $\,\tau\in[\tau]$. 
\vskip 0.3cm

Let us first look for 1-dimensional $\,\CG'_{D'}$-modules 
$\,\CE'_1=(E',\rho')\,$ satisfying the curvature 
constraint (\ref{F'}). 
By the above argument, we may take $\,E'\,$ as the line bundle over 
$\,Y'_{D'}\,$ that over each $\,P_i|_{\CC_\tau}\cong Y_{i\tau}\,$ restricts 
to $\,L_{i\tau}$. \,The problem is to define the isomorphism 
\qq
\rho':\,L'|_{{Y'}_{D'}^{[2]}}\otimes p_2^*E'
\ \longrightarrow\ p_1^*E' 
\label{rhod'}
\qqq 
that satisfies the associativity condition
\qq
\rho'\circ(\mu'\otimes id)\ =\ \rho'\circ({id}\otimes\rho')\,,
\label{vbi'}
\qqq
see (\ref{vbi}). Let $\,(y,y')\in {Y'}_{D'}^{[2]}$. \,This means that,
in the notations of Sect.\,\ref{sec:gnscg},
\qq
y=(g,\m h\gamma^{-1})\,,\quad\ zy'=(g,\m h{\gamma'}^{-1})\,,\quad\ 
y'=(z^{-1}g,\m h\m w_z{\gamma'_z}^{-1})
\nonumber
\qqq
with $\,g=h\m\ee^{2{\pi\si}\tau}\m h^{-1}\,$ for some 
$\,\tau\in[\tau]\cap\CA_{ij}\m$, $\,\gamma\in G_i\m$, 
$\,\gamma'\in G_j\,$ and $\,z\in Z$.
The equivalence classes
\qq
\hat\ell_{ij}\ =\ [\tilde\gamma,\tilde\gamma',u]_{ij}\,,\ \quad 
\hat\ell_{i\tau}\ =\ [\tilde\gamma,uv']_i\,,\ \quad\hat\ell_{j_z\tau_z}
\ = [\tilde\gamma'_z,v']_{j_z}\,,
\label{v'}
\qqq
with $\,\tau_z\equiv z^{-1}\tau\,$ and with $\,\tilde\gamma\in\tilde G_i\,$ 
projecting to $\,\gamma\,$ and
$\,\tilde\gamma'\in\tilde G_j\,$ to $\,\gamma'$, \,induce, respectively, 
the elements 
$\,\ell_{ij}\in L_{(y,zy')}=L'_{(y,y')}\m$, $\,\m\ell_{i\tau}\in
E'_{y}\,$ 
and $\m\,\ell_{j_z\tau_z}\in E'_{y'}$. \,Since the isomorphism 
$\,\rho'\,$ is required to preserve the connection, necessarily,
\qq
\rho'(\ell_{ij}\otimes\ell_{j_z\tau_z})\ =\ 
\CW^{ij}_{\tau;\m z}\,\ell_{i\tau}
\label{rho'}
\qqq
for $\,\CW^{ij}_{\tau;\m z}\in U(1)$. \,Besides, the conjugation
of $\,\rho'\,$ with an isomorphism of $\,E'\,$ reducing to 
multiplication by $\,\CY^i_\tau\in U(1)\,$ over the connected components 
$\,P_i|_{\CC_\tau}\,$ of $\,Y'_{D'}\m$, \,results in the change
\qq
\CW^{ij}_{\tau;\m z}\ \ \longmapsto\ \ \CY^i_\tau\,
\CW^{ij}_{\tau;\m z}\,{\CY^{j_z}_{\tau_z}}^{\hspace{-0.05cm}-1}\,.
\label{wtw}
\qqq
Such changes lead to isomorphic $\,\CG'|_{D'}$-modules and hence
to isomorphic $\,\CG'$-branes. 
\vskip 0.3cm

It is now easy to translate identity (\ref{vbi'}) to conditions
for coefficients $\,\CW^{ij}_{\tau;\m z}$. \,For a triple $\,(y,y',y'')\in 
{Y'}_{D'}^{[3]}\,$ as in (\ref{wmr0}-\ref{wmr2}) where 
$\,g=h\m\ee^{2{\pi\si}\tau}\m h^{-1}\,$ for some $\,\tau\in[\tau]
\cap\CA_{ijk}\m$, $\,\gamma\in G_i\m$, $\,\gamma'\in G_j\m$, $\,\gamma''\in 
G_k\,$ and $\,z,z'\in Z\m$, \,consider the equivalence classes
$\,\hat\ell_{ij}\m$, $\,\hat\ell_{j_zk_z}$ and $\,\hat\ell_{ik}\,$ 
generating elements $\,\ell_{ij}\m$, $\,\ell_{j_zk_z}$ and 
$\,\ell_{ik}\,$ as in (\ref{hatlll}) and (\ref{lll}). Besides, let
\qq
\hat\ell_{k_{zz'}\tau_{zz'}}=[(\tilde c_{z,z'}^{-1}
\tilde\gamma'')_{zz'},v'']_{k_{zz'}}\m,\quad\, 
\hat\ell'_{k_{zz'}\tau_{zz'}}=[(\tilde\gamma''_z)_{z'},\m
\chi_{k\tau}(\tilde c_{z,z'})^{-\sk}v'']_{k_{zz'}}\m.\ \ 
\label{ell'}
\qqq
It is easy to check employing the action (\ref{mapt}) of 
$\,G_{\tau_{zz'}}\,$ and the relations 
\qq
(\tilde\gamma''_z)_{z'}=(\tilde c_{z,z'}^{\m-1}\m\tilde\gamma''
\tilde c_{z,z'})_{zz'}\,,\quad\ \chi_{k_{zz'}\tau_{zz'}}
((\tilde c_{z,z'})_{zz'})=\chi_{k\tau}(\tilde c_{z,z'})
\nonumber
\qqq 
that both equivalence classes define the same 
element $\,\ell_{k_{zz'}\tau_{zz'}}\in E'_{y''}$.
\,Using Eqs.\,\,(\ref{mup}) and (\ref{rho'}) and the
first way of (\ref{ell'}) to represent $\,\ell_{k_{zz'}\tau_{zz'}}\m$, 
\,we infer that
\qq
&\rho'(\mu'(\ell_{ij},\ell_{j_zk_z})\otimes\ell_{k_{zz'}
\tau_{zz'}})\ =\ 
\ee^{2\pi\si\m\sk\,\tr\,\tau_k\m b_{z,z'}}\,
V_{z,z'}\,\,\rho'(\ell_{ik}\otimes\ell_{k_{zz'}\tau_{zz'}})&\cr\cr
&=\ 
\ee^{2\pi\si\m\sk\,\tr\,\tau_k\m b_{z,z'}}\,
V_{z,z'}\,\CW^{ik}_{\tau;\m zz'}\,\,\ell_{i\tau}&
\nonumber
\qqq
if $\,v'=u'v''\,$ in (\ref{v'}). On the other hand, using the second
way to represent $\,\ell_{k_{zz'}\tau_{zz'}}\m$, \,we obtain
\qq
&\rho'(\ell_{ij}\otimes\rho'(\ell_{j_zk_z}
\otimes\ell_{k_{zz'}\tau_{zz'}}))\ =\ 
\,\chi_{k\tau}(\tilde c_{z,z'})^{-\sk}\,\,
\rho'(\ell_{ij}\otimes \CW^{j_zk_z}_{\tau_z;\m z'}\,
\ell_{j_z\tau_z})&\cr\cr
&\ =\ \chi_{k\tau}
(\tilde c_{z,z'})^{-\sk}\,\m
\CW^{ij}_{\tau;\m z}\,\CW^{j_zk_z}_{\tau_z;\m z'}\,\,\ell_{i\tau}\,.&
\nonumber
\qqq
The associativity condition (\ref{vbi'}) requires that both sides be 
equal and, since $\,\chi_{k\tau}(\tilde c_{z,z'})=
\ee^{2\pi\si\m\sk\,\tr\,(\tau-\tau_k)b_{z,z'}}$, \,it is equivalent to the 
identity
\qq
\CW^{j_zk_z}_{\tau_z;\m z'}\,\,{\CW^{ik}_{\tau;
\m zz'}}^{\hspace{-0.3cm}-1}\,\,\CW^{ij}_{\tau;\m z}\ =\ \CV_{\tau;\m z,z'}
\label{dwv0}
\qqq
with 
\qq
\CV_{\tau;\m z,z'}\ =\ 
\ee^{2\pi\si\m\sk\,\tr\,\tau\m b_{z,z'}}\,V_{z,z'}
\label{cv}
\qqq
where $\,V_{z,z'}\,$ is the solution of Eq.\,\,(\ref{ce}) that
determines gerbe $\,\CG'$.
\,It is easy to verify that $\,\CV=(\CV_{\tau;\m z,z'})\,$ forms 
a $\,2$-cocycle on group $\,Z\,$ with values in the $\,Z$-module 
$\,U(1)^{[\tau]}\,$ of $\,U(1)$-valued functions on the 
$\,Z$-orbit $\,[\tau]\m$, \,i.e.\,\,that
\qq
(\delta\CV)_{\tau;\m z,z',z''}\,=\,\CV_{\tau_z;\m z',z''}\,
\CV_{\tau;\m zz',z''}^{-1}\,\CV_{\tau;\m z,z'z''}\,
\CV_{\tau;\m z,z'}^{-1}\,=\,1
\nonumber
\qqq  
for $\,\tau\in[\tau]\,$ and $\,z,z',z''\in Z$.
\,The cohomology class $\,[\CV]\in H^2(Z,U(1)^{[\tau]})\,$ is independent
of the choices of $\,b_{z,z'}$. \,Indeed, under the transformations
(\ref{changes}),
\qq
\CV\ \longmapsto (\delta\CW')\,\CV\qquad{\rm for}\qquad
\CW'_{\tau;z}\,=\,\ee^{2\pi\si\m\sk\,\tr\,\tau a_z}\,.
\label{ch1}
\qqq 
\vskip 0.3cm

Let us first show that if Eq.\,\,(\ref{dwv0}) has 
a solution then by a transformation
(\ref{wtw}), which maps solutions to solutions, one may achieve
that $\,\CW^{ij}_{\tau;\m z}\,$ does not depend
on $\,(i,j)$. \,We shall use the fact that 
$\,\CV_{\tau;\m z,1}=\CV_{\tau;\m 1,z'}\equiv1$.
\,For $\,z,z'=1\m$, \,Eq.\,\,(\ref{dwv0}) reduces to
\qq
\CW^{jk}_{\tau;\m 1}\,\m{\CW^{ik}_{\tau;\m 1}}^{\hspace{-0.13cm}-1}\,
\CW^{ij}_{\tau;\m 1}\ =\ 1
\nonumber
\qqq
which means that, for each $\,\tau\m$,  $\,(\CW^{ij}_{\tau;\m 1})\,$ is a 
$\,U(1)$-valued $\check{\rm C}$ech $\,1$-cocycle for the covering
$\,(\CA_i)\,$ of the point $\,\{\tau\}$. \,Such a 1-cocycle is necessarily 
a coboundary:
\qq
\CW^{ij}_{\tau;\m 1}\ =\ {\CY^i_\tau}^{-1}\,\CY^j_\tau
\nonumber
\qqq
for $\,\CY^i_\tau\in U(1)$. \,By the transformation (\ref{wtw}),
we assure that $\,\CW^{ij}_{\tau;\m 1}=1$. \,Now, specifying  
Eq.\,\,(\ref{dwv0}) to $\,z=1$ or to $\,z'=1\m$, \,we infer that 
\qq
\CW^{jk}_{\tau;\m z'}\ =\ \CW^{ik}_{\tau;\m z'}\,,\quad\ 
\CW^{ij}_{\tau;\m z}\ =\ \CW^{ik}_{\tau;\m z}\,,
\nonumber
\qqq
i.e.\,\ that $\,\CW^{ij}_{\tau;\m z}\,$ is independent of $\,(i,j)$.
\,Note in passing that the condition $\,\CW^{ij}_{\tau;\m1}=1\,$ means 
that the isomorphism $\,\rho'\,$ restricts to $\,\rho\,$ over 
$\,Y_D^{[2]}\subset {Y'_{D'}}^{\hspace{-0.2cm}[2]}\,$ for $\,D=\CC_\tau\,$ 
with $\,\tau\subset[\tau]$.
\vskip 0.3cm
 
Without the $\,(i,j)$-dependence, $\,(\CW_{\tau;\m z})\equiv\CW\,$ 
forms a 1-cochain on group $\,Z\,$ with values in the $\,Z$-module 
$\,U(1)^{[\tau]}$. \,The identity (\ref{dwv0}) 
reduces then to the cohomological equation
\qq
\delta\CW\ =\ \CV
\label{dwv}
\qqq
where $\,(\delta\CW)_{\tau;\m z,z'}=\CW_{\tau_z;\m z'}
\,\CW_{\tau;\m zz'}^{-1}\,\CW_{\tau;\m z}$. \m Eq.\,\,(\ref{dwv}) 
has a solution if and only if the cohomology class 
$$\,[\CV]\ \in\ H^2(Z,U(1)^{[\tau]})\ \cong\  
H^2(Z_{[\tau]},U(1))\,$$ is trivial where, as above, $\,Z_{[\tau]}
\subset Z\,$ denotes the stabilizer subgroup of $\,\tau\in[\tau]$. \,
Vanishing of the cohomology class $\,[\CV]\,$ is a necessary and 
sufficient condition for the existence of a rank 1 symmetric 
$\,\CG'$-brane supported by the integrable conjugacy class $\,\CC'_{[\tau]}$. 
\,Two solutions of Eq.\,\,(\ref{dwv}) lead to isomorphic 
$\,\CG'$-branes if and only if they differ by a coboundary $\,\delta\CY\,$ 
with $\,(\delta\CY)_{\tau;\m z}=\CY_{\tau_z}\CY_\tau^{-1}\,$ for
$\,\CY\in U(1)^{[\tau]}$. In general, however, they may differ by
a $\,1$-cocycle on $\,Z\,$ with values in $\,U(1)^{[\tau]}$.
\,We infer that the cohomology group $$\,H^1(Z,U(1)^{[\tau]})\ \cong\  
H^1(Z_{[\tau]},U(1))\ =\ Z_{[\tau]}^*\m,$$ where $\,Z_{[\tau]}^*\,$ is the 
character group of $\,Z_{[\tau]}\m$, \,acts freely and transitively on 
the set of the isomorphism classes of symmetric rank 1 
$\,\CG'$-branes supported by $\,D'=\CC'_{[\tau]}\m$, \,if this set is non-empty. 
\,Recall that $\,Z_{[\tau]}\,$ is the fundamental group of the conjugacy 
class $\,\CC'_{[\tau]}\subset G'\,$ so that $\,Z_{[\tau]}^*\,$ may be
viewed as the group of isomorphism classes of flat line bundles over 
$\,\CC'_{[\tau]}$. \,Let $\,\CE'_1(h)\m,\ h=1,\dots,Z_{[\tau]}\m$,
\,denote the non-isomorphic 1-dimensional $\,\CG'_{D'}$-modules obtained 
from non-equivalent solutions $\,\CW\,$ of Eq.\,\,(\ref{dwv}).
They give rise to non-isomorphic rank 1 $\,\CG'$-branes
$\,\CD_1(h)=(D',\CE'_1(h))$. \,In such a situation, any $\,N$-dimensional 
$\,\CG'|_D'$-module $\,\CE'_N\,$ giving rise to a symmetric brane 
is equal, up to isomorphism, to a product of $\,\CE'_1(h)\,$ 
(for any $\,h$) by a flat vector bundle over $\,D'=\CC'_{[\tau]}$. 
\,Such vector bundles are, again up to isomorphism, direct sums of flat 
line bundles so we may set
\qq
\CE'_N\ =\ \CE'_1(h_1)\oplus\cdots\oplus\CE'_1(h_N)\,.
\nonumber
\qqq
We infer that if the cohomology class $\,[\CV]\,$ is trivial then
any rank $N$ $\,\CG'$-brane $\,\CD'_N\,$ supported by $\,\CC'_{[\tau]}\,$ 
is isomorphic to a stack of $\,N\,$ (in general, different) 1-dimensional 
branes $\,\CD'_1(h_n),\ h_n\in\{1,\dots,Z_{[\tau]}\}$. \,This is always 
the situation if $\,Z_{[\tau]}\,$ is a cyclic group since then 
$\,H^2(Z_{[\tau]},U(1))=\{1\}$.
\vskip 0.3cm

Let us describe more precisely the solutions $\,\CW\,$ 
of Eq.\,\,(\ref{dwv}). First note that if two solutions differ by a 
$\,2$-cocycle $\,\CW'\,$ that is a coboundary, $\,\CW'=\delta\CY\m$, 
\,then they 
coincide when restricted to $\,Z_{[\tau]}$: $\,\CW'_{\tau;\m z_0}=1\,$ 
(here and below, we use subscript zero for elements of $\,Z_{[\tau]}$). 
\,Conversely, if two solutions coincide if restricted to $\,Z_{[\tau]}\,$ 
then, setting $\,\CY_{\hat\tau_z}=\CW'_{\hat\tau,z}\,$ for a fixed 
$\,\hat\tau\in[\tau]\m$, \,we have $\,\CW'=\delta\CY$. \,A solution of 
Eq.\,\,(\ref{dwv}) satisfies
\qq
\CW_{\hat\tau_z;z_0}\ =\ \CW_{\hat\tau;zz_0}\,\CW_{\hat\tau;z}^{-1}
\,\CV_{\hat\tau;z,z_0}\ =\ \CW_{\hat\tau;z_0}\,\CV_{\hat\tau;z_0,z}^{-1}\,
\CV_{\hat\tau;z,z_0}
\label{afr}
\qqq 
so that its restriction to $\,Z_{[\tau]}\,$ is determined by the solution 
of the restricted equation with fixed $\,\hat\tau$
\qq
(\delta\CW)_{\hat\tau;z_0,z'_0}\ =\  \CW_{\hat\tau;z'_0}\,
\CW_{\hat\tau;z_0z'_0}^{-1}\,\CW_{\hat\tau;z_0}\ =\ \CV_{\hat\tau;z_0,z'_0}
\label{dwv1}
\qqq
involving the simpler cohomology of group $\,Z_{[\tau]}\,$ with values 
in $\,U(1)$. \,The general solution of the last equation has the form
\qq
\CW_{\hat\tau;z_0}\ =\ \CW^{\hat\tau}_{\hat\tau;z_0}\,\hat\chi(z_0)\,,
\nonumber
\qqq
where $\,\CW^{\hat\tau}_{\hat\tau;z_0}\,$ is a particular solution and
$\,\hat\chi\,$ is a character of $\,Z_{[\tau]}$. 
\vskip 0.3cm

Any solution of the restricted equation (\ref{dwv1}) may be extended to 
a solution of (\ref{dwv}) in the following way. First, fix a map 
$\,\,[\tau]\ni\tau\longmapsto z(\tau)\in Z\,\,$
such that $\,z(\tau)\tau=\hat\tau\,$ and, for $\,\tau\not=\hat\tau\m$, 
\,choose 
$\,\CW_{\hat\tau;z(\tau)}\in U(1)\,$ arbitrarily. Setting
\qq
\CW_{\hat\tau;z_0z(\tau)}\ =\ \CW_{\hat\tau;z(\tau)}\,\CW_{\hat\tau;z_0}\,
\CV_{\hat\tau;z_0,z(\tau)}^{-1}\,,
\nonumber
\qqq
one defines $\,\CW_{\hat\tau;z}\,$ for all $\,z\in Z\,$ in such a way
that the above equation holds also if $\,z(\tau)\,$ is replaced by any
$\,z\in Z$. \,Finally, with the help of the relation
\qq
\CW_{\hat\tau_z;z'}\ =\ \CW_{\hat\tau;zz'}\,\CW_{\hat\tau;z}^{-1}\,
\CV_{\hat\tau;z,z'}
\nonumber
\qqq
one defines a solution $\,\CW\,$ of Eq.\,\,(\ref{dwv}) which, in particular,
satisfies the identity (\ref{afr}). 
\vskip 0.3cm

If another fixed element $\,{\hat\tau}'=\hat\tau_z\,$ of the orbit 
$\,[\tau]\,$ is used in the construction above with  
\qq
\CW_{{\hat\tau}',z_0}\ =\ \CW^{{\hat\tau}'}_{{\hat\tau}',z_0}\,\hat\chi'(z_0)
\nonumber
\qqq
as the solution of the restricted equation
then the resulting solutions of the complete equation (\ref{dwv})
coincides with the one described previously if and only if
\qq
\hat\chi'(z_0)\ =\ \hat\phi_{\hat\tau;z}(z_0)\m\,\hat\chi(z_0)
\nonumber
\qqq
for 
\qq
\hat\phi_{\hat\tau;z}(z_0)\ =\ \CW^{\hat\tau}_{\hat\tau;z_0}\,
{\CW^{\hat\tau_z}_{\hat\tau_z;z_0}}^{\hspace{-0.2cm}-1}\,
\CV_{\hat\tau;z,z_0}\,\CV_{\hat\tau;z_0,z}^{-1}\,,
\label{phi}
\qqq
as may be easily seen with the help of relation (\ref{afr}). Note that 
$\,\hat\phi\,$ has the following properties:
\qq
&\hat\phi_{\hat\tau;z_0z}(z'_0)\,=\,\hat\phi_{\hat\tau;z}(z'_0)\,,\ \ \quad
\hat\phi_{\hat\tau;z}(z_0z'_0)\,=\,\hat\phi_{\hat\tau;z}(z_0)\,
\hat\phi_{\hat\tau;z}(z'_0)\,,&\cr \\
&\hat\phi_{\hat\tau_z;z'}(z_0)\,\hat\phi_{\hat\tau;zz'}(z_0)^{-1}\,
\hat\phi_{\hat\tau;z}(z_0)\,=\,1\,.&
\nonumber
\qqq
They mean that $\,\hat\phi\,$ defines a $\,1$-cocycle on group 
$\,Z\,$ with values in the module of $\,Z_{[\tau]}^*$-valued
functions on the orbit $\,[\tau]\,$ which, due to the first relation, 
descends to a cocycle on the quotient group $\,Z/Z_{[\tau]}$. 
\,The cocycle $\,\hat\phi\,$ is not unique. 
If we multiply the special solutions $\,\CW^{\hat\tau}_{\hat\tau,z_0}\,$ 
by $\,\hat\tau$-dependent characters $\,\hat\chi_{\hat\tau}(z_0)\,$ of
$\,\CZ_{[\tau]}\,$ then \qq \hat\phi_{\hat\tau;z}(z_0)\ \longmapsto\
\hat\phi_{\hat\tau;z}(z_0)\,\,
\hat\chi_{\hat\tau}(z_0)\,\chi_{\hat\tau_z}(z_0)^{-1}\,.
\nonumber
\qqq 
\vskip 0.3cm

Given cocycle $\,\hat\phi\m$, \,we may identify the set of symmetric rank 1 
$\,\CG'$-branes supported by $\,\CC'_{[\tau]}\,$ with the set 
of equivalence classes $\,[\hat\tau,\hat\chi]\,$ of pairs $\,(\hat\tau,
\hat\chi)\,$ 
with $\,\hat\tau\in[\tau]$ and $\,\hat\chi\in Z_{[\tau]}^*\,$ such that
\qq
(\hat\tau,\hat\chi)\ \ \sim\ \ ({\hat\tau}',\hat\chi')\ \ \ {\rm if}\ \ 
{\hat\tau}'=\hat\tau_z\ \ {\rm and}\ \ \hat\chi'=
\hat\phi_{\hat\tau;z}(\,\cdot\,)
\,\hat\chi\,.
\nonumber
\qqq
Such a description of branes agrees with the general one conjectured in
\cite{FHSSW,SchwFW} for conformal field theories of the simple current 
extension type. The general classification of the
branes proposed there, based on consistency considerations,
involved equivalence classes of primary fields and characters of
their ``central stabilizers''. For rank 1 branes of the WZW models, 
the latter reduce to the ordinary stabilizer subgroups $\,\CZ_{[\tau]}$ 
in the simple current group $\,\CZ$. 
\vskip 0.3cm

In Appendix A we list solutions of Eq.\,\,(\ref{dwv}) giving
rise to all (up to isomorphism) symmetric rank 1 $\,\CG'$-branes supported 
by the conjugacy classes of non-simply connected groups $\,G'=G/Z\,$
with covering groups $\,G\,$ running through the Cartan list.
We use the description of the Weyl alcoves and of choices of 
the 2-chains $\,b\,$ and $\,V\,$ entering Eqs.\,\,(\ref{bzz'}) 
and (\ref{cv}) taken from \cite{GR2}\footnote{$\,V_{z,z'}\,$ 
is equal to $\,\ee^{-2\pi\si\m\sk\,\tr\,\tau_{zz'0}\m b_{z,z'}}\,u_{z,z'}\,$
with $\,u_{z,z'}\,$ defined and calculated in \cite{GR2}}. \,We specify:
\vskip 0.25cm

\noindent\ \ \ \,the action of the center of  $\,G\,$ on the Weyl alcove, 

\noindent\ \ \ \,the restrictions on $\m n_i\geq0$, 
$\m\sum\limits_{i=0}^rn_i=\sk$, \,such that $\,\sum\limits_{i=0}^r
n_i\tau_i=\sk\tau\equiv\lambda\,$ is a weight,

\noindent\ \ \ \,the subgroups $\,Z\,$ and the restrictions on the level 
$\,\sk\,$ of gerbes $\,\CG'\,$ over $\,G/Z\m$,
\vskip 0.15cm

\noindent\ \ \ \,the 2-chains $\,V\m$, 
\vskip 0.15cm

\noindent\ \ \ \,the 2-cocycles $\,\CV\m$, 
\vskip 0.15cm

\noindent\ \ \ \,possible stabilizer subgroups of $\,Z\m$,
\vskip 0.15cm

\noindent\ \ \ \,the special solutions $\,\CW^{\hat\tau}\,$ 
of Eq.\,\,(\ref{dwv}) for different $\,Z$-orbits $\,[\tau]\m$,
\vskip 0.15cm

\noindent\ \ \ \,general solutions leading to non-isomorphic $\,\CG'$-branes.  
\vskip 0.25cm
 
\noindent Describing 2-cocycle $\,\CV\,$ we use the 
the monodromy charge $\,Q_z(\lambda)\,$ defined 
for integrable weights $\,\lambda\,$ by
\qq
Q_z(\lambda)\ =\ h(\lambda)\,+\,h(k\tau_{z0})\,-\,h(z\lambda)
\ \,{\mathrm mod}\,\m1\ =\ \tr\,\lambda\tau_{0_z}\  
{\mathrm mod}\,\m1
\nonumber
\qqq
where $\,h(\lambda)\,$ is the conformal weight of the chiral primary field
with weight $\,\lambda\,$ and, by definition, $\,z\lambda=\sk\m z\tau\,$
if $\,\lambda=\sk\tau$. \,In most cases, the special solutions 
$\,\CW^{\hat\tau}_{\tau;
\m z,z'}\,$ are independent of $\,\hat\tau,\tau\in[\tau]\,$ as may be seen
from the restrictions on coefficients $\,n_i\,$ together with
the relation $\,\sum n_i=\sk\,$ combined with the restrictions on $\,\sk$.
Such independence implies that the cocycles $\,\hat\phi\,$ 
of Eq.\,\,(\ref{phi}) are trivial and the description of branes
by the equivalence classes $\,[\hat\tau,\hat\chi]\,$ is superfluous. 
The exception is the group $\,G'=Spin(4n)/({\NZ_2
\hspace{-0.06cm}\times\hspace{-0.06cm}\NZ_2})
=SO(4n)/\NZ_2\,$ where several cases give rise to non-trivial cocycle 
$\,\hat\phi$. \,Also for that group, which is the only one with non-cyclic 
fundamental group $\,Z\cong{\NZ_2\hspace{-0.06cm}\times
\hspace{-0.06cm}\NZ_2}\,$ and multiple (double) choice of 
stably-non-isomorphic gerbes $\,\CG'$, \,for one of those gerbes 
there are no (scalar) solutions of Eq.\,\,(\ref{dwv}) for single-point 
orbits $\,[\tau]\,$ (i.e.\,\,if $\,Z_{[\tau]}\cong{\NZ_2\hspace{-0.06cm}
\times\hspace{-0.06cm}\NZ_2}$). \,Consequently, there are no symmetric 
rank 1 $\,\CG'$-branes supported by the corresponding (integrable) 
conjugacy classes $\,\CC'_{[\tau]}\subset G'$. \,We shall devote 
the next section to the analysis of higher-rank $\,\CG'$-branes 
supported by such conjugacy classes.

\nsection{Non-Abelian symmetric branes for $\,SO(4n)/\NZ_2$}  
\label{sec:nasb}
  
For group $\,G'=Spin(2r)/Z\,$ with even rank $\,r=4,6,\dots\,$ and
$\,Z\cong{\NZ_2\hspace{-0.06cm}\times\hspace{-0.06cm}\NZ_2}\m$, 
\,the integrable conjugacy classes $\,\CC'_{[\tau]}\,$ with $\,Z_{[\tau]}
\cong{\NZ_2\hspace{-0.06cm}\times\hspace{-0.06cm}\NZ_2}\,$ correspond 
to 1-point orbits $\,[\tau]\,$ with
\qq
&{\displaystyle\ \ \sk\tau\ =\ \sum\limits_{i=0}^rn_i\tau_i\,,
\qquad\qquad\ n_i\geq0\,,\qquad\qquad\ \sum\limits_{i=0}^rn_i\ =\ \sk\,,}&\cr
&{\displaystyle n_0=n_1=n_{r-1}=n_r\in\NZ\,,\quad\  
n_i=n_{r-i}\in2\NZ\m,\ \,i=2,\dots,r-2\,.}
\nonumber
\qqq
Such conjugacy classes exist if and only if $\,\sk\,$ is even. 
The two inequivalent choices $\,\CG'_\pm\,$ for the gerbe over
$\,G'\,$ correspond to the upper and lower choices of signs in
the 2-chain $\,V\,$ that  enters the formula (\ref{mup}) for the groupoid
multiplication, 
\qq
V_{z,z'}\ =\ \cases{\ \hbox to 2cm{$\pm 1$\hfill}\hbox to 2cm{{\rm if}\hfill
$(z,z')=\,$}(z_2,z_1),(z_2,z_1z_2),\cr
\ \hbox to 4cm{\ \hfill}\m(z_1z_2,z_1),(z_1z_2,z_1z_2)\,,\cr
\ \hbox to 2cm{$\,\,1$\hfill}{\rm otherwise}}
\label{vzzp}
\qqq
where $\,z_1,z_2\,$ are the generators of $\,Z\m$, \,see Sec.\,4.4.II of 
\cite{GR2}. \,For the 1-point orbits $\,[\tau]\m$, \,the 2-cocycle 
$\,\CV\,$ of Eq.\,\,(\ref{cv}) reduces to the same expression
\qq
\CV_{\tau;\m z,z'}\ =\ V_{z,z}
\label{evod}
\qqq
and defines for the lower choice of signs in \ref{vzzp} a non-trivial 
cohomology class 
$\,[\CV]\in H^2({\NZ_2\hspace{-0.06cm}\times\hspace{-0.06cm}\NZ_2},U(1))
\cong\NZ_2\,$ which is an obstruction to the existence of solutions of 
Eq.\,\,(\ref{dwv}) and of  symmetric rank 1
$\,\CG'_-$-branes supported by the conjugacy class $\,\CC'_{[\tau]}$.
The simplest instances with $\,\sk>0\,$ occur for group
$\,G'=Spin(8)/({\NZ_2\hspace{-0.06cm}\times\hspace{-0.06cm}\NZ_2})
=SO(8)/\NZ_2\,$  for $\,\tau=\tau_2\,$ at levels $\,\sk=2,4,\dots\,$ and 
for $\,\tau={1\over 4}(\tau_0+\tau_1+\tau_3+\tau_4)\,$ at $\,\sk=4,6,\dots$ 
in conventions of \cite{GR2}.
\vskip 0.3cm

We shall search now for symmetric rank $N$ $\,\CG'_-$-branes 
supported by the integrable conjugacy classes $\,\CC'_{[\tau]}\,$ with 
1-point orbits $\,[\tau]$. \,As in the rank 1 case,
such a brane $\,\CD'_N=(D',\CE'_N)\,$ induces by restriction
a brane $\,(D,\CE_N)\,$ for the $\,Spin(2r)\,$ theory with the support 
$\,D\,$ given by the conjugacy class $\,\CC_\tau\subset Spin(2r)$. \,If 
$\,\CE'_N=(E',\rho')\,$ then $\,\CE_N=(E,\rho)\,$ with coinciding 
vector bundles $\,E'=E\,$ over $\,Y'_{D'}=Y_D$. \,Since the 
symmetric rank $N$ $\,\CG$-branes are isomorphic to stacks of $\,N\,$
rank 1 branes, we may assume, passing at most to
an isomorphic $\,\CG'_-|_{D'}$-module $\,\CE'_N\m$, \,that 
over each connected component $\,P_i|_{\CC_\tau}\cong Y_{i\tau}\,$ 
of $\,Y'_{D'}\m$, \,see (\ref{ydp}), the  vector bundle 
$\,E'\,$ restricts to the bundle $\,L_{i\tau}\otimes\NC^N$. 
\,Elements $\,\ell_{i\tau}\in L_{i\tau}\otimes\NC^N\,$ in the fiber 
over $\,y=(g,h\gamma^{-1})\in P_i|_{\CC_\tau}\,$ are now determined 
by the equivalence classes $\,\hat\ell_{i\tau}=[\gamma,v]_i\,$ defined 
as in (\ref{eri}) but with $\,v\in\NC^N\,$ rather than $\,v\in\NC$. 
\,With this modification, the discussion around Eqs.\,\,(\ref{rhod'}) 
to (\ref{cv}) may be repeated word by word with 
$\,\CW^{ij}_{\tau;\m z}\,$ and $\,\CY^i_\tau\,$ being now unitary 
$\,N\times N\,$ matrices so that their order in the formulae, previously
irrelevant, becomes important. The same argument as before shows that 
$\,\CW^{ij}_{\tau;\m z}\,$ may be chosen $\,(i,j)$-independent. 
\,Eq.\,\,(\ref{dwv0}) reduces then to the matrix relation
\qq  
\CW_{\tau;\m z}\,\CW_{\tau;\m z'}\ =\ \CV_{\tau;\m z,z'}\,\,\CW_{\tau;\m zz'}
\label{dwvna}
\qqq
which, with the use of Eqs.\,\,(\ref{evod}) and (\ref{vzzp}) with the lower 
choice for the signs, implies that $\,\CW_{\tau;\m 1}=1\,$ and that
\qq
\CW_{\tau;\m z_1}^2\,=\,1\,=\,\CW_{\tau;\m z_2}^2\,,\quad\ 
\CW_{\tau;\m z_1}\,\CW_{\tau;\m z_2}\,+\,\CW_{\tau;\m z_2}\,
\CW_{\tau;\m z_1}\,=\,0
\nonumber
\qqq
with $\,\CW_{\tau;\m z_1z_2}=\CW_{\tau;\m z_1}\m\CW_{\tau;\m z_2}$.
\,In other words, matrices $\,\CW_{\tau;\m z}\,$ define an
$\,N$-dimensional representation of the 2-dimensional Clifford 
algebra. Unitarily equivalent representations correspond to isomorphic 
$\,\CG'_-|_{D'}\,$ modules i.e. to isomorphic $\,\CG'_-$-branes and 
{\it vice versa}. The lowest dimensional representation is of dimension 
2 and is given by the Pauli matrices 
\qq
\CW_{\tau;\,z_1}\ =\ \left(\matrix{0&1\cr 1&0}\right)\,,\quad\ \ 
\CW_{\tau;\m z_2}\ =\ \left(\matrix{0&-\si\cr\si&0}\right)\,,\quad\ 
\CW_{\tau;\m z_1z_2}\ =\ \si\left(\matrix{1&0\cr 0&-1}\right).
\label{pauli}
\qqq
It is unique up to unitary equivalence. It gives rise to a 
2-dimensional $\,\CG'_-|_{D'}$-module $\,\CE'_2\m$, \,hence to a rank 2
$\,\CG'_-$-brane $\,\CD'_2$. \,The higher dimensional 
representations exist in even dimensions $\,N\,$ and are unitarily 
equivalent to direct sums of the 2-dimensional representation so that,
up to isomorphism,
\qq
\CE'_N\ =\ \mathop{\CE'_2\oplus\cdots\oplus\CE'_2}\limits_{N/2\ {\rm times}}\,.
\nonumber
\qqq
\vskip 0.3cm

We infer that for the gerbe $\,\CG'_-\,$ over $\,SO(4n)/\NZ_2\,$ 
corresponding to the lower choice of signs in (\ref{vzzp}) and
even level $\,\sk\m$, \,the symmetric rank $N$ $\,\CG'_-$-branes supported 
by the integrable conjugacy classes $\,\CC'_{[\tau]}\,$ with 1-point
orbits $\,[\tau]\,$ exist only for $\,N\,$ even and are isomorphic
to a stack of $\,N/2\,$ rank 2 non-Abelian $\,\CG'_-$-branes $\,\CD'_2\,$ 
determined by the solution (\ref{pauli}) of Eq.\,\,(\ref{dwvna}).
In the algebraic classification of branes \cite{FHSSW,SchwFW,Walch}, see also
\cite{Schellkac}, based on the simple current technique, those are the 
cases where the ``central stabilizer'' subgroup of the simple current group 
$\,Z\,$ is trivial hence smaller than the ordinary stabilizer $\,Z_{[\tau]}$. 
\,That the branes corresponding to such situations should be counted among
the boundary conditions follows by demanding that the total number of such 
conditions be equal to the dimension $\,I\,$ of the set of the Ishibashi 
states in the bulk sector of the theory \cite{FHSSW,SchwFW}. The simplest 
example of such a count is the case of group $\,G'=SO(8)/\NZ_2\,$
at level $\,\sk=2\,$ where the space of Ishibashi states 
has dimension $\,I_+=11\,$ and $\,I_-=8\,$ for the bulk theories related to
gerbes $\,\CG'_\pm\m$, \,respectively. \,There are five orbits $\,[\tau]\,$
in the Weyl alcove $\,\CA_{_W}\,$ such that $\,\sk\tau\,$ is a weight. 
\,With the choice of $\,\CA_{_W}\,$ and its vertices $\,\tau_i\,$ 
as in \cite{GR2}
leading to the $Z$-action $\,z_1\tau_i=\tau_{4-i}\,$ $\,z_2\tau_0=\tau_1\m$,
$\,z_2\tau_2=\tau_2\m$, $\,z_2\tau_3=\tau_4\m$, \,these orbits are:
\qq
&\{\tau_0,\,\tau_1,\,\tau_3,\,\tau_4\}\m,&\cr
&\{\hf\tau_0+\hf\tau_1,\,\hf\tau_3+\hf\tau_4\}\m,\quad\{\hf\tau_0+\hf\tau_3,
\,\hf\tau_1+\hf\tau_4\}\m,\quad\{\hf\tau_0+\hf\tau_4,\,\hf\tau_1+\hf\tau_3\}\m,&\cr
&\{\tau_2\}\,.&
\nonumber       
\qqq
For the boundary theory related to gerbe $\,\CG'_+\m$, \,each 
of those corresponds
to integrable conjugacy class carrying $\,|Z_{[\tau]}|\,$ non-isomorphic
symmetric rank 1 branes, so altogether we obtain $\,1+2+2+2+4=11\,$ branes.
For the theory related to $\,\CG'_-\m$, \,the conjugacy class corresponding
to the 1-point orbit $\,\{\tau_2\}\,$ gives rise to a rank 2 brane unique
up to isomorphism and the brane count becomes $\,1+2+2+2+1=8$.

\nsection{Boundary partition functions and operator product}
\label{sec:bpfop}

\subsection{Boundary states for simply connected target groups}
\label{sec:bssctg}

Although obtained by classical considerations, the geometric classification 
of branes in the WZW model permits to elucidate the quantized 
theory by providing structures that manifest themselves 
directly on the quantum level. The spaces of states in the quantum WZW 
theory may be realized with the use of geometric quantization
as spaces of sections of vector bundles over the group path spaces.
Such vector bundles are canonically associated to the gerbe $\,\CG\,$ 
on the group and to pairs of $\,\CG$-branes $\,\CD^s$, $\,s=0,1$. \,More 
concretely, for $\,\CL\,$ the line bundle over the loop group $\,\LG\,$ 
and $\,\CE_{\hspace{-0.02cm}\CD^0}^{\CD^1}\,$ the vector bundle over 
the space of open paths $\,\IG_{\hspace{-0.05cm}D^0}^{D^1}\,$ in 
the group extending between the brane supports, 
see Sect.\,\ref{sec:asacymf}, the spaces of sections\footnote{For 
positivity of energy, the level $\,\sk\,$ of the theory
has to be taken positive}
\qq
\NH\ =\ \Gamma(\CL)\,,\qquad\quad\NH_{\CD^0}^{\m\CD^1}\ 
=\ \Gamma(\CE_{\hspace{-0.02cm}\CD^0}^{\CD^1})
\label{gr}
\qqq
provide, respectively, the spaces of bulk and of boundary states
of the theory. The first one carries the action of a left-right current
algebra $\,\hat\Ng\oplus\hat\Ng\,$ whereas the second one that of
the chiral current algebra $\,\hat\Ng$. \,The analysis of the 
representation content (spectrum) of the space of bulk states for
the WZW models with all compact, simple but possibly non-simply connected 
groups has been performed in \cite{FGK}. For the space of boundary states, 
the decomposition into the irreducible highest weight representations
of the current algebras has the form
\qq
\NH_{\CD^0}^{\m\CD^1}\ \cong\ \mathop{\oplus}
\limits_{\lambda}\,\Big(M_{\CD^0\lambda}^{\,\CD^1}
\otimes \hat V_{\bar\lambda}\Big)
\label{decc}
\qqq
where the direct sum is over 
the integrable weights $\,\lambda\,$ and spaces $\,\hat V_{\bar\lambda}\,$ 
carry the irreducible representations of the current algebra $\,\hat\Ng\,$
of level $\,\sk\,$ and highest weight $\,\bar\lambda$\footnote{$\bar\lambda$ 
denotes the highest weight of the irreducible representation of $\,\Ng\,$ 
complex conjugate to the one with the highest weight $\,\lambda$.}.
For the theory with the simply connected target group $\,G\,$ and
rank $\,N_s\,$ symmetric $\,\CG$-branes $\,\CD^s\,$ that are stacks
of rank 1 branes supported by the integrable conjugacy classes 
$\,C_{\tau^s}\,$ corresponding to weights $\,\lambda^s=\sk\tau^s\m$, 
\,the multiplicity spaces take the product form  
\qq
M_{\CD^0\lambda}^{\,\CD^1}\ =\ 
\NC^{N_0}\otimes\overline{\NC^{N_1}}\otimes M_{\lambda^0\lambda}^{\,
\lambda^1}\,.
\label{decc1}
\qqq 
The dimensions of the multiplicity spaces 
$\,M_{\lambda^0\lambda}^{\,\lambda^1}\,$ for the rank 1
branes are equal to the fusion coefficients 
$\,N_{\lambda^0\lambda}^{\,\lambda^1}\,$ so
that the {\bf boundary partition functions} of the WZW theory with a simply
connected target group are given by the expressions
\qq
Z_{\hspace{-0.02cm}\CD^0}^{\CD^1}(T)\ \equiv\ 
{\mathrm Tr}_{\NH_{\CD^0}^{\m\CD^1}}
\,\ee^{-T\m L_0}\ 
=\ N_0N_1\sum\limits_\lambda N_{\lambda^0\lambda}^{\,\lambda^1}\ 
\hat\chi_{_{\hspace{-0.02cm}\bar\lambda}}({_{\si T}\over^{2\pi}})
\nonumber
\qqq
where the (restricted) affine characters 
$\,\hat\chi_{_\lambda}({{\si T}\over{2\pi}})=
{\mathrm Tr}_{\hat V_\lambda}\ee^{-T\m L_0}\,$ with $\,L_0\,$ denoting 
the Virasoro generator given by the Sugawara construction.
\vskip 0.3cm

The multiplicity spaces $\,M_{\lambda^0\lambda}^{\,\lambda^1}\,$ 
may be thought of in different ways, depending on the situation.
Firstly, they may be naturally identified \cite{GTT,GawGH} with
the spaces of 3-point genus zero conformal blocks of the bulk, group 
$\,G\,$ WZW theory with insertions of the primary fields corresponding 
to the integrable weights $\,\lambda^1$, $\,\bar\lambda^0\,$ 
and $\,\bar\lambda\,$ or, in other words, as spaces of properly 
defined intertwiners of the current algebra $\,\hat\Ng\,$ actions 
in the spaces $\,\hat V_{\lambda^1}\,$ and $\,\hat V_{\lambda^0}\otimes
\hat V_{\lambda}$. \,Secondly, they may be identified with the 
``fusion rule'' subspace of the intertwiners of the Lie algebra $\,\Ng\,$ 
action 
\qq
{\mathrm Hom}^{\mathrm FR}_{\m\Ng}\hspace{-0.02cm}(V_{\lambda^1},\m 
V_{\lambda^0}\otimes V_\lambda)\ \subset\ {\mathrm Hom}_{\m\Ng}(V_{\lambda^1},
\m V_{\lambda^0}\otimes V_\lambda)
\nonumber
\qqq
where $\,V_\lambda\,$ stands for the space of the irreducible 
highest weight $\,\lambda\,$ representation of $\,\Ng$. \,The relation 
with the previous picture is that the Lie algebra intertwiners 
in the fusion rule subspace are the ones that extend to the current 
algebra intertwiners (such extensions are unique). Finally, the spaces 
$\,M_{\lambda^0\lambda}^{\,\lambda^1}\,$ may be thought of as composed 
of the (``good'') intertwiners of the action of the quantum deformation
$\,\CU_q(\Ng)\,$ of the enveloping algebra of $\,\Ng\,$ in its highest
weight modules. 
\vskip 0.3cm

Different realizations of spaces $\,M_{\lambda^0\lambda}^{\,\lambda^1}\,$ 
are more convenient in different contexts. We shall need still another
realization that is derived from the one based 
on the fusion rule intertwiners of the Lie algebra action.
We shall need a more concrete description of the spaces 
$\,{\mathrm Hom}^{\mathrm FR}_{\m\Ng}\hspace{-0.02cm}(V_{\lambda^1},\m
V_{\lambda^0}\otimes V_\lambda)$. \,Consider the linear mapping
\qq
{\mathrm Hom}_{\m\Ng}(V_{\lambda^1},\m V_{\lambda^0}\otimes V_\lambda)\,
\ni\,\psi\ \longmapsto\ |\psi\rangle\,\in\, V_\lambda
\nonumber
\qqq
such that for all $\,|v\rangle\in V_\lambda\,$ and the highest weight vectors
$\,|\lambda^s\rangle\in V_{\lambda^s}\,$ annihilated by
the step generators $\,e_\alpha\,$ for positive roots $\,\alpha\,$
of $\,\Ng\m$,
\qq
\langle v\,|\,\psi\rangle\ = \langle\lambda^0\otimes v\,|\,\psi\,|\m\lambda^1
\rangle\,.
\nonumber
\qqq
It is easy to see that $\,|\m\psi\rangle\,$ determines
the intertwiner $\,\psi\,$ uniquely so that 
$\,{\mathrm Hom}_{\m\Ng}(V_{\lambda^1},\m
V_{\lambda^0}\otimes V_\lambda)\,$ and its fusion rule subspace
may be identified with subspaces in $\,V_\lambda$. \,The latter is 
characterized by the conditions 
\qq
&&\hbox to 5cm{$t\,|\m\psi\rangle\ =\ \tr\,t\m(\lambda^1-\lambda^0)\,|
\m\psi\rangle$\hfill}{\rm for}\quad t\in\Nt\,,\label{cint1}\\
&&\hbox to 5cm{$e_{-\alpha_i}^{\tr\,\alpha_i^\vee\lambda^1\m+\m1}\,
|\m\psi\rangle\ =\ 0$\hfill}{\rm for}\quad i=1,\dots,r\,,
\label{cint2}\\
&&\hbox to 5cm{$e_{\phi}^{\sk\m-\m\tr\,\phi^\vee\lambda^1\m+\m1}\,|\m
\psi\rangle\ =\ 0$\hfill}
\label{cint3}
\qqq
whereas for the former, the $3^{\rm rd}$ condition should be dropped.
Above $\,\Nt\,$ denotes the Cartan subalgebra of $\,\Ng\m$,  
$\,\alpha_i^\vee$ the simple coroots and $\,\phi,\,\phi^\vee\,$ the highest 
root of $\,\Ng\,$ and its coroot. \,Equivalently, one may 
replace conditions (\ref{cint2}) and (\ref{cint3}) by
\qq
&&\hspace{-0.07cm}\hbox to 5cm{$e_{\alpha_i}^{\tr\,\alpha_i^\vee\lambda^0\m
+\m1}\,|\m\psi\rangle\ =\ 0$\hfill}{\rm for}\quad i=1,\dots,r\,,
\cr
&&\hspace{-0.07cm}\hbox to 5cm{$e_{-\phi}^{\sk\m-\m\tr\,
\phi^\vee\lambda^0\m+\m1}\,|\m
\psi\rangle\ =\ 0\,.$\hfill}
\nonumber
\qqq
Below, we shall identify the multiplicity space $\,M_{\lambda^0
\lambda}^{\,\lambda^1}\,$ with the subspace
of $\,V_\lambda\,$ composed of vectors satisfying (\ref{cint1}),
(\ref{cint2}) and (\ref{cint3}):
\qq
M_{\lambda^0\lambda}^{\,\lambda^1}\ \subset\ V_\lambda\,.
\label{wlb}
\qqq
$M_{\lambda^0\lambda}^{\,\lambda^1}\,$ inherits this way the scalar 
product from $\,V_\lambda$. 
\vskip 0.3cm

The latter description of the multiplicity spaces is particularly natural
in the geometric realization (\ref{gr}) of the spaces of states where 
it may be obtained by evaluating sections of the vector bundle 
$\,\CE_{\hspace{-0.02cm}\CD^0}^{\CD^1}\,$ on the special path
\qq
[0,\pi]\,\ni\,x\ \mapsto\ g_{_{\hspace{-0.03cm}\tau^0\tau^1}}(x)=
\ee^{2\pi\si\m(\tau^0+x(\tau^1-\tau^0)/\pi)}
\nonumber
\qqq
in $\,G\,$ extending between the brane supports 
$\,D^0\,$ and $D^1$. \,Such evaluation
takes values in the fiber of $\,\CE_{\hspace{-0.02cm}\CD^0}^{\CD^1}$
over $\,g_{_{\hspace{-0.03cm}\tau^0\tau^1}}\,$ which may be canonically 
identified with $\,\NC^{N_0}\otimes\overline{\NC^{N_1}}$. \,The 
composition of natural injections
\qq
&\displaystyle{{\mathrm Hom}_{\m\hat\Ng}\left(\hat V_{\bar\lambda},
\,\Gamma(\CE_{\hspace{-0.02cm}\CD^0}^{\CD^1})\right)\ 
\,\hookrightarrow\,\ {\mathrm Hom}_{\m\Ng}\left(V_{\bar\lambda},
\,\Gamma(\CE_{\hspace{-0.02cm}\CD^0}^{\CD^1})\right)\ \,\hookrightarrow
\,\ \Gamma(\CE_{\hspace{-0.02cm}\CD^0}^{\CD^1})
\otimes V_{\lambda}}\cr\label{emb}\\
&\displaystyle{\hookrightarrow\ \,\NC^{N_0}\otimes\overline{\NC^{N_1}}
\otimes V_\lambda\,,}&
\nonumber
\qqq
where the last but one injection uses the identification between the dual
space of $\,V_{\bar\lambda}\,$ and $\,V_{\lambda}\,$ and the last one
evaluates the sections of $\,\CE_{\hspace{-0.02cm}\CD^0}^{\CD^1}$ at 
$\,g_{\tau^0\tau^1}\m$, \,embeds the multiplicity space into
$\,\overline{\NC^{N_1}}\otimes\NC^{N_0}\otimes V_\lambda$.

\subsection{Boundary states for non-simply connected target groups}
\label{sec:bsnsctg}

As pointed out in Sect.\,9.2 of \cite{GR1}, the geometric realization
(\ref{gr}) of the boundary spaces of states implies a simple relation
between those spaces for the WZW models with a simply-connected
target group $\,G\,$ and with the quotient target $\,G'=G/Z$.
\,In one phrase, the latter space of states is composed of the 
$\,Z$-invariant states of the former. More exactly, for symmetric 
$\,\CG'$-branes $\,{\CD'}^s$, $s=0,1$, \,supported by the 
integrable conjugacy classes $\,\CC'_{[\tau^s]}\subset 
G'$, \,let $\,\CD^s\,$ run through the $\,\CG$-branes 
supported by the conjugacy classes $\,\CC_{\tau^s}\,$ with 
$\,\tau^s\in[\tau^s]\,$ obtained by the pullback of branes $\,{\CD'}^s$.
There is one such $\,\CG$-brane $\,\CD^s\,$ for each $\,\tau^s\in[\tau^s]$. 
\,We shall denote by $\,\CD^s_{\hspace{-0.03cm}z}\,$ the brane supported by 
$\,z^{-1}\CC_{\tau^s}=\m\CC_{\tau^s_z}\,$ if $\,\CD^s\,$ is supported 
by $\,\tau^s$. \,As has been
discussed in \cite{GR1}, the $\,{\CD'^s}\,$ brane structures allow to lift 
naturally the action of the subgroup $\,Z\,$ of the center of $\,G\,$ to 
the vector bundles $\,\CE_{\hspace{-0.02cm}\CD^0}^{\CD^1}\m$:
\qq
&&\CE_{\hspace{-0.02cm}\CD^0_z}^{\CD^1_z}\quad\ \,\,\,
\mathop{\longrightarrow}\limits^{\hat z}
\quad\CE_{\hspace{-0.02cm}\CD^0}^{\CD^1}\cr
&&\hspace{0.23cm}\downarrow\hspace{2.25cm}\downarrow\cr
&&\IG_{\hspace{-0.05cm}D^0_z}^{D^1_z}\quad
\mathop{\longrightarrow}\limits^{z}\quad
\IG_{\hspace{-0.05cm}D^0}^{D^1}
\nonumber
\qqq
with the lower line given by multiplication by $\,z$. \,The action 
of $\,Z\,$ on the vector bundles $\,\CE_{\hspace{-0.02cm}\CD^0}^{\CD^1}\,$ 
satisfies $\,\hat z\m\hat z'=\widehat{zz'}\,$ and induces 
the action on the sections:
\qq
(\NU(z)\Psi)(g)\ =\ \hat z\,\Psi(z^{-1}g)
\nonumber
\qqq
for $\,\Psi\in\Gamma(\CE_{\hspace{-0.02cm}\CD^0_z}^{\CD^1_z})$. \,One obtains
in this way a representation $\,\NU\,$ of $\,Z\,$ in the space
\qq
\widetilde\NH'^{{\m\CD'}^1}_{{\CD'}^0}\ 
=\ \mathop{\oplus}\limits_{(\CD^0,\CD^1)}
\NH_{\CD^0}^{\m\CD^1}
\nonumber
\qqq
containing all the $\,G$-theory states compatible with the ones in
$\,\NH'^{\m{\CD'}^1}_{{\CD'}^0}$. Operators $\,\NU(z)\,$ map 
$\,\NH_{\CD^0_z}^{\m\CD^1_z}\,$ to $\,\NH_{\CD^0}^{\m\CD^1}$.
\,They commute with the action of the current algebra
so that through the decomposition (\ref{decc}) they induces the maps
\qq
U_{\CD^0\lambda}^{\,\CD^1}(z)\,:\ 
\NC^{N_0}\otimes\overline{\NC^{N_1}}\otimes
M_{\lambda^0_z\lambda}^{\,\lambda^1_z}
\ \longrightarrow\ \NC^{N_0}\otimes\overline{\NC^{N_1}}
\otimes M_{\lambda^0\lambda}^{\,\lambda^1}
\nonumber
\qqq
on the multiplicity spaces where
$\,z\lambda=\sk\m z\tau\,$ for $\,\lambda=\sk\tau\,$ and 
$\,\lambda_z=\sk\tau_z=\sk\m z^{-1}\tau$.
\,A closer inspection of the the embeddings (\ref{emb})  
shows that, in the realization (\ref{wlb}) of spaces 
$\,M_{\lambda^0\lambda}^{\,\lambda^1}\m$,
\qq
U_{\CD^0\lambda}^{\,\CD^1}(z)\ =\ 
\CW^0_{\lambda^0;\m z}\ \otimes\ \overline{\CW^1_{\lambda^1;\m z}}\ 
\otimes\ w_z
\label{Uz}
\qqq
where $\,\CW^s_{\lambda^s;z}\,$ are the 
solutions of the cohomological equation (\ref{dwv}) (with values in 
$\,N_s\times N_s\,$ matrices) relabeled, for notational convenience,
by the weights $\,\lambda^s=\sk\tau^s$. \,Recall that the solutions
$\,\CW^s\,$ define the $\,\CG'|_{{\CD'}^s}$-modules $\,{\CE'}^s\,$ such 
that $\,{\CD'}^s=(\CC'_{[\tau^s]},{\CE'}^s)$. \,
The elements $\,w_z\in G\,$ are as in (\ref{wz}). Their action in
$\,V_{\lambda}\,$ intertwines the multiplicity spaces:
\qq
w_z\m:\ M_{\lambda^0_z\lambda}^{\,\lambda^1_z}\ 
\longrightarrow\ M_{\lambda^0\lambda}^{\,\lambda^1}\,.
\label{wzW}
\qqq
A direct check of this fact may be found in Appendix B. 
The action (\ref{wzW}) is a special case of the action 
of outer current-algebra automorphisms on spaces of conformal 
blocks studied in \cite{FS}, see also \cite{AMW}.
\vskip 0.3cm

The composition rule that assures that $\,\NU\,$ is a representation 
of $\,Z\m$,
\qq
U_{\CD^0\lambda}^{\,\CD^1}(z)\ \m
U_{\CD_z^0\lambda}^{\,\CD_z^1}(z')\ =\ 
U_{\CD^0\lambda}^{\,\CD^1}(zz') \,,
\nonumber
\qqq
follows easily from the relation $\,w_z\m w_z'=\ee^{2\pi\si\,b_{z.z'}}\m
w_{zz'}\,$ and the fact that $\,\ee^{2\pi\si\,b_{z,z'}}\,$ acts on
on $\,M_{\lambda^0\lambda}^{\,\lambda^1}\subset V_\lambda\,$
as the multiplication by $\,\ee^{2\pi\si\,\tr\,(\lambda_1-\lambda_0)b_{z,z'}}
=\CV_{\lambda^0;z,z'}\m\,\overline{\CV_{\lambda^1;z,z'}}\m$, \,see 
Eqs.\,\,(\ref{cint1}) and (\ref{cv}). \,Operators (\ref{Uz}) are 
unitary. They behave naturally under the changes 
of the solutions $\,\CW^s\,$ by coboundaries. Such changes lead 
to isomorphic modules $\,{\CE'}^s\,$ and to isomorphisms of vector 
bundles $\,\CE_{\hspace{-0.02cm} \CD^0}^{\CD^1}\,$ and of
multiplicity spaces $\,M_{\CD^0\lambda}^{\,\CD^1}\m$. \,Operators 
$\,\NU(z)\,$ and $\,U_{\CD^0\lambda}^{\CD^1}(z)\,$ change by conjugation 
with unitary operators induced by such isomorphisms. Similarly, a change 
$\,w_z\mapsto\ee^{2\pi\si\m a_z}w_z\,$ for $\,a_z\in\Nt\,$ induces 
the change $\,\CW_{\lambda;z}\mapsto\ee^{2\pi\si\,\tr\,\lambda a_z}
\m\CW_{\lambda;z}\m$, see Eq.\,\,(\ref{ch1}), that compensates that 
of $\,w_z\,$ in Eq.\,\,(\ref{Uz}). 
\vskip 0.3cm

As indicated in \cite{GR1}, the space of boundary states 
$\,\NH'^{\m{\CD'}^1}_{\m{\CD'}^0}\,$ of the WZW model
with the $\,G'\,$ target may be identified with
the $\,Z$-invariant subspace of $\,\widetilde
\NH'^{\m{\CD'}^1}_{\m{\CD'}^0}\m$:
\qq
\NH'^{\m{\CD'}^1}_{\m{\CD'}^0}\ =\ 
{_1\over^{|Z|}}\,\sum\limits_{z\in Z}\NU(z)\ \,
\widetilde \NH'^{\m{\CD'}^1}_{\m{\CD'}^0}\ \equiv\ 
\NP\ \,\widetilde \NH'^{\m{\CD'}^1}_{\m{\CD'}^0}
\label{Hb'}
\qqq
where $\,\NP\,$ is the orthogonal projector on the $\,Z$-invariant states.
The scalar product in $\,\NH'^{\m{\CD'}^1}_{\m{\CD'}^0}$ should be divided
by $\,|Z|\,$ with respect to the one inherited from the subspace
of $\,\widetilde\NH'^{\m{\CD'}^1}_{\m{\CD'}^0}$. 
\,In terms of the decomposition into the highest weight 
representations of the current algebra $\,\hat\Ng\m$,
\qq
\tilde\NH'^{\m{\CD'}^1}_{\m{\CD'}^0}\ =\ {\tilde M}'^{\,{\CD'}^1}_{
{\CD'}^0\lambda}\otimes\hat V_{\bar\lambda}
\label{tdecc}
\qqq
with the multiplicity spaces
\qq
{\tilde M}'^{\,{\CD'}^1}_{{\CD'}^0\lambda}\ =\ 
\mathop{\oplus}\limits_{(\CD^0,\CD^1)}M_{\CD^0\lambda}^{\,\CD^1}\ 
=\ \mathop{\oplus}\limits_{(\lambda^0,\lambda^1)\atop\lambda^s\in[\lambda^s]}
\NC^{N_0}\otimes\overline{\NC^{N_1}}\otimes 
M_{\lambda^0\lambda}^{\lambda^1}
\label{tms}
\qqq
where $\,[\lambda]\,$ denotes the $\,Z$-orbit of weight $\,\lambda\,$
under the action $\,\lambda\mapsto z\lambda$. \,The representation 
$\,\NU\,$ induces the operators  
\qq
U'^{\,{\CD'}^1}_{\hspace{-0.02cm}{\CD'}^0\lambda}(z)\,\ 
=\ \mathop{\oplus}\limits_{(\CD^0,\CD^1)}\hspace{-0.08cm}
U_{\CD^0\lambda}^{\,\CD^1}(z)
\label{UD}
\qqq
providing a unitary representations of $\,Z\,$ in the multiplicity spaces 
$\,{\tilde M}'^{\,{\CD'}^1}_{{\CD'}^0\m
\lambda}$. \,The relation (\ref{Hb'}) and the decomposition (\ref{tdecc}) 
imply then that
\qq
\NH'^{\m{\CD'}^1}_{\m{\CD'}^0}\ =\ M'^{\,{\CD'}^1}_{{\CD'}^0\lambda}
\otimes\hat V_{\bar\lambda}
\nonumber
\qqq
where 
\qq
M'^{\,{\CD'}^1}_{{\CD'}^0\lambda}\ 
=\ {_1\over^{|Z|}}\sum\limits_{z\in Z}
U'^{\,{\CD'}^1}_{\hspace{-0.02cm}{\CD'}^0\lambda}(z)\ \,
{\tilde M}'^{\,{\CD'}^1}_{{\CD'}^0\lambda}
\ \equiv\ P_\lambda\  
{\tilde M}'^{\,\CD^1}_{\CD^0\lambda}\,.
\label{pl}
\qqq
In short, the multiplicity spaces for the group $\,G'\,$ theory are
the $\,Z$-invariant subspaces of the direct sum of the multiplicity 
spaces for all group $\,G\,$ branes compatible with those of the
$\,G'\,$ theory.  
\vskip 0.3cm

Together with the representation (\ref{Uz}), the above constructions 
lead to the following expression for the boundary partition functions 
of the group $\,G'\,$ WZW theory:
\qq
Z'^{\CD'_1}_{\CD'_0}(T)&\,\equiv\,&{\mathrm Tr}_{
\NH'^{\CD'_1}_{\m{\CD'}^0}}\ee^{-T\m L_0}\ =\ \sum\limits_\lambda
N'^{{\,\CD'}^1}_{{\CD'}^0\lambda}\ \,\hat\chi_{\bar\lambda}({_{\si T}
\over^{2\pi}})
\nonumber
\qqq
where
\qq
N'^{\,{\CD'}^1}_{{\CD'}^0\lambda}\,&=&\,{\mathrm dim}
\left(M'^{\,{\CD'}^1}_{{\CD'}^0\lambda}\m\right)\ 
=\ {1\over|Z|}\sum\limits_{z\in Z}\,
{\mathrm Tr}\,\,U'^{\,{\CD'}^1}_{\hspace{-0.02cm}{\CD'}^0\lambda}(z)\cr
&=&\,{1\over|Z|}
\sum\limits_{\lambda^s\in[\lambda^s]}
\,\sum\limits_{z\in Z_{[\lambda^0]}\cap Z_{[\lambda^1]}}\sum\limits_\lambda
\ \,\m{\rm Tr}\m\,\CW_{\lambda^0;z}\ \,{\rm Tr}\m\,
\overline{\CW_{\lambda^1;z}}\ \,\,
{\mathrm Tr}_{M_{\lambda^0
\lambda}^{\,\lambda^1}}w_z\,.
\nonumber
\qqq
The non-negative integers $\,N'^{\,{\CD'}^1}_{{\CD'}^0\lambda}\,$ 
are the entries of matrices $\,N'_\lambda\,$ that provide a ``NIM 
{\bf representation}'' of the fusion algebra \cite{BPPZ,PZ}. 
The relation of the above formulae to those of \cite{FHSSW,SchwFW} was 
discussed in \cite{GR1}. It is based on the Verlinde type expressions 
conjectured in \cite{FS} for the traces of the action of outer automorphisms 
on the spaces of conformal blocks, see Eq.\,\,9.62 in \cite{GR1}. 
It should be possible to provide 
a direct proof of such trace formulae by studying the action of 
$\,w_z\,$ on $\,M_{\lambda^0\lambda}^{\,\lambda}\subset V_\lambda\,$
(or on the quantum group $\,\CU_q(\Ng)\,$ realization thereof).

\subsection{Boundary operator product}
\label{sec:bop}

As for the boundary operator product for the WZW theory, it
is provided by a linear mapping
\qq
\NA_{\hspace{-0.02cm}\CD^0\CD^1}^{\,\CD^1\CD^2}:\ 
\NH_{\CD^0}^{\m\CD_1}\m\otimes\m\,
\NH_{\CD^1}^{\m\CD_2}\ \mathop{\longrightarrow}\ \,
\NH_{\CD^0}^{\m\CD_2}\,.
\nonumber
\qqq
In terms of the multiplicity spaces of the decomposition
(\ref{decc}), it is determined by the maps 
\qq
A_{\CD^0\CD^1\lambda\m\mu}^{\,\CD^1\CD^2\,\nu}\,:\ 
M_{\CD^0\lambda}^{\,\CD^1}\m\otimes M_{\CD^1\mu}^{\,\CD^2}
\ \longrightarrow\ M_{\CD^0\nu}^{\,\CD^2}
\,\otimes\,{\mathrm Hom}^{\mathrm FR}_{\m\Ng}(V_\nu,\m V_\lambda\otimes V_\mu)
\label{ope}
\qqq
which upon a choice of basis become ``{\bf operator product coefficients}''.
The last factor in (\ref{ope}) determines the current-algebra intertwiner 
from $\,\hat V_{\bar\lambda}\otimes\hat V_{\bar\mu}\,$ to 
$\,\hat V_{\bar\nu}\,$ and plays in what follows a spectator role.
For simply connected target groups $\,G\m$ and rank 1 branes supported
by integral conjugacy classes $\,\CC_{\tau^s}\,$ with $\,\lambda^s=\sk
\tau^s$, \,the maps (\ref{ope}) reduce to
\qq
A^{\,\lambda^1\lambda^2\,\nu}_{\lambda^0\lambda^1\lambda\,\mu}
\,:\ M_{\lambda^0\lambda}^{\,\lambda^1}\,\otimes\,M_{\lambda^1
\mu}^{\,\lambda^2}\ \mathop{\longrightarrow}\ M_{\lambda^0
\nu}^{\,\lambda^2}\,\otimes\,M_{\lambda\m\mu}^{\,\nu}
\label{ope1}   
\qqq 
or, in the realization of the latter spaces as fusion rule Lie algebra
intertwiners, to the maps
\qq
F_{_{\hspace{-0.04cm}\lambda^1\nu}}\hspace{-0.07cm}
\Big[\matrix{_{\hspace{-0.08cm}\lambda^0}&\hspace{-0.35cm}_{\lambda^2}
\cr\hspace{-0.08cm}^{\lambda}&\hspace{-0.35cm}^{\mu}}\Big]\,:
&&{\mathrm Hom}^{\mathrm FR}_{\m\Ng}
(V_{\lambda^1},\m V_{\lambda^0}\otimes 
V_\lambda)\,\otimes\,{\mathrm Hom}^{\mathrm FR}_{\m\Ng}(V_{\lambda^2},\m 
V_{\lambda^1}\otimes V_\mu)\cr\cr
\longrightarrow\quad&&
{\mathrm Hom}^{\mathrm FR}_{\m\Ng}(V_{\lambda^2},\m 
V_{\lambda^0}\otimes V_\nu)
\,\otimes\,{\mathrm Hom}^{\mathrm FR}_{\m\Ng}(V_\nu,\m V_\lambda\otimes V_\mu)
\nonumber
\qqq
that may be identified with the fusing $F$-matrices of the bulk WZW
theory \cite{MS,Runkel1,FFFS}. Finally, in the realization of the spaces 
$\,M_{\lambda\m\mu}^{\,\nu}\,$ as intertwiners of the quantum group 
$\,\CU_q(\Ng)\,$ action, the maps (\ref{ope1}) become \cite{AGS} the quantum 
``$\,6j\,$ symbols'' \cite{CFS} identifying two ways to decompose triple 
tensor products of highest weight representations.
\vskip 0.3cm

For stacks of rank 1 branes in the model with a simply connected target, 
the maps (\ref{ope}) for the multiplicity spaces given by (\ref{decc1}) 
are obtained from those in (\ref{ope1}) by tensoring with the contraction
\qq
J\,:\ \NC^{N_0}\otimes\overline{\NC^{N_1}}\otimes\NC^{N_1}\otimes
\overline{\NC^{N_2}}\ \longrightarrow\ \NC^{N_0}\otimes\overline{\NC^{N_2}}
\nonumber
\qqq
of the middle factors employing the standard scalar product on $\,\NC^{N_1}$.
\vskip 0.3cm
 
As pointed out in \cite{GR1}, the picture of the boundary states of the WZW 
model with the $\,G'=G/Z\,$ target as $\,Z$-invariant states in the 
$\,G\,$ target model leads to a simple relation between the boundary
operator products in both theories. In short, one should project the
operator product of the $\,G\,$ model to the $\,Z$-invariant sector.
More precisely, consider the space 
\qq
{\widetilde\NH}'^{\m{\CD'}^1{\CD'}^2}_{\m{\CD'}^0{\CD'}^1}\ \,=\ 
\mathop{\oplus}\limits_{(\CD^0,\CD^1,\CD^2)}\NH_{\CD^0}^{\m
\CD^1}\,\otimes\,\NH_{\CD^1}^{\,\CD^2}\ \,\subset\,\,\ 
{\widetilde\NH}'^{\m{\CD'}^1}_{\m{\CD'}^0}\,\otimes\,
{\widetilde\NH}'^{\m{\CD'}^2}_{\m{\CD'}^1}
\nonumber
\qqq
where the direct sum is over the pullback branes in the $\,G\,$ model 
and let
\qq
\widetilde\NP\,:\ {\widetilde\NH}'^{\m{\CD'}^1}_{\m{\CD'}^0}\,\otimes
\,{\widetilde\NH}'^{\m{\CD'}^2}_{\m{\CD'}^1}\ \,\longrightarrow\ 
\,\,{\widetilde\NH}'^{\m{\CD'}^1{\CD'}^2}_{\m{\CD'}^0{\CD'}^1}
\nonumber
\qqq
be the orthogonal projection. Then the boundary operator product of the
$\,G'\,$ model is given by
\qq
\NA'^{\,{\CD'}^1{\CD'}^2}_{{\CD'}^0{\CD'}^1}\ =\ 
\,\NP\ \Big(\mathop{\oplus}\limits_{(\CD^0,\CD^1,\CD^2)}
\hspace{-0.2cm}\NA_{{\CD}^0{\CD}^1}^{\,{\CD}^1{\CD}^2}\Big)\ \widetilde\NP\,
\Big|_{\,{\NH'^{\m{\CD'}^1}_{\m{\CD'}^0}\otimes\,\m
\NH'^{\m{\CD'}^2}_{\m{\CD'}^1}}}
\label{A'}
\qqq
where the operator in parenthesis maps from 
$\,{\widetilde\NH}'^{\,{\CD'}^1{\CD'}^2}_{\m{\CD'}^0{\CD'}^1}\,$ to 
$\,\,{\widetilde\NH}'^{\m{\CD'}^2}_{\m{\CD'}^0}$. \,The projector
$\,\NP\,$ above is spurious, since the result lands anyway in 
$\,\NH'^{\m{\CD'}^2}_{\m{\CD'}^0}\,$ due to the symmetry property
\qq
\NA_{{\CD}^0{\CD}^1}^{\,{\CD}^1{\CD}^2}\ \NU(z)\otimes\NU(z)\ =\ 
\NU(z)\,\,\NA_{{\CD}_z^0{\CD}_z^1}^{\,{\CD}_z^1{\CD}_z^2}
\label{As}
\qqq
of the operator product. Formula (\ref{A'}) descends
to the multiplicity spaces. Let
\qq
{\tilde M}'^{\,{\CD'}^1{\CD'}^2}_{\hspace{-0.02cm}
{\CD'}^0{\CD'}^1\lambda\,\mu}\ =\ 
\mathop{\oplus}\limits_{(\CD^0,\CD^1,\CD^2)}\hspace{-0.1cm}
M_{\CD^0\lambda}^{\,\CD^1}\,\otimes\,M_{\CD^1\mu}^{\,\CD^2}\ \,\subset\ \,\,
{\tilde M}'^{\,{\CD'}^1}_{\m{\CD'}^0\lambda}\,\otimes\,
{\tilde M}'^{\,{\CD'}^2}_{\m{\CD'}^1\mu}
\nonumber
\qqq
and let $\,\tilde P\,$ denote the orthogonal projection from
$\,{\tilde M}'^{\,{\CD'}^1}_{{\CD'}^0\lambda}\otimes
{\tilde M}'^{\,{\CD'}^2}_{{\CD'}^1\mu}\,$ to 
$\,{\tilde M}'^{\,{\CD'}^1{\CD'}^2}_{{\CD'}^0{\CD'}^1\lambda\,\mu}$. \,Then
\qq
A'^{\,{\CD'}^1{\CD}'^2\m\nu}_{{\CD'}^0{\CD'}^1\lambda\,\mu}
\ =\ \,P_\nu\ \Big(J\,\otimes\hspace{-0.2cm}
\mathop{\oplus}\limits_{(\lambda^0,\lambda^1,
\lambda^2)\atop{\lambda^s\in[\lambda^s]}}\hspace{-0.2cm}A^{\,\lambda^1
\lambda^2\m\nu}_{\lambda^0\lambda^1\lambda\,\mu}}\Big)\ \tilde P
\,\Big|_{\,M'^{\,{\CD'}^1}_{{\CD'}^0\lambda}\otimes
\,M'^{\,{\CD'}^2}_{{\CD'}^1\mu}\,.
\label{A'm}
\qqq
The projector $\,P_\nu\m$, \,see Eq.\,\,(\ref{pl}), may again be dropped
due to the symmetry property (\ref{As}) that on the multiplicity spaces
in the realization (\ref{wlb}) reduces to the relation
\qq
A^{\,\lambda^1\lambda^2\m\nu}_{\lambda^0\lambda^1\lambda\,\mu}\ w_z\otimes
w_z\ =\ w_z\otimes I\ A^{\,\lambda_z^1\lambda_z^2\m\nu}_{\lambda_z^0
\lambda_z^1\lambda\,\mu}\,,
\label{Asm}
\qqq
see (\ref{ope1}) and (\ref{wzW}). To further elucidate the above formulae, 
in Appendix C we work out their 
details for the case of the $\,SO(3)\,$ WZW model.

\nsection{Conclusions}

We have discussed in this paper how to set up carefully the Lagrangian 
approach to the WZW models with boundary using the concepts of gerbes 
and gerbe modules. The possible boundary conditions involving  the 
Chan-Paton coupling to gauge fields were described in terms of 
D-branes in the target group carrying gerbe modules. The corresponding 
gerbes are obtained by the restriction to the D-branes of the gerbe 
on the target group with the curvature proportional to the invariant 
3-form $\,\tr\,(g^{-1}dg)^3$. \,In particular, we have discussed the 
(maximally) symmetric branes that preserve the diagonal current-algebra 
symmetry of the WZW model with a compact simple group as the target. 
Such branes are supported by the integrable conjugacy classes in the 
target group, i.e.\,\,the ones that contain elements $\,\ee^{2\pi\si\m
\lambda/\sk}\,$ where $\,\sk\,$ is the level of the model and 
$\,\lambda\,$ is the (integrable) highest weight of a chiral current 
algebra primary field. 
\vskip 0.3cm

For a simply connected target group $\,G\m$, \,up to isomorphism, there 
is exactly one 1-dimensional gerbe module that leads to a current-algebra 
symmetric boundary condition over each integrable conjugacy 
class $\,\CC_\tau\subset G\,$ with $\,\sk\tau=\lambda\m$, \,in agreement 
with Cardy's classification of boundary conditions \cite{Cardy}. 
The $N$-dimensional gerbe modules with the same property are direct sums 
of the 1-dimensional one. In particular, the symmetric branes in the WZW 
models with a simply connected target carry only Abelian (twisted) gauge 
fields. 
\vskip 0.3cm

For the non-simply connected target groups $\,G'=G/Z\m$, \,the integrable
conjugacy classes $\,\CC'_{[\tau]}\subset G'\,$ correspond to
$\,Z$-orbits $\,[\tau]\,$ of rescaled weights $\,\tau\m$, \,with $\,z\CC_\tau
\equiv\CC_{z\tau}\subset G\,$ projecting to the same conjugacy class in 
$\,G'\,$ as $\,\CC_\tau\subset G$. \,The construction of gerbe modules 
over $\,\CC'_{[\tau]}\,$ that define symmetric branes required solving 
a cohomological equation $\,\delta\CW=\CV\,$ with $\,\CV\,$ an 
explicit $\,2$-cocycle on $\,Z\,$ taking values in the $\,Z$-module of 
$\,U(1)$-valued functions on the orbit $\,[\tau]$. \,The scalar 
solutions $\,\CW\m$, \,if existent, lead to 1-dimensional gerbe modules 
over $\,\CC'_{[\tau]}\,$ with non-isomorphic choices labeled by the characters 
of the fundamental group of $\,\CC'_{[\tau]}$. \,The $N$-dimensional gerbe 
modules are then direct sums of the 1-dimensional ones, resulting again 
in symmetric branes carrying Abelian gauge fields. \,The exceptional
case involves the groups $\,Spin(4n)/({\NZ_2\hspace{-0.06cm}\times
\hspace{-0.06cm}\NZ_2})=SO(4n)/\NZ_2\,$ that admit two non-equivalent 
gerbes differing by discrete torsion. For one of those choices and 
the integrable conjugacy classes $\,\CC'_{[\tau]}\,$ corresponding to 
1-point orbits $\,[\tau]\,$ there are no scalar solutions of the 
cohomological equation $\,\delta\CW=\CV\,$ and no 1-dimensional gerbe 
modules leading to current-algebra symmetric boundary conditions. 
There exist, however, matrix-valued solutions $\,\CW$. \,The one
of minimal rank is provided by the Pauli matrices. It gives rise to 
a 2-dimensional gerbe module over $\,\CC'_{[\tau]}$. \,All other gerbe 
modules resulting in symmetric branes supported by such conjugacy classes 
are direct sums of the 2-dimensional one. The corresponding branes carry 
a (projectively flat, twisted) non-Abelian $\,U(2)\,$ gauge field. 
The spontaneous enhancement of the gauge symmetry is even more common 
in the boundary WZW models with non-simple targets \cite{FKLLSW,Walch}. 
We plan to return to this subject, and to the geometric classification 
of the so called twisted-symmetric branes \cite{BFS} in the WZW models 
and of branes in general coset theories \cite{GawGH,ES} in the future.
Another theme left for future is the geometric treatment along 
similar lines of the WZW amplitudes on non-oriented worldsheets.
\vskip 0.3cm

In the last section of the paper we have indicated how the solutions 
of the cohomological equations defining symmetric branes in the WZW models 
with non-simply connected targets enter directly the expressions for 
the boundary partition functions and operator product coefficients of 
those models. This demonstrated the utility of the geometric 
considerations which provide a simple and unified view on both classical 
and quantum WZW theory. The general discussion was illustrated by the 
detailed computation in Appendix C for the WZW model with the $\,SO(3)\,$
group target called, in the algebraic approach, the non-diagonal 
$\,D_{{\sk\over2}+2}\,$ series of the WZW models 
with $\,\hat{su}_2\,$ symmetry. 
\vskip 0.3cm

An interesting question left aside concerns the stability 
\cite{BDS,Pawel} of the symmetric branes discussed in this paper, see 
also Sect.\,\,4.3 of \cite{Schom}. The natural guess is that, in the
present context, stable branes are the ones with gerbe modules 
of the minimal dimension whose number is equal to the dimension of 
the bulk Ishibashi states \cite{Walch}. We plan in the future to 
address the stability issue and its geometric formulation. This question 
is also related to that of the Ramond-Ramond charges conserved in the 
brane condensation processes in the supersymmetric version of the WZW 
model \cite{WittKT,BCMMS,FSchom,BDR,BR,GG2}.  
\vfill
\eject

\nappendix{A\ \ List of 1-dimensional 
symmetric $\,\CG'$-branes}
\label{sec:appA}
\vskip 0.7cm
\noindent $================================================$
\vskip 0.2cm
\noindent\hbox to 4.3cm{\Large\bf Covering Group:\hfill}{\Large$\,\bf SU(r+1)$}
\vskip 0.2cm
\noindent $--------------------------------------$
\vskip 0.6cm

\noindent\hbox to 4.3cm{\bf Center:\hfill}$\,\cong\,\NZ_{r+1}$
\vskip 0.4cm

\noindent\hbox to 4.3cm{\bf Generator:\hfill}$\,z\,,\ \,z^{r+1}=1$
\vskip 0.4cm

\noindent\hbox to 4.3cm{{\bf Action on} $\,\CA_{_W}\,$:\hfill}$\,z\tau_i
=\tau_{i+1}\,,\ \,i=0,1,\dots,r-1\,,\quad\,z\tau_r=\tau_0$
\vskip 0.4cm

\noindent\hbox to 4.3cm{{\bf weight} $\,\lambda\equiv\sk\tau\,$:\hfill}$\,
\sum\limits_{i=0}^r n_i\tau_i\,,\ \ n_i\in\NZ\m$ 
\vskip 0.5cm

\noindent $--------------------------------$
\vskip 0.3cm

\noindent\hbox to 4.3cm{{\bf Subgroup} $\,Z\,$:\hfill}$\,
\{\,z^n\ |\ n\ \,{\rm divisible\ by}\ \,N''=(r+1)/N'\,\}
\,\cong\,\NZ_{N'}$
\vskip 0.4cm

\noindent\hbox to 4.3cm{\bf Quotient group:\hfill}$\,SU(r+1)/\NZ_{N'}$ 
\vskip 0.3cm

\noindent $--------------------------------$
\vskip 0.3cm

\noindent\hbox to 4.3cm{\bf level:\hfill}$\,\sk\in\NZ\,\,\,\,$ if 
$\,N'\,$ odd or $\,N''\,$ even,

\noindent\hbox to 4.3cm{\ \hfill}$\,\sk\in2\NZ\,$ if $\,N'\,$ 
even and $\,N''\,$ odd
\vskip 0.5cm

\noindent\hbox to 4.3cm{2-{\bf chain} $\,V\,$:\hfill}$\,
V_{z^n,z^m}\ \equiv\ 1$
\vskip 0.5cm

\noindent\hbox to 4.3cm{2-{\bf cocycle} $\,\CV\,$:\hfill}$\,
\CV_{\tau;\m z^n,z^m}\ =\ \cases{\ \hbox to 4.9cm{$1$\hfill}{\rm if}
\ \ n+m\leq r\,,\cr\ \hbox to 4.9cm{$(-1)^{r\hspace{-0.05cm}
\sum\limits_{i=0}^ri\m n_i}=\ee^{\pi\si\, r(r+1)\,
Q_z(\lambda)}$\hfill}{\rm if}\ \ n+m>r}$
\vskip 0.2cm

\noindent\hbox to 4.3cm{\ \hfill}\,for $\,n,m\,$ divisible 
by $\,N''$, $\,0\leq n,m\leq r$
\vskip 0.6cm

\noindent\hbox to 4.3cm{\bf Stabilizers:\hfill}$\,Z_{[\tau]}\cong
\NZ_{n'}\,$ if $\,n'\,$ is the largest integer dividing $\,N'\,$ such that

\noindent\hbox to 4.3cm{\ \hfill}$\m\,n_i=n_{j}\,$ 
for $\,j=i+(r+1)/n'\,\,{\rm mod}\,(r+1)\m$,

\noindent\hbox to 4.3cm{\ \hfill}\m\,possible if $\,n'\,$ divides $\,\sk$ 
\vskip 0.7cm

\noindent\hbox to 4.3cm{\bf Special solution:\hfill}$\,
\CW^{\hat\tau}_{\tau;\m z^n}\ =\ \ee^{\pi\si\,\sigma\,n/(r+1)}$
\vskip -0.1cm

\noindent\hbox to 4.3cm{\ \hfill}\,where $\,\sigma=0,1\,$ is the parity of 
$\,\,\,r\hspace{-0.07cm}\sum\limits_{i=0}^ri\m n_i=r(r+1)\,Q_z(\lambda)\,\,
{\mathrm mod}\,2$ 

\noindent\hbox to 4.3cm{\ \hfill}\,constant on $\ [\tau]$
\vskip 0.5cm

\noindent\hbox to 4.3cm{{\bf cocycle} $\,\hat\phi\,$:\hfill}$
\hat\phi_{\hat\tau;\m z^n}(z_0)\ \equiv\ 1$
\vskip 0.6cm

\noindent\hbox to 4.3cm{\bf General solution:\hfill}$\,
\CW_{\tau;\m z^n}\ =\ \ee^{2\pi\si\m np/(r+1)}\,\m\CW^{\hat\tau}_{\tau;
\m z^n}$ \quad\ for \ $\,p=0,1,\dots,n'-1$

\noindent\hbox to 9.75cm{\ \hfill}if \ $\ Z_{[\tau]}\cong\NZ_{n'}$

\vskip 1.5cm
\vfill
\eject

\noindent $================================================$
\vskip 0.2cm
\noindent\hbox to 4.3cm{\Large\bf Covering group:\hfill}{\Large$\,\bf 
Spin(2r+1)$}
\vskip 0.2cm
\noindent $--------------------------------------$
\vskip 0.6cm

\noindent\hbox to 4.3cm{\bf Center:\hfill}$\,\cong\,\NZ_{2}$
\vskip 0.4cm

\noindent\hbox to 4.3cm{\bf Generator:\hfill}$\,z\,,\ \,z^2=1$
\vskip 0.4cm

\noindent\hbox to 4.3cm{{\bf Action on} $\,\CA_{_W}\,$:\hfill}$\,z\tau_0
=z\tau_1\,,\quad \,z\tau_1=\tau_0\,,\quad \,z\tau_i=\tau_i\,,\ 
\,i=2,\dots,r$
\vskip 0.4cm

\noindent\hbox to 4.3cm{{\bf weight} $\,\lambda\equiv\sk\tau\,$:\hfill}$\,
\sum\limits_{i=0}^r n_i\tau_i\m,\ \,n_0,n_1,n_r\in\NZ\m,
\ \,n_i\in2\NZ\ \,{\rm for}\ \,i=2,\dots,r-1$
\vskip 0.5cm

\noindent $--------------------------------$
\vskip 0.3cm

\noindent\hbox to 4.3cm{{\bf Subgroup} $\,Z\,$:\hfill}$\,\{1,z\}\cong\NZ_{2}$
\vskip 0.4cm

\noindent\hbox to 4.3cm{\bf Quotient group:\hfill}$\,SO(2r+1)$
\vskip 0.3cm

\noindent $--------------------------------$
\vskip 0.6cm

\noindent\hbox to 4.3cm{\bf level:\hfill}$\,\sk\in\NZ\,$
\vskip 0.5cm

\noindent\hbox to 4.3cm{2-{\bf chain} $\,V\,$:\hfill}$\,
V_{z^n,z^m}\ \equiv\ 1$
\vskip 0.5cm

\noindent\hbox to 4.3cm{2-{\bf cocycle} $\,\CV\,$:\hfill}$\,
\CV_{\tau;\m z^n,z^m}\ =\ \cases{\ \hbox to 4.9cm{$1$\hfill}{\rm if}
\ \ n+m\leq 1\,,\cr\ \hbox to 4.9cm{$(-1)^{r n_r}
=\ee^{2\pi\si\, r\,Q_z(\lambda)}$\hfill}{\rm if}\ \ n+m=2}$
\vskip 0.2cm

\noindent\hbox to 4.3cm{\ \hfill}\,for $\,n,m=0,1$
\vskip 0.7cm

\noindent\hbox to 4.3cm{\bf Stabilizers:\hfill}$\,Z_{[\tau]}=\{1,z\}\,$
\ if $\,\,n_0=n_1\,$ and \ $\,Z_{[\tau]}=\{1\}\,$ otherwise
\vskip 0.7cm

\noindent\hbox to 4.3cm{\bf Special solution:\hfill}$\,
\CW^{\hat\tau}_{\tau;\m z^n}\ =\ \ee^{\pi\si\m\sigma n/2}$
\vskip 0.1cm

\noindent\hbox to 4.3cm{\ \hfill}\,where $\,\sigma=0,1\,$ is the parity of 
$\,\,rn_r=2\m r\,Q_z(\lambda)\,\m{\mathrm mod}\,2\ $ 

\noindent\hbox to 4.3cm{\ \hfill}\,constant on $\ [\tau]$
\vskip 0.6cm

\noindent\hbox to 4.3cm{{\bf cocycle} $\,\hat\phi\,$:\hfill}$
\hat\phi_{\hat\tau;\m z^n}(z_0)\ \equiv\ 1$
\vskip 0.6cm

\noindent\hbox to 4.3cm{\bf General solution:\hfill}$\,
\CW_{\tau;\m z^n}\ =\ (-1)^{np}\,\m\CW^{\hat\tau}_{\tau;\m z^n}$
\quad\ for \ $\,p=0,1\,$ if $\,Z_{[\tau]}=\{1,z\}$
\vskip 1.5cm
\vfill
\eject

\noindent $================================================$
\vskip 0.2cm
\noindent\hbox to 4.3cm{\Large\bf Covering group:\hfill}{\Large$\,\bf Sp(2r)$}
\vskip 0.2cm
\noindent $--------------------------------------$
\vskip 0.6cm

\noindent\hbox to 4.3cm{\bf Center:\hfill}$\,\cong\,\NZ_{2}$
\vskip 0.4cm

\noindent\hbox to 4.3cm{\bf Generator:\hfill}$\,z\,,\ \,z^2=1$
\vskip 0.4cm

\noindent\hbox to 4.3cm{{\bf Action on} $\,\CA_{_W}\,$:\hfill}$\,
z\tau_i=\tau_{r-i}\,,\ 
\,i=0,\dots,r$
\vskip 0.4cm

\noindent\hbox to 4.3cm{{\bf weight} $\,\lambda\equiv\sk\tau\,$:\hfill}$\,
\sum\limits_{i=0}^r n_i\tau_i\,,\ \ n_i\in\NZ\ \,{\rm for}\ 
\,i=0,\dots,r$
\vskip 0.6cm

\noindent $--------------------------------$
\vskip 0.3cm

\noindent\hbox to 4.3cm{{\bf Subgroup} $\,Z\,$:\hfill}$\,\{1,z\}\cong\NZ_{2}$
\vskip 0.4cm

\noindent\hbox to 4.3cm{\bf Quotient group:\hfill}$\,Sp(2r)/\NZ_2$
\vskip 0.3cm

\noindent $--------------------------------$
\vskip 0.5cm

\noindent\hbox to 4.3cm{\bf level:\hfill}$\,\sk\in\NZ\,\,$ if $\,\,r\,\,$ 
even and $\,\,\sk\in2\NZ\,\,$ if $\,\,r\,\,$ is odd 
\vskip 0.5cm

\noindent\hbox to 4.3cm{2-{\bf chain} $\,V\,$:\hfill}$\,
\CV_{\tau;\m z^n,z^m}\ \equiv\ 1$
\vskip 0.5cm

\noindent\hbox to 4.3cm{2-{\bf cocycle} $\,\CV\,$:\hfill}$\,
\CV_{\tau;\m z^n,z^m}\ =\ \cases{\ \hbox to 4.9cm{$1$\hfill}{\rm if}
\ \ n+m\leq 1\,,\cr\ \hbox to 4.9cm{$(-1)^{\sum\limits_{i=0}^r i\m 
n_i}=\ee^{2\pi\si\,Q_z(\lambda)}$\hfill}{\rm if}\ \ n+m=2}$
\vskip 0.2cm

\noindent\hbox to 4.3cm{\ \hfill}\,for $\,n,m=0,1$
\vskip 0.7cm

\noindent\hbox to 4.3cm{\bf Stabilizers:\hfill}$\,Z_{[\tau]}=\{1,z\}\,$
\ if $\,n_i=n_{r-i}\m$, $\,i=0,\dots,r\m$,

\noindent\hbox to 4.3cm{\ \hfill}$\,Z_{[\tau]}=\{1\}\,$ otherwise
\vskip 0.7cm

\noindent\hbox to 4.3cm{\bf Special solution:\hfill}$\,
\CW^{\hat\tau}_{\tau;\m z^n}\ =\ \ee^{\pi\si\m\sigma n/2}$
\vskip 0.1cm

\noindent\hbox to 4.3cm{\ \hfill}\,where $\,\sigma=0,1\,$ is the parity of 
$\,\sum\limits_{i=0}^r i\m n_i=2\,Q_z(\lambda)\,\,{\mathrm mod}\,2\ $ 

\noindent\hbox to 4.3cm{\ \hfill}\,constant on $\ [\tau]$ 
\vskip 0.6cm

\noindent\hbox to 4.3cm{{\bf cocycle} $\,\hat\phi\,$:\hfill}$
\hat\phi_{\hat\tau;\m z^n}(z_0)\ \equiv\ 1$
\vskip 0.6cm

\noindent\hbox to 4.3cm{\bf General solution:\hfill}$\,
\CW_{\tau;\m z^n}\ =\ (-1)^{np}\,\m\CW^{\hat\tau}_{\tau;\m z^n}$
\quad\ for \ $\,p=0,1\,$ if $\,Z_{[\tau]}=\{1,z\}$
\vskip 1.5cm
\vfill
\eject

\noindent $================================================$
\vskip 0.2cm
\noindent\hbox to 4.3cm{\Large\bf Covering group:\hfill}{\Large$\,
\bf Spin(2r)\,$ for odd $\,r$}
\vskip 0.2cm
\noindent $--------------------------------------$
\vskip 0.6cm

\noindent\hbox to 4.3cm{\bf Center:\hfill}$\,\cong\,\NZ_{4}$
\vskip 0.4cm

\noindent\hbox to 4.3cm{\bf Generator:\hfill}$\,z\,,\ \,z^4=1$
\vskip 0.4cm

\noindent\hbox to 4.3cm{{\bf Action on} $\,\CA_{_W}\,$:\hfill}$\,
z\tau_0=\tau_{r-1}\,,\quad\, z\tau_1=\tau_r\,,\quad\, z\tau_{i}
=\tau_{r-i}\,,\ \,i=2,\dots,r$
\vskip 0.4cm

\noindent\hbox to 4.3cm{{\bf weight} $\,\lambda\equiv\sk\tau\,$:\hfill}$\,
\sum\limits_{i=0}^r n_i\tau_i\,,\ \ n_0,n_1,n_{r-1},n_r\in\NZ\,,$

\noindent\hbox to 4.3cm{\ \hfill}$\,n_i\in2\NZ\ \,{\rm for}\ \,i=2,\dots,r-2$
\vskip 0.6cm

\noindent $--------------------------------$
\vskip 0.3cm

\noindent\hbox to 4.3cm{{\bf Subgroup} $\,Z\,$:\hfill}$\,\{1,z^2\}\cong\NZ_{2}$
\vskip 0.4cm

\noindent\hbox to 4.3cm{\bf Quotient group:\hfill}$\,SO(2r)$
\vskip 0.3cm

\noindent $--------------------------------$
\vskip 0.5cm

\noindent\hbox to 4.3cm{\bf level:\hfill}$\,\sk\in\NZ\,$
\vskip 0.5cm
\noindent\hbox to 4.3cm{2-{\bf chain} $\,V\,$:\hfill}$\,
V_{z^{2n},z^{2m}}\ \equiv\ 1$
\vskip 0.5cm

\noindent\hbox to 4.3cm{2-{\bf cocycle} $\,\CV\,$:\hfill}$\,
\CV_{\tau;\m z^{2n},z^{2m}}\ =\ \cases{\ \hbox to 4.9cm{$1$\hfill}{\rm if}
\ \ n+m\leq 1\,,\cr
\ \hbox to 4.9cm{$(-1)^{n_{r-1}+n_{r}}=\ee^{4\pi\si\,Q_z(\lambda)}
$\hfill}{\rm if}\ \ n+m=2}$
\vskip 0.2cm

\noindent\hbox to 4.3cm{\ \hfill}\,for $\,n,m=0,1$
\vskip 0.7cm

\noindent\hbox to 4.3cm{\bf Stabilizers:\hfill}\hbox to 3cm{$\,
Z_{[\tau]}=\{1,z^2\}\,$\hfill}if $\,n_0=n_1\,$ and $\,n_{r-1}=n_r\m$,

\noindent\hbox to 7.3cm{\ \hfill}possible if $\,\sk\in2\NZ\m$,
\vskip 0.1cm

\noindent\hbox to 4.3cm{\ \hfill}\hbox to 3cm{$\,Z_{[\tau]}=\{1\}\,$
\hfill}otherwise
\vskip 0.7cm

\noindent\hbox to 4.3cm{\bf Special solution:\hfill}$\,
\CW^{\hat\tau}_{\tau;\m z^{2n}}\ =\ \ee^{\pi\si\m\sigma n/2}$
\vskip 0.1cm

\noindent\hbox to 4.3cm{\ \hfill}\,where $\,\sigma=0,1\,$ is 
the parity of $\,n_{r-1}+n_r=4\,Q_z(\lambda)\,\,{\mathrm mod}\,2\ $ 

\noindent\hbox to 4.3cm{\ \hfill}\,constant on $\ [\tau]$
\vskip 0.6cm

\noindent\hbox to 4.3cm{{\bf cocycle} $\,\hat\phi\,$:\hfill}$
\hat\phi_{\hat\tau;\m z^{2n}}(z_0)\ \equiv\ 1$
\vskip 0.6cm

\noindent\hbox to 4.3cm{\bf General solution:\hfill}$\,
\CW_{\tau;\m z^{2n}}\ =\ (-1)^{np}\,\m\CW^{\hat\tau}_{\tau;\m z^{2n}}$
\quad\ for \ $\,p=0,1\,$ if $\,Z_{[\tau]}=\{1,z^2\}$
\vskip 0.6cm
\vfill
\eject

\noindent $--------------------------------$
\vskip 0.3cm

\noindent\hbox to 4.3cm{{\bf Subgroup} $\,Z\,$:\hfill}$\,\cong\,\NZ_{4}$
\vskip 0.4cm

\noindent\hbox to 4.3cm{\bf Quotient group:\hfill}$\,SO(2r)/\NZ_2$
\vskip 0.3cm

\noindent $--------------------------------$
\vskip 0.5cm

\noindent\hbox to 4.3cm{\bf level:\hfill}$\,\sk\in2\NZ\,$
\vskip 0.5cm

\noindent\hbox to 4.3cm{2-{\bf chain} $\,V\,$:\hfill}$\,
V_{z^{n},z^{m}}\ \equiv\ 1$
\vskip 0.5cm

\noindent\hbox to 4.3cm{2-{\bf cocycle} $\,\CV\,$:\hfill}$\,
\CV_{\tau;\m z^{n},z^{m}}\ =\ \cases{\ \hbox to 4.9cm{$1$\hfill}{\rm if}
\ \ n+m<4\,,\cr
\ \hbox to 4.9cm{$(-1)^{n_{r-1}+n_{r}}=\ee^{4\pi\si\,Q_z(\lambda)}
$\hfill}{\rm if}\ \ n+m\geq 4}$
\vskip 0.2cm

\noindent\hbox to 4.3cm{\ \hfill}\,for $\,n,m=0,1,2,3$
\vskip 0.7cm

\noindent\hbox to 4.3cm{\bf Stabilizers:\hfill}\hbox to 4cm{$\,
Z_{[\tau]}\cong\NZ_4\,$\hfill}if $\,n_0=n_1=n_{r-1}=n_r\,$

\noindent\hbox to 4.3cm{\ \hfill}\hbox to 4cm{\ \hfill}and 
$\,n_i=n_{r-i},\ i=2,\dots,r-2$\m,

\noindent\hbox to 8.3cm{\ \hfill}possible if $\,\,\sk\in4\NZ\m$, 
\vskip 0.2cm

\noindent\hbox to 4.3cm{\ \hfill}\hbox to 4cm{$\,Z_{[\tau]}=\{1,z^2\}
\cong\NZ_2$\hfill}if $\,n_0=n_1\,$\ and\ $\,n_{r-1}=n_r\,$

\noindent\hbox to 4.3cm{\ \hfill}\hbox to 4cm{\ \hfill}but 
$\,Z_{[\tau]}\not\cong\NZ_4$\m, 
\vskip 0.2cm

\noindent\hbox to 4.3cm{\ \hfill}\hbox to 4cm{$\,Z_{[\tau]}=\{1\}$
\hfill}otherwise
\vskip 0.7cm

\noindent\hbox to 4.3cm{\bf Special solution:\hfill}$\,
\CW^{\hat\tau}_{\tau;\m z^{n}}\ =\ \ee^{\pi\si\m\sigma n/4}$
\vskip 0.1cm

\noindent\hbox to 4.3cm{\ \hfill}\,where $\,\sigma=0,1\,$ is the parity 
of $\,n_{r-1}+n_r=4\,Q_z(\lambda)\,\,{\mathrm mod}\,2$ 

\noindent\hbox to 4.3cm{\ \hfill}\,constant on $\ [\tau]$
\vskip 0.6cm

\noindent\hbox to 4.3cm{{\bf cocycle} $\,\hat\phi\,$:\hfill}$
\hat\phi_{\hat\tau;\m z^n}(z_0)\ \equiv\ 1$
\vskip 0.6cm

\noindent\hbox to 4.3cm{\bf General solution:\hfill}$\,
\CW_{\tau;\m z^{n}}\ =\ \ee^{\pi\si\m pn/2}\,\m\CW^{\hat\tau}_{\tau;\m z^{n}}$
\quad\ for \ $\,p=0,\dots,n'-1\,$ 
if $\,Z_{[\tau]}\cong\NZ_{n'}$
\vskip 1.5cm
\vfill
\eject

\noindent $================================================$
\vskip 0.2cm
\noindent\hbox to 4.3cm{\Large\bf Covering group:\hfill}{\Large$\,
\bf Spin(2r)\,$ for even $\,r$}
\vskip 0.2cm
\noindent $--------------------------------------$
\vskip 0.6cm

\noindent\hbox to 4.3cm{\bf Center:\hfill}$\,\cong\,{\NZ_2\hspace{-0.06cm}\times\hspace{-0.06cm}\NZ_2}$
\vskip 0.4cm

\noindent\hbox to 4.3cm{\bf Generators:\hfill}$\,z_1,\,z_2\,,\ \,z_1^2=1=z_2^2$
\vskip 0.4cm

\noindent\hbox to 4.3cm{{\bf Action on} $\,\CA_{_W}\,$:\hfill}$\,
z_1\tau_i=\tau_{r-i}\,,\ i=0,\dots,r\,,$
\vskip 0.17cm

\noindent\hbox to 4.3cm{\ \hfill}$\,z_2\tau_0=\tau_1\,,\quad \,
z_2\tau_1=\tau_0\,,\quad\,z_2\tau_i=\tau_i\,,
\ \,i=2,\dots,r-2\,,$

\noindent\hbox to 4.3cm{\ \hfill}$\,z_2\tau_{r-1}=\tau_r\,,\quad 
\,z_2\tau_r=\tau_{r-1}$
\vskip 0.4cm

\noindent\hbox to 4.3cm{{\bf weight} $\,\lambda\equiv\sk\tau\,$:\hfill}$\,
\sum\limits_{i=0}^r n_i\tau_i\,,\ \ 
n_0,n_1,n_{r-1},n_r\in\NZ\,,$

\noindent\hbox to 4.3cm{\ \hfill}$\,n_i\in2\NZ\ \,{\rm for}\ \,i=2,\dots,r-2$
\vskip 0.6cm

\noindent $--------------------------------$
\vskip 0.3cm

\noindent\hbox to 4.3cm{{\bf Subgroup} $\,Z\,$:\hfill}$\,\{1,z_1\}
\cong\NZ_{2}$
\vskip 0.4cm

\noindent\hbox to 4.3cm{\bf Quotient group:\hfill}$\,Spin(2r)/\{1,z_1\}$
\vskip 0.3cm

\noindent $--------------------------------$
\vskip 0.4cm

\noindent\hbox to 4.3cm{\bf level:\hfill}$\,\sk\in\NZ\ \,$ \,if $\,r/2\,$  
is even,

\noindent\hbox to 4.3cm{\ \hfill}$\,\sk\in2\NZ\,$ if $\,r/2\,$ is odd
\vskip 0.5cm

\noindent\hbox to 4.3cm{2-{\bf chain} $\,V\,$:\hfill}$\,
V_{z_1^{n},z_1^{m}}\ \equiv\ 1$
\vskip 0.5cm

\noindent\hbox to 4.3cm{2-{\bf cocycle} $\,\CV\,$:\hfill}$\,
\CV_{\tau;\m z_1^{n},z_1^{m}}\,=\, 
\cases{\,\hbox to 4.9cm{$1$\hfill}{\rm if}\ \ n+m\leq1\,,\cr
\,\hbox to 4.9cm{$(-1)^{{r\over2}(n_{r-1}+n_r)}
=\ee^{\pi\si\,r\,Q_{z_{_2}}(\lambda)}
$\hfill}{\rm if}\ \ n+m=2}$
\vskip 0.2cm

\noindent\hbox to 4.3cm{\ \hfill}\,for $\,n,m=0,1$
\vskip 0.6cm

\noindent\hbox to 4.3cm{\bf Stabilizers:\hfill}\hbox to 2.7cm{$\,
Z_{[\tau]}=\{1,z_1\}\,$\hfill}if $\,n_i=n_{r-i}\m$, $\,i=0,\dots,r\m$,
\ \,possible if $\,\,\sk\in2\NZ\m$,

\noindent\hbox to 4.3cm{\ \hfill}\hbox to 2.7cm{$\,
Z_{[\tau]}=\{1\}\,$\hfill}otherwise
\vskip 0.6cm

\noindent\hbox to 4.3cm{\bf Special solution:\hfill}$\,
\CW^{\hat\tau}_{\tau;\m z_1^{n}}\ =\ \ee^{\pi\si\m\sigma n/2}$
\vskip 0.2cm

\noindent\hbox to 4.3cm{\ \hfill}\,where $\,\sigma=0,1\,$ is 
the parity of $\,{r\over2}(n_{r-1}+n_r)=r\,Q_{z_{_2}}
(\lambda)\,\,{\mathrm mod}\,2$ 

\noindent\hbox to 4.3cm{\ \hfill}\,constant on $\ [\tau]$
\vskip 0.6cm

\noindent\hbox to 4.3cm{{\bf cocycle} $\,\hat\phi\,$:\hfill}$
\hat\phi_{\hat\tau;\m z_1^n}(z_0)\ \equiv\ 1$
\vskip 0.6cm

\noindent\hbox to 4.3cm{\bf General solution:\hfill}$\,
\CW_{\tau;\m z_1^{n}}\ =\ (-1)^{np}\,\m\CW^{\hat\tau}_{\tau;\m z_1^{n}}$
\quad\ for \ $\,p=0,1\,$ if $\,Z_{[\tau]}=\{1,z_1\}$
\vfill
\eject

\noindent $--------------------------------$
\vskip 0.3cm

\noindent\hbox to 4.3cm{{\bf Subgroup} $\,Z\,$:\hfill}$\,\{1,z_2\}
\cong\NZ_{2}$
\vskip 0.4cm

\noindent\hbox to 4.3cm{\bf Quotient group:\hfill}$\,SO(2r)$
\vskip 0.3cm

\noindent $--------------------------------$
\vskip 0.5cm

\noindent\hbox to 4.3cm{\bf level:\hfill}$\,\sk\in\NZ\,$
\vskip 0.5cm

\noindent\hbox to 4.3cm{2-{\bf chain} $\,V\,$:\hfill}$\,
V_{z_2^{n},z_2^{m}}\ \equiv\ 1$
\vskip 0.5cm

\noindent\hbox to 4.3cm{2-{\bf cocycle} $\,\CV\,$:\hfill}$\,
\CV_{\tau;\m z_2^{n},z_2^{m}}\,=\, 
\cases{\,\hbox to 4.9cm{$1$\hfill}{\rm if}\ \ n+m\leq1\m,\cr
\,\hbox to 4.9cm{$(-1)^{n_{r-1}+n_r}=\ee^{2\pi\si\,Q_{z_{_2}}(\lambda)}
$\hfill}{\rm if}\ \ n+m=2}$
\vskip 0.2cm

\noindent\hbox to 4.3cm{\ \hfill}\,for $\,n,m=0,1$
\vskip 0.6cm

\noindent\hbox to 4.3cm{\bf Stabilizers:\hfill}\hbox to 2.7cm{$\,
Z_{[\tau]}=\{1,z_2\}\,$\hfill}if $\,\,n_0=n_1\,\ {\rm and}\ \,n_{r-1}=n_r\m$,
\ possible if $\,\,\sk\in2\NZ\m$,   

\noindent\hbox to 4.3cm{\ \hfill}\hbox to 2.7cm{$\,Z_{[\tau]}=\{1\}
\,$\hfill}otherwise
\vskip 0.6cm

\noindent\hbox to 4.3cm{\bf Special solution:\hfill}$\,
\CW^{\hat\tau}_{\tau;\m z_2^{n}}\ =\ \hbox to 2.4cm{$\,\ee^{\pi\si\m
\sigma n/2}$\hfill}$
\vskip 0.2cm

\noindent\hbox to 4.3cm{\ \hfill}\,where $\,\sigma=0,1\,$ is 
the parity of $\,n_{r-1}+n_r=2\,Q_{z_{_2}}(\lambda)\,\,{\mathrm mod}\,2$ 

\noindent\hbox to 4.3cm{\ \hfill}\,constant on $\ [\tau]$
\vskip 0.6cm

\noindent\hbox to 4.3cm{{\bf cocycle} $\,\hat\phi\,$:\hfill}$
\hat\phi_{\hat\tau;\m z_2^n}(z_0)\ \equiv\ 1$
\vskip 0.6cm

\noindent\hbox to 4.3cm{\bf General solution:\hfill}$\,
\CW_{\tau;\m z_2^{n}}\ =\ (-1)^{np}\,\m\CW^{\hat\tau}_{\tau;\m z_2^{n}}$
\quad\ for \ $\,p=0,1\,$ if $\,Z_{[\tau]}=\{1,z_2\}$
\vskip 0.6cm
\vfill
\eject

\noindent $--------------------------------$
\vskip 0.3cm

\noindent\hbox to 4.3cm{{\bf Subgroup} $\,Z\,$:\hfill}$\,\{1,z_1z_2\}
\cong\NZ_{2}$
\vskip 0.4cm

\noindent\hbox to 4.3cm{\bf Quotient group:\hfill}$\,Spin(2r)/\{1,z_1z_2\}$
\vskip 0.3cm

\noindent $--------------------------------$
\vskip 0.5cm

\noindent\hbox to 4.3cm{\bf level:\hfill}$\,\sk\in\NZ\ \,\,$ if $\,r/2\,$  
is even,

\noindent\hbox to 4.3cm{\ \hfill}$\,\sk\in2\NZ\,$ if $\,r/2\,$ is odd
\vskip 0.5cm

\noindent\hbox to 4.3cm{2-{\bf chain} $\,V\,$:\hfill}$\,
V_{(z_1z_2)^{n},(z_1z_2)^{m}}\ \equiv\ 1$
\vskip 0.5cm

\noindent\hbox to 4.3cm{2-{\bf cocycle} $\,\CV\,$:\hfill}$\,
\CV_{\tau;\m(z_1z_2)^n,(z_1z_2)^{m}}\hspace{-0.05cm}=\hspace{-0.07cm}
\cases{\hbox to 4.74cm{$\hspace{-0.07cm}1$\hfill}{\rm if}\ n+m\leq1\m,\cr
\hbox to 4.74cm{$\hspace{-0.07cm}(-1)^{{r\over2}(n_{r-1}+n_r)}
\hspace{-0.05cm}=\ee^{\pi\si\,r\,Q_{z_{_2}}(\lambda)}$\hfill}{\rm if}\ n+m=2}$
\vskip 0.2cm

\noindent\hbox to 4.3cm{\ \hfill}\,for $\,n,m=0,1$
\vskip 0.6cm

\noindent\hbox to 4.3cm{\bf Stabilizers:\hfill}\hbox to 3cm{$\,
Z_{[\tau]}=\{1,z_1z_2\}\,$\hfill}if \ $\,n_0=n_{r-1}\m$, $\,n_1=n_r\m$,

\noindent\hbox to 7.3cm{\ \hfill}and \,$\,n_i=n_{r-i}\,$ for $\,i=2,
\dots,r-2\m,$ 

\noindent\hbox to 7.3cm{\ \hfill}possible if $\ \sk\in2\NZ\m$,
\vskip 0.1cm

\noindent\hbox to 4.3cm{\ \hfill}\hbox to 3cm{$\,
Z_{[\tau]}=\{1\}\,$\hfill}otherwise
\vskip 0.6cm

\noindent\hbox to 4.3cm{\bf Special solution:\hfill}$\,
\CW^{\hat\tau}_{\tau;\m(z_1z_2)^{n}}\ =\ \ee^{\pi\si\m\sigma n/2}$
\vskip 0.2cm

\noindent\hbox to 4.3cm{\ \hfill}\,where $\,\sigma=0,1\,$ is 
the parity of $\,{r\over2}(n_{r-1}+n_r)=r\,Q_{z_{_2}}(\lambda)\,\,
{\mathrm mod}\,2$ 

\noindent\hbox to 4.3cm{\ \hfill}\,constant on $\ [\tau]$
\vskip 0.6cm

\noindent\hbox to 4.3cm{{\bf cocycle} $\,\hat\phi\,$:\hfill}$
\hat\phi_{\hat\tau;\m(z_1z_2)^n}(z_0)\ \equiv\ 1$
\vskip 0.6cm

\noindent\hbox to 4.3cm{\bf General solution:\hfill}$\,
\CW_{\tau;\m (z_1z_2)^{n}}=(-1)^{np}\,\m\CW^{\hat\tau}_{\tau;
\m(z_1z_2)^{n}}$\ \ \,for \,$\,p=0,1\,$ if $\,Z_{[\tau]}=\{1,z_1z_2\}$
\vskip 0.6cm
\vfill
\eject

\noindent $--------------------------------$
\vskip 0.3cm

\noindent\hbox to 4.3cm{{\bf Subgroup} $\,Z\,$:\hfill}$\,\{1,z_1,z_2,z_1z_2\}
\m\cong\m{\NZ_2\hspace{-0.06cm}\times\hspace{-0.06cm}\NZ_2}$
\vskip 0.4cm

\noindent\hbox to 4.3cm{\bf Quotient group:\hfill}$\,SO(2r)/\NZ_2$
\vskip 0.3cm

\noindent $--------------------------------$
\vskip 0.5cm

\noindent\hbox to 4.3cm{\bf level:\hfill}$\,\sk\in\NZ\ \,\,$ if $\,r/2\,$  
is even,

\noindent\hbox to 4.3cm{\ \hfill}$\,\sk\in2\NZ\,$ if $\,r/2\,$ is odd
\vskip 0.5cm

\noindent\parbox{4.3cm}{\ \hfill\\
2-{\bf chain}\ $\,V\,$:\hfill\\
$\sk\,$\ odd\hfill}$\,
V_{z,z'}=\cases{\,\hbox to 2.55cm{$\mp\si$\hfill}
{\rm if}\ \ (z,z')=(z_2,z_1)\,,\cr
\,\hbox to 2.55cm{$\pm\si$\hfill}
{\rm if}\ \ (z,z')=(z_2,z_1z_2),(z_1z_2,z_1),\cr
\hspace{4.62cm}(z_1z_2,z_1z_2)\,,\cr
\,\hbox to 2.55cm{$-1$\hfill}
{\rm if}\ \ (z,z')=(z_1z_2,z_2)\,,\cr
\,\hbox to 2.55cm{$\, 1$\hfill}{\rm otherwise}}$
\vskip 0.6cm

\noindent\parbox{4.3cm}{\ \hfill\\
2-{\bf chain}\ $\,V\,$:\hfill\\
$\sk\,$\ even\hfill}$\,
V_{z,z'}=\cases{\,\hbox to 2.55cm{$\pm1$\hfill}
{\rm if}\ \ \hbox to 4.92cm{$(z,z')=(z_2,z_1),(z_2,z_1z_2)\m,$\hfill}\cr
\hspace{4.62cm}(z_1z_2,z_1),(z_1z_2,z_1z_2)\m,\cr
\,\hbox to 2.55cm{$\, 1$\hfill}{\rm otherwise}}$
\vskip 0.6cm

\noindent\parbox{4.3cm}{\ \hfill\\
2-{\bf cocycle}\ $\,\CV\,$:\hfill\\
$\sk\,$\ odd\hfill}$\,
\CV_{\tau;\m z,z'}=\cases{\,\hbox to 2.55cm{$\mp\si(-1)^{n_{r-1}+n_r}$\hfill}
{\rm if}\ \ (z,z')=(z_2,z_1)\,,\cr
\,\hbox to 2.55cm{$\, (-1)^{n_{r-1}+n_r}$\hfill}
{\rm if}\ \ (z,z')=(z_2,z_2)\,,\cr
\,\hbox to 2.55cm{$\pm \si$\hfill}
{\rm if}\ \ (z,z')=(z_2,z_1z_2),(z_1z_2,z_1z_2)\,,\cr
\,\hbox to 2.55cm{$\pm\si(-1)^{n_{r-1}+n_r}$\hfill}
{\rm if}\ \ (z,z')=(z_1z_2,z_1)\,,\cr
\,\hbox to 2.55cm{$-(-1)^{n_{r-1}+n_r}$\hfill}
{\rm if}\ \ (z,z')=(z_1z_2,z_2)\,,\cr
\,\hbox to 2.55cm{$\, 1$\hfill}{\rm otherwise}}$
\vskip 0.6cm

\noindent\parbox{4.3cm}{\ \hfill\\
2-{\bf cocycle}\ $\,\CV\,$:\hfill\\
$\sk\,$\ even, $\,r/2\,$ even\hfill}$\,
\CV_{\tau;\m z,z'}=\cases{\,\hbox to 2.55cm{$\pm(-1)^{n_{r-1}+n_r}$\hfill}
{\rm if}\ \ \hbox to 4.92cm{$(z,z')=(z_2,z_1),(z_1z_2,z_1)\m,$\hfill}\cr
\,\hbox to 2.55cm{$\, (-1)^{n_{r-1}+n_r}$\hfill}
{\rm if}\ \ \hbox to 4.92cm{$(z,z')=(z_2,z_2),(z_1z_2,z_2)\m,$\hfill}\cr
\,\hbox to 2.55cm{$\pm1$\hfill}
{\rm if}\ \ \hbox to 5.22cm{$(z,z')=(z_2,z_1z_2),(z_1z_2,z_1z_2)\m,$\hfill}\cr
\,\hbox to 2.55cm{$\, 1$\hfill}{\rm otherwise}}$
\vskip 0.6cm

\noindent\parbox{4.3cm}{\ \hfill\\
2-{\bf cocycle}\ $\,\CV\,$:\hfill\\
$\sk\,$\ even, $\,r/2\,$ odd\hfill}$\,
\CV_{\tau;\m z,z'}=\cases{\,\hbox to 2.55cm{$\, (-1)^{n_{r-1}+n_r}$\hfill}
{\rm if}\ \ \hbox to 4.92cm{$(z,z')=(z_1,z_1),(z_1,z_1z_2),$\hfill}\cr
\hbox to 4.62cm{\ \hfill}(z_2,z_2),(z_1z_2,z_2)\m,\cr
\,\hbox to 2.55cm{$\pm(-1)^{n_{r-1}+n_r}$\hfill}
{\rm if}\ \ \hbox to 4.92cm{$(z,z')=(z_2,z_1),(z_1z_2,z_1z_2)\m,$\hfill}\cr
\,\hbox to 2.55cm{$\pm1$\hfill}
{\rm if}\ \ \hbox to 4.92cm{$(z,z')=(z_2,z_1z_2),(z_1z_2,z_1)\m,$\hfill}\cr
\,\hbox to 2.55cm{$\, 1$\hfill}{\rm otherwise}}$
\vskip 0.5cm

\noindent with $\,(-1)^{n_{r-1}+n_r}\hspace{-0.04cm}=\hspace{-0.01cm}
\ee^{2\pi\si\,Q_{z_{_2}}(\lambda)}\,$ and the signs corresponding 
to two choices $\,\CG'_\pm\,$ of the\nobreak\ \nobreak gerbe
\vfill
\eject

\noindent$------------------$
\vskip 0.5cm

\noindent\hbox to 4.3cm{\bf Stabilizers:\hfill}\hbox to 3cm{$\,
Z_{[\tau]}=\{1\}\,$\hfill}\hbox to 0.8cm{if\hfill}$\,n_i\not=n_{r-i}\,$ 
for some $\,i=0,\dots,r\,$

\noindent\hbox to 7.3cm{\ \hfill}\hbox to 0.8cm{and\hfill}$\,n_0\not=
n_1\,$ or $\,n_{r-1}\not=n_r$

\noindent\hbox to 7.3cm{\ \hfill}\hbox to 0.8cm{and\hfill}$\,n_0\not=
n_{r-1}\,$ or $\,n_1\not=n_r\,$ 
or $\,n_i\not=n_{r-i}$

\noindent\hbox to 7.3cm{\ \hfill}\hbox to 0.8cm{\ \hfill}\,for some 
$\,i=2,\dots,r-2$
\vskip 0.1cm

\noindent\hbox to 4.3cm{\bf Special solution:\hfill}$\,
\CW^{\hat\tau}_{\hat\tau_z;\m z'}\ =\ \CV_{\hat\tau;\m z,z'}$
\vskip 0.5cm

\noindent\hbox to 4.3cm{{\bf cocycle} $\,\hat\phi\,$:\hfill}$
\,\hat\phi_{\hat\tau;z}(1)\ \equiv\ 1$
\vskip 0.8cm

\noindent$------------------$
\vskip 0.5cm

\noindent\hbox to 4.3cm{\bf Stabilizers:\hfill}\hbox to 3cm{$\,
Z_{[\tau]}=\{1,z_1\}\,$\hfill}if \quad $\,n_i=n_{r-i}\,$ for $\,i=0,
\dots,r\,$ but $\,n_0\not=n_1\m$,

\noindent\hbox to 7.3cm{\ \hfill}possible if $\ \sk\in2\NZ$
\vskip 0.4cm

\noindent\parbox{4.3cm}{\ \hfill\\
{\bf Special solution:\hfill}\\
$r/2\,$ even}
$\,
\CW^{\hat\tau}_{\hat\tau_z;\m z'}\ =\ 
\CV_{\hat\tau;z,z'}$
\vskip 0.3cm

\noindent\parbox{4.3cm}{\ \hfill\\
{\bf Special solution:\hfill}\\
$r/2\,$ odd}
$\,\CW^{\hat\tau}_{\hat\tau_z;\m z'}\ =\   
\cases{\,\hbox to 1.8cm{$\ee^{\pi\si
\m \sigma/2}$\hfill}\hbox to 0.7cm{{\rm if}\hfill}(z,z')=(1,z_1),(1,z_1z_2),
\cr
\hbox to 4.05cm{\ \hfill}\hbox to 4.04cm{$(z_1,z_1),(z_1,z_1z_2),$\hfill}\cr
\,\hbox to 1.8cm{$\pm\ee^{-\pi\si
\m \sigma/2}$\hfill}\hbox to 0.7cm{{\rm if}\hfill}(z,z')=(z_2,z_1),
(z_1z_2,z_1),\cr
\,\hbox to 1.8cm{$(-1)^{\sigma}$\hfill}\hbox to 0.7cm{{\rm if}
\hfill}(z,z')=(z_2,z_2),(z_1z_2,z_2),\cr
\,\hbox to 1.8cm{$\pm\ee^{\pi\si
\m \sigma/2}$\hfill}\hbox to 0.7cm{{\rm if}\hfill}(z,z')=(z_2,z_1z_2),
(z_1z_2,z_1z_2),\cr
\,\hbox to 1.8cm{$1$\hfill}{\rm otherwise}}$
\vskip 0.2cm

\noindent\hbox to 4.3cm{\ \hfill}\,where $\ \sigma=0,1\ $ is the parity of 
$\ n_{r-1}+n_r=2\,Q_{z_{_2}}(\lambda)\,\,{\mathrm mod}\,2$ 

\noindent\hbox to 4.3cm{\ \hfill}\,constant on $\ [\tau]$
\vskip 0.5cm

\noindent\hbox to 4.3cm{{\bf cocycle} $\,\hat\phi\,$:\hfill}$\,
\hat\phi_{\hat\tau;z}(z_0)\ 
=\ \cases{\ \hbox to 1.93cm{$\pm(-1)^{\sigma}$\hfill}
{\rm if}\ \ (z,z_0)=(z_2,z_1),(z_1z_2,z_1)\,,\cr
\ \hbox to 1.93cm{$1$\hfill}{\rm otherwise}}$
\vskip 0.5cm

\noindent\hbox to 4.3cm{\bf General solution:\hfill}$\,
\CW_{\tau;\m z_1^nz_2^m}\ =\ (-1)^{np}\,\m\CW^{\hat\tau}_{\tau;
\m z_1^nz_2^m}$ \quad\ for \ $\,p=0,1$ 
\vskip 0.8cm

\noindent$------------------$
\vskip 0.5cm

\noindent\hbox to 4.3cm{\bf Stabilizers:\hfill}\hbox to 3cm{$\,
Z_{[\tau]}=\{1,z_2\}\,$\hfill}\hbox to 0.8cm{if\hfill}$\,n_0=n_1\,$ 
and $\,n_{r-1}=n_r$

\noindent\hbox to 7.3cm{\ \hfill}but $\,n_i\not=n_{r-i}\,$ for some 
$\,i=0,\dots,r\m$,

\noindent\hbox to 7.3cm{\ \hfill}possible if $\ \sk\in2\NZ$
\vskip 0.4cm

\noindent\hbox to 4.3cm{\bf Special solution:\hfill}$\,
\CW^{\hat\tau}_{\hat\tau_z;\m z'}\ =\ \cases{\ \hbox to 1.6cm{$\pm1$\hfill}
{\rm if}\ \ (z,z')=(1,z_1z_2),(z_1,z_2),\cr
\noindent\hbox to 3.73cm{\ \hfill}(z_2,z_1z_2),(z_1z_2,z_2)\,,\cr
\ \hbox to 1.6cm{$1$\hfill}{\rm otherwise}}$
\vskip 0.4cm

\noindent\hbox to 4.3cm{{\bf cocycle} $\,\hat\phi\,$:\hfill}$\,
\hat\phi_{\hat\tau;z}(z_0)\ =\ \cases{\ \hbox to 1.75cm{$\pm1$
\hfill}
{\rm if}\ \ (z,z_0)=(z_1,z_2),(z_1z_2,z_2)\,,\cr
\ \hbox to 1.75cm{$1$\hfill}{\rm otherwise}}$
\vskip 0.5cm

\noindent\hbox to 4.3cm{\bf General solution:\hfill}$\,
\CW_{\tau;\m z_1^nz_2^m}\ =\ (-1)^{mp}\,\m\CW^{\hat\tau}_{\tau;\m z_1^nz_2^m}$
\quad\ for \ $\,p=0,1$ 
\vfill
\eject

\noindent$------------------$
\vskip 0.5cm

\noindent\hbox to 4.3cm{\bf Stabilizers:\hfill}\hbox to 3cm{$\,
Z_{[\tau]}=\{1,z_1z_2\}\,$\hfill}\hbox to 0.8cm{if\hfill}$\,n_0=n_{r-1}
\not=n_1=n_r$

\noindent\hbox to 7.3cm{\ \hfill}and $\ n_i=n_{r-i}\,$ for $\,i=2,\dots,r-2\m$,

\noindent\hbox to 7.3cm{\ \hfill}possible if $\ \sk\in2\NZ$
\vskip 0.5cm

\noindent\parbox{4.3cm}{\ \hfill\\
{\bf Special solution:}\hfill\\
$r/2\,$\ even}$\,
\CW^{\hat\tau}_{\hat\tau_z;\m z'}=\cases{\,\hbox to 2.3cm{$
\left\{{1\atop\si}\right\}$\hfill}\hbox to 0.7cm{{\rm if}\hfill}(z,z')
=(1,z_1z_2),(z_1,z_2),\cr
\hbox to 4.55cm{\ \hfill}(z_2,z_2),(z_1z_1,z_1z_2),\cr
\,\hbox to 2.3cm{$(-1)^\sigma\left\{{1\atop-\si}\right\}
$\hfill}\hbox to 0.7cm{{\rm if}\hfill}(z,z')=(1,z_2),(z_1,z_1z_2),\cr
\hbox to 4.55cm{\ \hfill}(z_2,z_1z_2),(z_1z_2,z_2),\cr
\,\hbox to 2.3cm{$1$\hfill}{\rm otherwise}}$
\vskip 0.7cm

\noindent\parbox{4.3cm}{\ \hfill\\
{\bf Special solution:}\hfill\\
$r/2\,$\ odd}$\,
\CW^{\hat\tau}_{\hat\tau_z;\m z'}=\cases{\,\hbox to 2.3cm{$
\ee^{\pi\si\m \sigma/2}\left\{{1\atop-\si}\right\}
$\hfill}\hbox to 0.7cm{{\rm if}\hfill}(z,z')=(1,z_2),(z_1z_2,z_2),\cr
\,\hbox to 2.3cm{$\ee^{\pi\si\m \sigma/2}\left\{{1\atop\si}\right\}
$\hfill}\hbox to 0.7cm{{\rm if}\hfill}(z,z')=(1,z_1z_2),(z_1,z_2),\cr
\hbox to 4.55cm{\ \hfill}(z_2,z_2),(z_1z_2,z_1z_2),\cr
\,\hbox to 2.3cm{$(-1)^\sigma$\hfill}\hbox to 0.7cm{{\rm if}\hfill}
(z,z')=(z_1,z_1),(z_2,z_1),\cr
\,\hbox to 2.3cm{$\ee^{-\pi\si\m \sigma/2}
\left\{{1\atop-\si}\right\}$\hfill}\hbox to 0.7cm{{\rm if}\hfill}
(z,z')=(z_1,z_1z_2),(z_2,z_1z_2),\cr
\,\hbox to 2.3cm{$1$\hfill}{\rm otherwise}}$
\vskip 0.3cm

\noindent\hbox to 4.3cm{\ \hfill}\,where $\ \sigma=0,1\ $ is the parity of 
$\ n_{r-1}+n_r=2\,Q_{z_{_2}}(\lambda)\,\,{\mathrm mod}\,2$ 

\noindent\hbox to 4.3cm{\ \hfill}\,constant on $\ [\tau]$
\vskip 0.6cm

\noindent\hbox to 4.3cm{{\bf cocycle} $\,\hat\phi\,$:\hfill}$\,
\hat\phi_{\hat\tau;z}(z_0)\ =\ \cases{\ \hbox to 2.2cm{$\pm(-1)^{\sigma}$
\hfill}
{\rm if}\ \ (z,z_0)=(z_1,z_1z_2),(z_2,z_1z_2)\,,\cr
\ \hbox to 2.2cm{$1$\hfill}{\rm otherwise}}$
\vskip 0.6cm

\noindent\hbox to 4.3cm{\bf General solution:\hfill}$\,
\CW_{\tau;\m z_1^nz_2^m}\ =\ (-1)^{np}\,\m\CW^{\hat\tau}_{\tau;\m z_1^nz_2^m}$
\quad\ for \ $\,p=0,1$ 
\vskip 1cm

\noindent$------------------$
\vskip 0.5cm

\noindent\hbox to 4.3cm{\bf Stabilizers:\hfill}\hbox to 4.4cm{$\,
Z_{[\tau]}=\{1,z_1,z_2,z_1z_2\}\,$\hfill}\hbox to 0.8cm{if\hfill}$\,n_0=n_1
=n_{r-1}=n_r$ 

\noindent\hbox to 8.73cm{\ \hfill}and $\ n_i=n_{r-i}\,$ for $\,i=2,\dots,r-2\m$,

\noindent\hbox to 8.73cm{\ \hfill}possible if $\ \sk\in2\NZ$
\vskip 0.7cm

\noindent\hbox to 4.3cm{\bf Special solution:\hfill}$\,
\CW^{\hat\tau}_{\tau;\m z,z'}\ =\ \cases{\ $1$\,,\cr
\ {\rm no\ solution\ for\ the\ lower\ sign\ case}}$ 
\vskip 0.6cm

\noindent\hbox to 4.3cm{{\bf cocycle} $\,\hat\phi\,$:\hfill}$
\hat\phi_{\hat\tau;z}(z_0)\ \equiv\ 1$
\vskip 0.6cm

\noindent\hbox to 4.3cm{\bf General solution:\hfill}$\,
\CW_{\tau;\m z_1^nz_2^m}\ =\ (-1)^{np+mq}\,\m\CW^{\hat\tau}_{\tau;\m z_1^nz_2^m}$
\quad\ for \ $\,p,q=0,1$ 
\vskip 1.5cm
\vfill
\eject

\noindent $================================================$
\vskip 0.2cm
\noindent\hbox to 4.3cm{\Large\bf Covering group:\hfill}{\Large$\,\bf E_6$}
\vskip 0.2cm
\noindent $--------------------------------------$
\vskip 0.6cm

\noindent\hbox to 4.3cm{\bf Center:\hfill}$\,\cong\,\NZ_{3}$
\vskip 0.5cm

\noindent\hbox to 4.3cm{\bf Generator:\hfill}$\,z\,,\ \,z^3=1$
\vskip 0.5cm

\noindent\hbox to 4.3cm{{\bf Action on} $\,\CA_{_W}\,$:\hfill}$\,
z\tau_0=\tau_1\m,\ \ z\tau_1=\tau_5\m,\ \ z\tau_2=\tau_4\m,\ \   
z\tau_3=\tau_3\m,$

\noindent\hbox to 4.3cm{\ \hfill}$\,z\tau_4=\tau_6\m,\ \ z\tau_5=\tau_0\m,
\ \ z\tau_6=\tau_2$
\vskip 0.5cm

\noindent\hbox to 4.3cm{{\bf weight} $\,\lambda\equiv\sk\tau\,$:\hfill}$\,
\sum\limits_{i=0}^r n_i\tau_i\,,\ \ n_0,n_1,n_5\in\NZ\m,\ \,
n_2,n_4,n_6\in2\NZ\m,\ \,n_3\in3\NZ$
\vskip 0.6cm

\noindent $--------------------------------$
\vskip 0.3cm

\noindent\hbox to 4.3cm{{\bf Subgroup} $\,Z\,$:\hfill}$\,\{1,z,z^2\}
\cong\NZ_{3}$
\vskip 0.4cm

\noindent\hbox to 4.3cm{\bf Quotient group:\hfill}$\,E_6/\NZ_3$
\vskip 0.3cm

\noindent $--------------------------------$
\vskip 0.5cm

\noindent\hbox to 4.3cm{\bf level:\hfill}$\,\sk\in\NZ\,$ 
\vskip 0.5cm

\noindent\hbox to 4.3cm{2-{\bf chain} $\,V\,$:\hfill}$\,
V_{z^{n},z^{m}}\ \equiv\ 1$
\vskip 0.6cm

\noindent\hbox to 4.3cm{2-{\bf cocycle} $\,\CV\,$:\hfill}$\,
\CV_{\tau;\m z^n,z^m}\ =\ 1$
\vskip 0.6cm

\noindent\hbox to 4.3cm{\bf Stabilizers:\hfill}$\,Z_{[\tau]}
=\{1,z,z^2\}\,$
\ if \ $\,n_0=n_1=n_5\,$ and $\,n_2=n_4=n_6\m,$

\noindent\hbox to 7.26cm{\ \hfill}possible if $\ \sk\in3\NZ$
\vskip 0.1cm

\noindent\hbox to 4.3cm{\ \hfill}$\,Z_{[\tau]}=\{1\}$\,\quad\qquad otherwise
\vskip 0.6cm

\noindent\hbox to 4.3cm{\bf Special solution:\hfill}$\,
\CW^{\hat\tau}_{\tau;\m z^n}\ =\ 1$
\vskip 0.6cm

\noindent\hbox to 4.3cm{{\bf cocycle} $\,\hat\phi\,$:\hfill}$
\hat\phi_{\hat\tau;z}(z_0)\ \equiv\ 1$
\vskip 0.6cm

\noindent\hbox to 4.3cm{\bf General solution:\hfill}$\,
\CW_{\tau;\m z^n}\ =\ \ee^{2\pi\si\m np/3}$
\quad\ for \ $\,p=0,1,2\,\,$ if $\,\,Z_{[\tau]}=\{1,z,z^2\}$
\vfill
\eject

\noindent $================================================$
\vskip 0.2cm
\noindent\hbox to 4.3cm{\Large\bf Covering group:\hfill}{\Large$\,\bf E_7$}
\vskip 0.2cm
\noindent $--------------------------------------$
\vskip 0.6cm

\noindent\hbox to 4.3cm{\bf Center:\hfill}$\,\cong\,\NZ_{2}$
\vskip 0.5cm

\noindent\hbox to 4.3cm{\bf Generator:\hfill}$\,z\,,\ \,z^2=1$
\vskip 0.5cm

\noindent\hbox to 4.3cm{{\bf Action on} $\,\CA_{_W}\,$:\hfill}$\,
z\tau_0=\tau_1\m,\ \ z\tau_1=\tau_0\m,\ \ z\tau_i=\tau_{8-i}\ \,{\rm for}
\ \,i=2,\dots,6\m,\ \ z\tau_7=\tau_7$
\vskip 0.5cm

\noindent\hbox to 4.3cm{{\bf weight} $\,\lambda\equiv\sk\tau\,$:\hfill}$\,
\sum\limits_{i=0}^r n_i\tau_i\,,\ \ n_0,n_1\in\NZ\m,\ \,
n_2,n_6,n_7\in2\NZ\m,$

\noindent\hbox to 4.3cm{\ \hfill}$\,n_3,n_5\in3\NZ\m,\ \,n_4\in4\NZ$
\vskip 0.6cm

\noindent $--------------------------------$
\vskip 0.3cm

\noindent\hbox to 4.3cm{{\bf Subgroup} $\,Z\,$:\hfill}$\,\{1,z\}
\cong\NZ_{2}$
\vskip 0.5cm

\noindent\hbox to 4.3cm{\bf Quotient group:\hfill}$\,E_7/\NZ_2$
\vskip 0.3cm

\noindent $--------------------------------$
\vskip 0.5cm

\noindent\hbox to 4.3cm{\bf level:\hfill}$\,\sk\in2\NZ\,$ 
\vskip 0.5cm

\noindent\hbox to 4.3cm{2-{\bf chain} $\,V\,$:\hfill}$\,
V_{z^{n},z^{m}}\ \equiv\ 1$
\vskip 0.6cm

\noindent\hbox to 4.3cm{2-{\bf cocycle} $\,\CV\,$:\hfill}$\,
\CV_{\tau;\m z^n,z^m}\ =\ 
\cases{\ \hbox to 4.9cm{$\,1$\hfill}
{\rm if}\ \ n+m\leq 1\,,\cr
\ \hbox to 4.9cm{$(-1)^{n_1+n_3+{n_7\over2}}=\ee^{2\pi\si\,Q_z(\lambda)}
$\hfill}{\rm if}\ \ n+m=2}$
\vskip 0.2cm

\noindent\hbox to 4.3cm{\ \hfill}\,for $\,n,m=0,1$
\vskip 0.6cm

\noindent\hbox to 4.3cm{\bf Stabilizers:\hfill}$\,Z_{[\tau]}
=\{1,z\}\,$
\ if \ $\,n_0=n_1\m,$ $\,n_2=n_6\,$ and $\,n_3=n_5\,,$

\noindent\hbox to 4.3cm{\ \hfill}$\,Z_{[\tau]}=\{1\}
\,$ \ \ \ \hspace{0.03cm}\,otherwise
\vskip 0.6cm

\noindent\hbox to 4.3cm{\bf Special solution:\hfill}$\,
\CW^{\hat\tau}_{\tau;\m z^n}\ =\ \ee^{\pi\si\m\sigma n/2}$
\vskip 0.2cm

\noindent\hbox to 4.3cm{\ \hfill}\,where \ $\,\sigma=0,1\,\,$ is the
parity of $\,\,n_1+n_3+{n_7\over2}=2\,Q_z(\lambda)\,\,{\mathrm mod}\,2$ 

\noindent\hbox to 4.3cm{\ \hfill}\,constant on $\ [\tau]$ 
\vskip 0.6cm

\noindent\hbox to 4.3cm{{\bf cocycle} $\,\hat\phi\,$:\hfill}$
\hat\phi_{\hat\tau;z^n}(z_0)\ \equiv\ 1$
\vskip 0.6cm

\noindent\hbox to 4.3cm{\bf General solution:\hfill}$\,
\CW_{\tau;\m z^n}\ =\ (-1)^{np}\,\m\CW^{\hat\tau}_{\tau;\m z^n}$
\quad\ for \ $\,p=0,1\,$ if $\,Z_{[\tau]}=\{1,z\}$
\vfill
\eject

\nappendix{B}
\label{sec:appB}
\vskip 0.4cm

We provide here a direct check that the action of the group elements 
$\,w_z\,$ in $\,V_{\lambda}\,$ intertwines the multiplicity subspaces
$\,M_{\lambda^0\lambda}^{\,\lambda^1}$, \,see (\ref{wzW}). 
We have to show that under the action of $\,w_z\,$
in $\,V_\lambda\,$ an element $\,|\m\psi\rangle\in
M_{\lambda^0_z\lambda}^{\,\lambda^1_z}\subset V_\lambda\,$
lands in $\,M_{\lambda^0\lambda}^{\,\lambda^1}$.
\,One has to check then that $\,w_z\m|\m\psi\rangle\,$ satisfies
conditions (\ref{cint1}), (\ref{cint2}) and (\ref{cint3}). The first one
follows since
\qq
&t\,w_z\m|\m\psi\rangle\,=\,w_z\,w_z^{-1}t\,w_z\m|\m\psi\rangle
\,=\,[\tr\,(w_z^{-1}t\, w_z)(\lambda^1_z-\lambda^0_z)]
\m\,w_z\m|\m\psi\rangle&\cr
&=\,[\tr\,t(\lambda^1-\lambda^0)]\m\,
w_z\m|\m\psi\rangle\,.&
\nonumber
\qqq
For the second and the third condition, one has to use the relations
\qq
&&\hbox to 1.95cm{$w_z^{-1}\alpha_i\, w_z$\hfill}
\ =\ \cases{\,\hbox to 2.3cm{$\alpha_{i_z}$\hfill}
{\rm if}\quad i_z\not=0\,,\cr
\,\hbox to 2.3cm{$-\phi$\hfill}{\rm if}\quad i_z=0\,,}\cr
&&\hbox to 1.95cm{$w_z^{-1}\phi\,w_z$\hfill}\ =\ 
-\alpha_{_{\hspace{-0.02cm}0_z}}
\nonumber
\qqq
easy to check by computing the Killing form between both sides
and $\,\tau_i={1\over k_i^\vee}\lambda_i$. \,The above equalities 
imply that
\qq
&&\hbox to 1.95cm{$w_z^{-1}e_{-\alpha_i}\, w_z$\hfill}
\ =\ \cases{\,\hbox to 2.3cm{$e_{-\alpha_{i_z}}
$\hfill}
{\rm if}\quad i_z\not=0\,,\cr
\,\hbox to 2.3cm{$e_{\phi}$\hfill}{\rm if}\quad i_z=0\,,}\cr
&&\hbox to 1.95cm{$w_z^{-1}e_{\phi}\,w_z$\hfill}\ =\ 
e_{-\alpha_{_{\hspace{-0.02cm}0_z}}}
\nonumber
\qqq
and that 
\qq
&&\hbox to 1.95cm{$\tr\,\alpha_i^\vee\lambda^1$\hfill}
\ =\ \cases{\,\hbox to 2.3cm{$\tr\,\alpha_{i_z}
\lambda^1_z$\hfill}{\rm if}\quad i_z\not=0\,,\cr
\,\hbox to 2.3cm{$k\m-\m\tr\,\phi^\vee\lambda^1_z$\hfill}{\rm if}
\quad i_z=0\,,}\cr
&&\hbox to 1.95cm{$k\m-\m\tr\,\phi^\vee\lambda^1$\hfill}
\ =\ \tr\,\alpha_{0_z}^\vee\lambda^1_z\,.
\nonumber
\qqq
As a result,
\qq
&&\hbox to 3.15cm{$e_{-\alpha_i}^{\tr\,\alpha_i^\vee\lambda^1\m+\m1}
w_z\m|\m\psi\rangle$\hfill}\ =\ 
\cases{\,\hbox to 3.6cm{$w_z\,
e_{-\alpha_{i_z}}^{\tr\,\alpha_{i_z}^\vee\lambda_z^1\m+\m1}\m
|\m\psi\rangle$\hfill}{\rm if}\quad i_z\not=0\,,\cr
\,\hbox to 3.6cm{$w_z\,e_{\phi}^{k\m-\m\tr\,\phi^\vee\lambda^1_z\m+\m1}
\m|\m\psi\rangle$\hfill}{\rm if}
\quad i_z=0\,,}\cr
&&\hbox to 3.15cm{$
e_{\phi}^{k\m-\m\tr\,\phi^\vee\lambda^1_z\m+\m1}
w_z\m|\m\psi\rangle$\hfill}
\ =\ w_z\,e_{-\alpha_{_{\hspace{-0.02cm}0_z}}}^{\tr\,
\alpha_{0_z}^\vee\lambda^1_z\m+\m1}
\m|\m\psi\rangle
\nonumber
\qqq
and the right hand sides vanish since $\,|\m\psi\rangle\in
M_{\lambda^0_z\lambda}^{\,\lambda^1_z}$. \,Note that one of the 
consequences of the intertwining property of $\,w_z\,$ is the symmetry
$\,N_{\lambda^0_z\lambda}^{\,\lambda^1_z}
=N_{\lambda^0\lambda}^{\,\lambda^1}\,$ of the fusion coefficients
giving the dimensions of the spaces $\,M_{\lambda^0\lambda}^{\,\lambda^1}$.
\vfill
\eject

\nappendix{C}
\label{sec:appC}
\vskip 0.4cm

For the sake of illustration, we make explicit the formulae
of Sect.\,\ref{sec:bpfop} for the case of the WZW model with the
$\,SO(3)\,$ target. With the Cartan subalgebra of $\,su(2)\,$
spanned by the Pauli matrix $\,\sigma^3$, we label
the weights $\,j\sigma^3\,$ integrable at level $\,\sk\,$ by
spins $\,j=0,{_1\over^2},\dots,{\sk\over2}$. \,They correspond to
the integrable conjugacy classes $\,\CC_{j\sigma^3/\sk}\subset SU(2)\,$
supporting the rank 1 branes $\,\CD_1\,$ of the theory with the 
$\,SU(2)\,$ target that we shall also label by $\,j$. \,In the rank 
1 case, the multiplicity spaces of the decomposition (\ref{decc}) 
and (\ref{decc1}) take in the realization (\ref{wlb}) the form
\qq
M_{j^0j}^{j^1}\,=\,\cases{\,\hbox to 2.42cm{$\NC\,|j,j^1-j^0\rangle$\hfill}
{\rm if}\ \,j+j^0+j^1\in\NZ\ \,{\rm and}\ \,|j^0-j^1|\leq j\leq
{\mathrm min}({j^0+j^1\atop k-j^0-j^1})\,,\cr
\,\hbox to 2.42cm{$\{0\}$\hfill}{\rm otherwise}}\ \ \quad
\nonumber
\qqq
where $\,|j,m\rangle\,$ are the vectors of the standard orthonormal bases
of the spin $\,j\,$ representation $\,V_j\,$ labeled by the 
magnetic number $\,m=-j,-j+1,\dots,j$. \,The dimensions 
$\,N_{j^0j}^{\,j^1}\,$ are equal 1 or 0. \,In other words,
the multiplicity space $\,M_{j^0j}^{\,j^1}\,$ is spanned by the vector
\qq
e_{j^0j}^{\,j^1}\ \equiv\ N_{j^0j}^{\,j^1}\,\,|j,j^1-j^0\rangle\,\,\in\,V_j\,.
\nonumber
\qqq
\vskip 0.3cm

The group \,$SO(3)\,$ is the quotient of $\,SU(2)\,$ by the center 
$\,Z=\{1,-1\}$. \,The level $\,\sk\,$ of the $\,SO(3)\,$ model 
has to be even. The nontrivial element of $\,Z\,$ acts on the weights 
by $\,j\mapsto{\sk\over2}-j\equiv j_{_{-}}$. \,One has 
$\,N_{j^0_{_{-}}j}^{\,j^1_{_{-}}}=N_{j^0j}^{\,j^1}$.  \,We shall label the 
integrable conjugacy classes in $\,SO(3)\,$ by the corresponding 
$\,Z$-orbits of spins, with the two-point orbits $\,[j]=\{j,j_{_{-}}\}\m$, 
$\,j=0,{_1\over^2},\dots,{\sk\over4}-{1\over2}\m$, \,and the single-point
one $\,[{\sk\over4}]=\{{\sk\over4}\}$. \,With the choice 
$\,w_1=1\m$, $\,w_{-1}=(\matrix{
_0&\hspace{-0.2cm}_i\cr^i&\hspace{-0.2cm}^0})\,$ and 
$\,b_{-1,-1}={1\over2}\sigma^3\,$ as in \cite{GR2},
the special solution $\,\CW\,$ of the cohomological equation (\ref{dwv}), 
relabeled by spins $\,j\,$ such that $\,\sk\tau=j\sigma^3$, \,has the form
\qq
\CW_{j;1}\,=\,1\,,\qquad \CW_{j;-1}\,=\,\cases{\,1\quad{\rm if}\ \,2j
\ \,\m{\rm is\ even}\,,\cr
\,\,\si\quad\,{\rm if}\ \,2j
\ \,{\rm is\ odd}\,,}
\label{Wj}
\qqq
see Appendix A. \,It induces unique (up to isomorphism) rank 1 branes 
$\,\CD'_1\,$ of the $\,SO(3)\,$ model supported by the conjugacy classes 
corresponding to the two-point orbits $\,[j]$. \,To simplify notations, 
we shall label such branes by $\,[j]$. On the other hand, there are 
two non-isomorphic choices, induced by $\,\CW_{j;-1}\,$ of Eq.\,\,(\ref{Wj}) 
and by its negative, for the branes supported by the conjugacy class 
corresponding to the single-point orbit $\,[{\sk\over4}]$. \,We shall 
label them $\,[{\sk\over4}]_{_\pm}$.
\,Since $\,w_{-1}\,|j,m\rangle=\si^{2j}\m|j,-m\rangle\m$, \,for the solution
(\ref{Wj}) the map 
\qq 
U_{j^0j}^{\,j^1}(-1)\,:\ M_{j^0_{_{-}} 
j}^{\,j^1_{_{-}}}\ \longrightarrow\ M_{j^0j}^{\,j^1}
\nonumber
\qqq
of Eq.\,\,(\ref{Uz}) takes the form 
\qq
U_{j^0j}^{\,j^1}(-1)\,\,e_{j^0_{_{-}}j}^{\,j^1_{_{-}}}\ =\ 
u_{j^0j}^{\,j^1}\,\,e_{j^0j}^{\,j^1}
\nonumber
\qqq
where
\qq
u_{j^0j}^{\,j^1}\ =\ \cases{
\,\hbox to 1.3cm{$\si^{2j}$\hfill}{\rm if}\quad2j\quad\m
\ {\rm is\ \,even}\,,\cr
\,\hbox to 1.3cm{$\si^{2j-1}$\hfill}{\rm if}\quad2j^0\quad
{\rm is\ \,even\,\ and}\quad2j^1\quad{\rm is\,\ odd}\,,\cr
\,\hbox to 1.3cm{$\si^{2j+1}$\hfill}{\rm if}\quad2j^0
\quad{\rm is\,\,\ odd\ \ and}\quad2j^1\quad{\rm is\,\ even}\,.}
\label{Ul}
\qqq
Note that $\,u_{j^0j}^{\,j^1}=u_{j^0_{_{-}}j}^{\,j^1}
=u_{j^0j}^{j^1_{_{-}}}\,$ and that $\,(u_{j^0j}^{\,j^1})^2=1$.
\vskip 0.3cm

Given operators $\,U_{j^0j}^{\,j^1}(z)\m$, \,the multiplicity spaces 
in the boundary $\,SO(3)\,$ model are obtained using relations 
(\ref{tms}), (\ref{UD}) and (\ref{pl}). \,First, for 
$\,j^0,j^1<{\sk\over4}\m$, \,the space $\,{\tilde M}'^{\,
[j^1]}_{\m[j^0]\,j}\,$ is spanned by the vectors 
$\,\,e_{j^0j}^{\,j^1}\m,\ e_{j^0_{_{-}}j}^{\,j^1}\m,\ 
e_{j^0j}^{j^1_{_{-}}}\m,\ e_{j^0_{_{-}}}^{\,j^1_{_{-}}}\,\,$
so that, with $\,N\,$ denoting  the diagonal matrix with 
the entries $\,N_{j^0j}^{\,j^1},\,N_{j^0_{_{-}}j}^{\,j^1},\,
N_{j^0_{_{-}}j}^{\,j^1},\,N_{j^0j}^{\,j^1}\m$, \,we may identify
\qq
{\tilde M}'^{\,[j^1]}_{[j^0]\,j}\ \cong\ N\,\NC^4\,.
\nonumber
\qqq
The operator $\,U_{[j^0]\,j}^{\,[j^1]}(-1)\,$ becomes 
in this representation a ${\m4\times 4\m}$ anti-diagonal matrix with 
the entries given by Eq.\,\,(\ref{Ul}). \,The multiplicity
spaces $\,M'^{\,[j^1]}_{[j^0]\,j}\,$ are spanned by the vectors
\qq
e'^{\,[j^1]}_{[j^0]\,j}\ \,=\,{_1\over^{\sqrt{2}}}\,\Big(
\,e_{j^0j}^{\,j^1}\,+\,u_{j^0j}^{\,j^1}\,\,e_{j^0_{_{-}}j}^{\,j^1_{_{-}}}
\Big)\,,\qquad f'^{\,[j^1]}_{[j^0]\,j}\ \,=\,\ 
{_1\over^{\sqrt{2}}}\,\Big(\,e_{j^0_{_{-}}j}^{\,j^1}\,+\,u_{j^0j}^{\,j^1}
\,\,e_{j^0j}^{\,j^1_{_{-}}}\Big)
\nonumber
\qqq
and their dimension is
\qq
N'^{\,[j^1]}_{[j^0]\,j}
\ =\ N_{j^0\m j}^{\,j^1}+N_{j^0_-\,j}^{j^1}\,.
\nonumber
\qqq
Next, for $\,j^0={\sk\over4}\,$ and $\,j^1<{\sk\over4}\m$, \,the space 
$\,{\tilde M}'^{\,[j^1]}_{[{_\sk\over^4}]_{_\pm}j}\,$
is spanned by the vectors $\,\,e_{\ssk\m j}^{\,j^1}\m,\ 
e_{\ssk\m j}^{j^1_{_{-}}}
\,\,$ so that we may identify
\qq
{\tilde M}'^{\,[j^1]}_{[{_\sk\over^4}]_{_\pm}j}\ \cong\ 
N_{{_\sk\over^4}\m j}^{\,j^1}\,\NC^2\,.
\nonumber
\qqq
The operator $\,U_{[j^0]\,j}^{\,[{_\sk\over^4}]_{_\pm}}(-1)\,$
is represented by an anti-diagonal $\,2\times2\m$ matrix with the entries
given by Eq.\,\,(\ref{Ul}) multiplied by $\,\pm1\,$ for the 
$\,[{_\sk\over^4}]_{_\pm}\,$ branes. The multiplicity space 
$\,M'^{\,[{_\sk\over^4}]_{_\pm}}_{[j^0]\,j}\,$ is spanned by
the vector
\qq
e'^{\,[j^1]}_{[{_\sk\over^4}]_{_\pm}j}\ \,=\,\ {_1\over^{\sqrt{2}}}\,
\Big(\,e_{{_\sk\over^4}\m j}^{\,j^1}\,\pm\,
u_{{_\sk\over^4}\m j}^{\,j^1}\,\,e_{{_\sk\over^4}\m j}^{j^1_{_{-}}}\Big)
\nonumber
\qqq
and has the dimension
\qq
N'^{\,[j^1]}_{[{_\sk\over^4}]_{_\pm}j}\ 
=\ N_{{_\sk\over^4}\m j}^{\,j^1}\,.
\nonumber
\qqq
The case $\,j_0<{_\sk\over^4}\m$, $\,j^1={_\sk\over^4}\,$ is similar
with
\qq
{\tilde M}'^{\,[{_\sk\over^4}]_{_\pm}}_{\m[j^0]\,j}\ 
\cong\ N_{j^0\m j}^{\,{_\sk\over^4}}
\,\NC^2\,,
\nonumber
\qqq
the multiplicity space $\,M'^{\,[{_\sk\over^4}]_{_\pm}}_{\m[j^0]\,j}\,$
spanned by the vector
\qq
e'^{\,[{_\sk\over^4}]_{_\pm}}_{[j^0]\,j}\ \,=\,\ {_1\over^{\sqrt{2}}}\,\Big(\,
e_{j^0j}^{\,{_\sk\over^4}}\,\pm\,
u_{j^0j}^{\,{_\sk\over^4}}\,\,e_{j^0_{_{-}}j}^{{_\sk\over^4}}\Big)
\nonumber
\qqq
and its dimension equal to
\qq
N'^{\,[{_\sk\over^4}]_{_\pm}}_{[j^0]\,j}\ =\ N_{j^0\m j}^{\,{_\sk\over^4}}\,.
\nonumber
\qqq 
Finally, for $\,j^0=j^1={_\sk\over^4}\m$, \,the space
$\,{\tilde M}'^{\,[{_\sk\over^4}]_{_\pm}}_{\m[{_\sk\over^4}]_{_\pm}j}\,$ 
is spanned by the vector $\,e_{{_\sk\over^4}\m j}^{\,{_\sk\over^4}}$.
Note that $\,N_{{_\sk\over^4}\m j}^{\,{_\sk\over^4}}\,$ is equal to $\,1\,$
for integer $\,j\,$ and to $\,0\,$ for half-integer $\,j\,$
so that $\,e_{{_\sk\over^4}\m j}^{\,{_\sk\over^4}}=0\,$ for half-integer
$\,j$. \,The operator $\,U_{[{_\sk\over^4}]_{_\pm}j}^{\,
[{_\sk\over^4}]_{_\pm}}\,$ acts as multiplication by
$\,(\pm)(\pm)(-1)^j\,$
with the signs labeling the choices of the branes. 
The multiplicity space 
$\,M'^{\,[{_\sk\over^4}]_{_\pm}}_{\m[{_\sk\over^4}]_{_\pm}j}\,$
is spanned by the vector
\qq
e'^{\,[{_\sk\over^4}]_{_\pm}}_{[{_\sk\over^4}]_{_\pm}j}
\ \,=\,\ {_1\over^2}\,\Big(\,1+(\pm)(\pm)(-1)^j\Big)\,\,
e_{{_\sk\over^4}\m j}^{\,{_\sk\over^4}}
\nonumber
\qqq
and its dimension is
\qq
N'^{\,[{_\sk\over^4}]_{_\pm}}
_{[{_\sk\over^4}]_{_\pm}\m j}\,=\,{_1\over^2}\,\Big(\,1+(\pm)(\pm)(-1)^j\Big)
\,N_{{_\sk\over^4}\m j}^{\,{_\sk\over^4}}\,.
\nonumber
\qqq
In particular, for $\,j={_1\over^2}\,$ and $\,j^0,j^1=
0,{_1\over^2},\dots,{_\sk\over^4}-{_1\over^2}\m$, one obtains:
\qq
&&N'^{\,[j^1]}_{[j^0]{_1\over^2}}\ \ =\ \cases{\,\hbox to 1cm{$1$\hfill}
{\rm if}\quad|j^0-j^1|={1\over2}\,,\cr
\,\hbox to 1cm{$0$\hfill}{\rm otherwise}\,,}\qquad
N'^{\,[{_\sk\over^4}]_{_\pm}}_{[j^0]{_1\over^2}}\ =\ \cases{\,
\hbox to 1cm{$1$\hfill}{\rm if}\quad j^0={\sk\over4}-{1\over2}\,,\cr
\,\hbox to 1cm{$0$\hfill}{\rm otherwise}\,,}\cr\cr
&&N'^{\,[j^1]}_{[{_\sk\over^4}]_{_\pm}{_1\over^2}}\ =\ \cases{\,
\hbox to 1cm{$1$\hfill}{\rm if}\quad j^1={\sk\over4}-{1\over2}\,,\cr
\,\hbox to 1cm{$0$\hfill}{\rm otherwise}\,,}\qquad\ \ 
N'^{\,[{_\sk\over^4}]_{_\pm}}_{[{_\sk\over^4}]_{_\pm}{_1\over^2}}\ =\ 0
\nonumber
\qqq
so that the matrix $\,N'_{{_1\over^2}}\,$ may be identified with the
adjacency matrix of the $\,D_{{_\sk\over^2}+2}\,$ Dynkin diagram
\cite{BPPZ}. 
\vskip 0.3cm

For the $\,SU(2)\,$ theory, the operator product coefficients are given by
the map $\,A^{\,j^1j^2j''}_{j^0j^1j\,j'}\,$ of (\ref{ope1}) such that
\qq
A^{\,j^1j^2j''}_{j^0j^1j\,j'}\ \,e_{j^0j}^{\,j^1}\otimes 
e_{j^1j'}^{\,j^2}
\ =\ 
F_{_{\hspace{-0.04cm}j^1j''}}\hspace{-0.07cm}
\Big[\matrix{_{\hspace{-0.08cm}j^0}&\hspace{-0.35cm}_{j^2}
\cr\hspace{-0.08cm}^{j}&\hspace{-0.35cm}^{j'}}\Big]
\ \,e_{j^0j''}^{\,j^2}\otimes e_{j\,j'}^{\,j''}
\nonumber
\qqq
where the entries of the fusing $F$-matrix 
$\,F_{_{\hspace{-0.04cm}j^1j''}}\hspace{-0.07cm}
\Big[\matrix{_{\hspace{-0.08cm}j^0}&\hspace{-0.35cm}_{j^2}
\cr\hspace{-0.08cm}^{j}&\hspace{-0.35cm}^{j'}}\Big]\,$
vanish unless $\,N_{j^0j}^{j^1}N_{j^1j'}^{j^2}
N_{j^0j''}^{j^2}N_{j\m j'}^{j''}=1$.
\,The symmetry (\ref{Asm}) reduces to the relation
\qq
F_{_{\hspace{-0.04cm}j^1_{_{-}}j''}}\hspace{-0.07cm}
\Big[\matrix{_{\hspace{-0.08cm}j^0_{_{-}}}&
\hspace{-0.35cm}_{j^2_{_{-}}}
\cr\hspace{-0.08cm}^{j}&\hspace{-0.35cm}^{j'}}\Big]
\ =\ (-1)^{j''-j-j'}\,F_{_{\hspace{-0.04cm}j^1j''}}\hspace{-0.07cm}
\Big[\matrix{_{\hspace{-0.08cm}j^0}&\hspace{-0.35cm}_{j^2}
\cr\hspace{-0.08cm}^{j}&\hspace{-0.35cm}^{j'}}\Big]
\nonumber
\qqq
for the fusing matrices, see \cite{BSchom}. 
For the boundary operator product of the $\,SO(3)\,$
model, see \cite{Runkel2} for a related work for the minimal models and
\cite{MSchomS} for the case of level $\,\sk\,$ divisible by $\,4\m$, 
\,one obtains using the relation (\ref{A'm}) and the identity
\qq
(-1)^{j''-j-j'}\,u_{j^0j}^{\,j^1}\,u_{j^1j'}^{\,j^2}\ =\ 
u_{j^0j''}^{\,j^2}
\nonumber
\qqq
holding whenever $\,N_{j^0j}^{j^1}N_{j^1j'}^{j^2}
N_{j^0j''}^{j^2}N_{j\m j'}^{j''}=1\m$: 
\qq
&&\hbox to 4.71cm{$\displaystyle{A'^{\,[j^1][j^2]\,j''}_{[j^0][j^1]
\,j\,j'}\ \,e'^{\,[j^1]}_{[j^0]\,j}\otimes
e'^{\,[j^2]}_{[j^1]\,j'}}$\hfill}\ =\ {_1\over^{\sqrt{2}}}\,
F_{_{\hspace{-0.04cm}j^1j''}}\hspace{-0.07cm}
\Big[\matrix{_{\hspace{-0.08cm}j^0}&\hspace{-0.35cm}_{j^2}
\cr\hspace{-0.08cm}^{j}&\hspace{-0.35cm}^{j'}}\Big]\ 
\,e'^{[j^2]}_{[j^0]\,j''}\otimes e_{j\,j'}^{\,j''}\,,\cr\cr
&&\hbox to 4.71cm{$\displaystyle{A'^{\,[j^1][j^2]\,j''}_{[j^0][j^1]
\,j\,j'}\ \,f'^{\,[j^1]}_{[j^0]\,j}\otimes
f'^{\,[j^2]}_{[j^1]\,j'}}$\hfill}\ =\ {_1\over^{\sqrt{2}}}\,
u_{j^0j}^{\,j^1}\,\,
F_{_{\hspace{-0.04cm}j^1_{_{-}}j''}}\hspace{-0.07cm}
\Big[\matrix{_{\hspace{-0.08cm}j^0}&\hspace{-0.35cm}_{j^2}
\cr\hspace{-0.08cm}^{j}&\hspace{-0.35cm}^{j'}}\Big]\ 
\,e'^{[j^2]}_{[j^0]\,j''}\otimes e_{j\,j'}^{\,j''}\,,\cr\cr
&&\hbox to 4.71cm{$\displaystyle{A'^{\,[j^1][j^2]\,j''}_{[j^0][j^1]
\,j\,j'}\ \,e'^{\,[j^1]}_{[j^0]\,j}\otimes
f'^{\,[j^2]}_{[j^1]\,j'}}$\hfill}\ =\ {_1\over^{\sqrt{2}}}\,
u_{j^0j}^{\,j^1}\,\,
F_{_{\hspace{-0.04cm}j^1_{_{-}}j''}}\hspace{-0.07cm}
\Big[\matrix{_{\hspace{-0.08cm}j^0_{_{-}}}&\hspace{-0.35cm}_{j^2}
\cr\hspace{-0.08cm}^{j}&\hspace{-0.35cm}^{j'}}\Big]\ 
\,f'^{[j^2]}_{[j^0]\,j''}\otimes e_{j\,j'}^{\,j''}\,,\cr\cr
&&\hbox to 4.71cm{$\displaystyle{A'^{\,[j^1][j^2]\,j''}_{[j^0][j^1]
\,j\,j'}\ \,f'^{\,[j^1]}_{[j^0]\,j}\otimes
e'^{\,[j^2]}_{[j^1]\,j'}}$\hfill}\ =\ {_1\over^{\sqrt{2}}}\,
F_{_{\hspace{-0.04cm}j^1j''}}\hspace{-0.07cm}
\Big[\matrix{_{\hspace{-0.08cm}j^0_{_{-}}}&\hspace{-0.35cm}_{j^2}
\cr\hspace{-0.08cm}^{j}&\hspace{-0.35cm}^{j'}}\Big]\ 
\,f'^{[j^2]}_{[j^0]\,j''}\otimes e_{j\,j'}^{\,j''}\,,\cr\cr
&&\hbox to 4.71cm{$\displaystyle{A'^{\,[j^1][j^2]\,j''}_{[\ssk]_{_{\pm}}
[j^1]\,j\,j'}\ \,e'^{\,[j^1]}_{[\ssk]_{_{\pm}}j}\otimes
e'^{\,[j^2]}_{[j^1]\,j'}}$\hfill}\ =\ {_1\over^{\sqrt{2}}}\,
F_{_{\hspace{-0.04cm}j^1j''}}\hspace{-0.07cm}
\Big[\matrix{_{\hspace{-0.08cm}\ssk}&\hspace{-0.35cm}_{j^2}
\cr\hspace{-0.08cm}^{j}&\hspace{-0.35cm}^{j'}}\Big]\ 
\,e'^{[j^2]}_{[\ssk]_{_\pm}j''}\otimes e_{j\,j'}^{\,j''}\,,\cr\cr
&&\hbox to 4.71cm{$\displaystyle{A'^{\,[j^1][j^2]\,j''}_{[\ssk]_{_{\pm}}
[j^1]\,j\,j'}\ \,e'^{\,[j^1]}_{[\ssk]_{_{\pm}}j}\otimes
f'^{\,[j^2]}_{[j^1]\,j'}}$\hfill}\ =\ \pm{_1\over^{\sqrt{2}}}\,
u_{\ssk\m j}^{j^1}\,\,F_{_{\hspace{-0.04cm}j^1_{_{-}}j''}}\hspace{-0.07cm}
\Big[\matrix{_{\hspace{-0.08cm}\ssk}&\hspace{-0.35cm}_{j^2}
\cr\hspace{-0.08cm}^{j}&\hspace{-0.35cm}^{j'}}\Big]\ 
\,e'^{[j^2]}_{[\ssk]_{_\pm}j''}\otimes e_{j\,j'}^{\,j''}\,,\cr\cr
&&\hbox to 4.71cm{$\displaystyle{A'^{\,[\ssk]_{_\pm}[j^2]\,j''}_{[j^0]
[\ssk]_{_\pm}j\,j'}\ \,e'^{\,[\ssk]_{_\pm}}_{[j^0]\,j}\otimes
e'^{\,[j^2]}_{[\ssk]_{_\pm}j'}}$\hfill}\ =\ {_1\over^{\sqrt{2}}}\,
F_{_{\hspace{-0.04cm}\ssk\m j''}}\hspace{-0.07cm}
\Big[\matrix{_{\hspace{-0.08cm}j^0}&\hspace{-0.35cm}_{j^2}
\cr\hspace{-0.08cm}^{j}&\hspace{-0.35cm}^{j'}}\Big]\ 
\,e'^{[j^2]}_{[j^0]\,j''}\otimes e_{j\,j'}^{\,j''}\,\cr
&&\hspace{5.5cm}\pm\,
{_1\over^{\sqrt{2}}}\,u_{j^0j}^{\,\ssk}\,\,
F_{_{\hspace{-0.04cm}\ssk\m j''}}\hspace{-0.07cm}
\Big[\matrix{_{\hspace{-0.08cm}j^0_{_{-}}}&\hspace{-0.35cm}_{j^2}
\cr\hspace{-0.08cm}^{j}&\hspace{-0.35cm}^{j'}}\Big]\ 
\,f'^{[j^2]}_{[j^0]\,j''}\otimes e_{j\,j'}^{\,j''}\,,\cr\cr
&&\hbox to 4.71cm{$\displaystyle{A'^{\,[j^1][\ssk]_{_\pm}j''}_{[j^0]
[j^1]\,j\,j'}\ \,e'^{\,[j^1]}_{[j^0]\,j}\otimes
e'^{\,[\ssk]_{_\pm}}_{[j^1]\,j'}}$\hfill}\ =\ {_1\over^{\sqrt{2}}}\,
F_{_{\hspace{-0.04cm}j^1j''}}\hspace{-0.07cm}
\Big[\matrix{_{\hspace{-0.08cm}j^0}&\hspace{-0.35cm}_{\ssk}
\cr\hspace{-0.08cm}^{j}&\hspace{-0.35cm}^{j'}}\Big]\ 
\,e'^{\,[\ssk]_{_\pm}}_{[j^0]\,j''}\otimes e_{j\,j'}^{\,j''}\,,\cr\cr
&&\hbox to 4.71cm{$\displaystyle{A'^{\,[j^1][\ssk]_{_\pm}j''}_{[j^0]
[j^1]\,j\,j'}\ \,f'^{\,[j^1]}_{[j^0]\,j}\otimes
e'^{\,[\ssk]_{_\pm}}_{[j^1]\,j'}}$\hfill}\ =\ \pm\,{_1\over^{\sqrt{2}}}\,
u_{j^0j''}^{\,\ssk}\,\,
F_{_{\hspace{-0.04cm}j^1j''}}\hspace{-0.07cm}
\Big[\matrix{_{\hspace{-0.08cm}j^0_-}&\hspace{-0.35cm}_{\ssk}
\cr\hspace{-0.08cm}^{j}&\hspace{-0.35cm}^{j'}}\Big]\ 
\,e'^{\,[\ssk]_{_\pm}}_{[j^0]\,j''}\otimes e_{j\,j'}^{\,j''}\,,\cr\cr
&&\hbox to 4.71cm{$\displaystyle{A'^{\,[\ssk]_{_\pm}[j^2]
\,j''}_{[\ssk]_{_\pm}[\ssk]_{_\pm}j\,j'}\ 
\,e'^{\,[\ssk]_{_\pm}}_{[\ssk]_{_\pm}j}\otimes
e'^{\,[j^2]}_{[\ssk]_{_\pm}j'}}$\hfill}\ =\ {_1\over^2}\,
\Big(\,1+(\pm)(\pm)(-1)^j\Big)\,\cr
&&\hspace{5.5cm}\cdot\,\,F_{_{\hspace{-0.04cm}\ssk\m j''}}\hspace{-0.07cm}
\Big[\matrix{_{\hspace{-0.08cm}\ssk}&\hspace{-0.35cm}_{j^2}
\cr\hspace{-0.08cm}^{j}&\hspace{-0.35cm}^{j'}}\Big]\ 
\,e'^{\,[j^2]}_{[\ssk]_{_\pm}j''}\otimes e_{j\,j'}^{\,j''}\,,\cr\cr
&&\hbox to 4.71cm{$\displaystyle{A'^{\,[\ssk]_{_\pm}[\ssk]_{_\pm}
j''}_{[j^0][\ssk]_{_\pm}j\,j'}\ 
\,e'^{\,[\ssk]_{_\pm}}_{[j^0]\,j}\otimes
e'^{\,[\ssk]_{_\pm}}_{[\ssk]_{_\pm}j'}}$\hfill}\ 
=\ {_1\over^2}\,\Big(\,1+(\pm)(\pm)(-1)^{j'}\Big)\,\cr
&&\hspace{5.5cm}\cdot\,\,F_{_{\hspace{-0.04cm}\ssk\m j''}}\hspace{-0.07cm}
\Big[\matrix{_{\hspace{-0.08cm}j^0}&\hspace{-0.35cm}_{\ssk}
\cr\hspace{-0.08cm}^{j}&\hspace{-0.35cm}^{j'}}\Big]\ 
\,e'^{\,[\ssk]_{_\pm}}_{[j^0]\,j''}\otimes e_{j\,j'}^{\,j''}\,,\cr\cr
&&\hbox to 4.71cm{$\displaystyle{A'^{\,[j^1][\ssk]_{_\pm}
j''}_{[\ssk]_{_\pm}[j^1]\,j\,j'}\ 
\,e'^{\,[j^1]}_{[\ssk]_{_\pm}j}\otimes
e'^{\,[\ssk]_{_\pm}}_{[j^1]\,j'}}$\hfill}\ 
=\ F_{_{\hspace{-0.04cm}j^1j''}}\hspace{-0.07cm}
\Big[\matrix{_{\hspace{-0.08cm}\ssk}&\hspace{-0.35cm}_{\ssk}
\cr\hspace{-0.08cm}^{j}&\hspace{-0.35cm}^{j'}}\Big]\ 
\,e'^{\,[\ssk]_{_\pm}}_{[\ssk]_{_\pm}j''}\otimes 
e_{j\,j'}^{\,j''}\,,\cr\cr
&&\hbox to 4.71cm{$\displaystyle{A'^{\,[\ssk]_{_\pm}[\ssk]_{_\pm}
j''}_{[\ssk]_{_\pm}[\ssk]_{_\pm}j\,j'}\ 
\,e'^{\,[\ssk]_{_\pm}}_{[\ssk]_{_\pm}j}\otimes
e'^{\,[\ssk]_{_\pm}}_{[\ssk]_{_\pm}j'}}$\hfill}\ 
=\ {_1\over^2}\,\Big(\,1+(\pm)(\pm)(-1)^{j}\Big)\,\cr
&&\hspace{5.5cm}\cdot\,\,F_{_{\hspace{-0.04cm}\ssk\m j''}}\hspace{-0.07cm}
\Big[\matrix{_{\hspace{-0.08cm}\ssk}&\hspace{-0.35cm}_{\ssk}
\cr\hspace{-0.08cm}^{j}&\hspace{-0.35cm}^{j'}}\Big]\ 
\,e'^{\,[\ssk]_{_\pm}}_{[\ssk]_{_\pm}j''}\otimes e_{j\,j'}^{\,j''}\,.
\nonumber
\qqq
\vfill
\eject

\end{document}